\documentclass[letterpaper,twocolumn,10pt]{article}

\usepackage{amsmath}
\usepackage{url}
\usepackage{booktabs}
\usepackage{graphicx}
\usepackage{float}
\usepackage{subfigure}
\usepackage{wrapfig}
\usepackage{picinpar}
\usepackage{cutwin}
\usepackage{multirow}
\usepackage{multicol}
\usepackage{makecell}
\usepackage{subcaption}
\usepackage{mathptmx}
\usepackage[T1]{fontenc}
\usepackage[utf8]{inputenc}
\usepackage{pslatex}
\usepackage{cite}
\usepackage{xcolor}
\usepackage{listings}
\usepackage[pagebackref=true,breaklinks=true,colorlinks,bookmarks=false]{hyperref}
\usepackage[labelfont={small,bf},textfont={small,bf}]{caption}
\usepackage{balance}
\usepackage
[
    a4paper, 
    left=1.5cm,
    right=1.5cm,
    top=2.5cm,
    bottom=2.5cm,
    columnsep=1cm,
]{geometry}
\usepackage{titlesec}
\usepackage[hang,flushmargin]{footmisc}
\usepackage{titling}

\newcommand{\mypara}[1]{\noindent\textbf{#1:}}

\newcommand{\decoder}{\mathcal{E}^{-1}}
\newcommand{\singleencoder}{\mathcal{E}}
\newcommand{\nlpencoder}{\mathcal{E}_{1}}
\newcommand{\cvencoder}{\mathcal{E}_{2}}
\newcommand{\nlpextractor}{\mathcal{F}_{\mathit{NLP}}}
\newcommand{\cvextractor}{\mathcal{F}_{\mathit{CV}}}
\newcommand{\linear}{\mathcal{F}_{l}}
\newcommand{\hijackingdataset}{\mathcal{D}_{h}}
\newcommand{\originaldataset}{\mathcal{D}_{o}}
\newcommand{\fuseddataset}{\mathcal{D}_{f}}
\newcommand{\containerdataset}{\mathcal{D}_{c}}
\newcommand{\adapter}{\mathcal{A}}

\newcommand{\visualloss}{\mathcal{L}_{v}}
\newcommand{\semanticloss}{\mathcal{L}_{s}}
\newcommand{\hijackingsample}{{x}_{h}}

\newcommand{\fusedsample}{{x}_{f}}
\newcommand{\containersample}{{x}_{c}}
\newcommand{\parameters}{\theta}

\hypersetup{
  colorlinks,
  linkcolor={blue!70!green},
  citecolor={green!70!blue},
  urlcolor={orange!70!red}
}
\let\oldbibliography\thebibliography
\renewcommand{\thebibliography}[1]{%
  \oldbibliography{#1}%
  \setlength{\itemsep}{1pt}%
}

\titleformat*{\section}{\large\bfseries}
\titleformat*{\subsection}{\large\bfseries}
\titleformat*{\subsubsection}{\large\bfseries}
\titlespacing*{\section}{0cm}{*3}{0.2cm}
\titlespacing*{\subsection}{0cm}{*2}{0.15cm}
\begin{document}

\date{}

\title{Vera Verto: Multimodal Hijacking Attack}

\author{
Minxing Zhang\textsuperscript{1}\ \ \
Ahmed Salem\textsuperscript{2}\ \ \
Michael Backes\textsuperscript{1}\ \ \
Yang Zhang\textsuperscript{1}
\\
\\
\textsuperscript{1}\textit{CISPA Helmholtz Center for Information Security} \ \ \ 
\textsuperscript{2}\textit{Microsoft}
}

\maketitle

\begin{abstract}
The increasing cost of training machine learning (ML) models has led to the inclusion of new parties to the training pipeline, such as users who contribute training data and companies that provide computing resources.
This involvement of such new parties in the ML training process has introduced new attack surfaces for an adversary to exploit.
A recent attack in this domain is the model hijacking attack, whereby an adversary hijacks a victim model to implement their own -- possibly malicious -- hijacking tasks.
However, the scope of the model hijacking attack is so far limited to the homogeneous-modality tasks.
In this paper, we transform the model hijacking attack into a more general \emph{multimodal} setting, where the hijacking and original tasks are performed on data of different modalities.
Specifically, we focus on the setting where an adversary implements a natural language processing (NLP) hijacking task into an image classification model.
To mount the attack, we propose a novel encoder-decoder based framework, namely the Blender, which relies on advanced image and language models.
Experimental results show that our \emph{modal hijacking attack} achieves strong performances in different settings.
For instance, our attack achieves $94\%$, $94\%$, and $95\%$ attack success rate when using the Sogou news dataset to hijack STL10, CIFAR-10, and MNIST classifiers.
Our code is available \href{https://github.com/minxingzhang/ModalHijacking}{here}.\footnote{Vera Verto is a magical transformation incantation used in the fictional wizarding world of J.K. Rowling's Harry Potter series, which transforms NLP sentences into CV images in the paper.}
\end{abstract}

\section{Introduction}
\label{section: introduction}

As a critical component of various applications, machine learning (ML) models have been increasingly expensive to train.
Hence, the training of ML models has transformed gradually into a joint process, e.g., new parties are included in the training of the model either by providing data or computational resources.
However, the involvement of these new parties has created new attack surfaces against ML models, e.g., poisoning and backdoor attacks~\cite{SHNSSDG18,CLLLS17}.

Recently, the model hijacking attack has been proposed by Salem et al. in this domain~\cite{SBZ22}, where the adversary is able to stealthily implement their own hijacking task into a target victim model.
Concretely, assuming the same ability as in poisoning attacks, the adversary first camouflages their own hijacking dataset for stealthiness to have a similar visual appearance as the target model's training dataset.
Then the adversary poisons the training dataset of the target model with their camouflaged hijacking dataset.
This attack could induce two different risks.
The first one is about accountability which is the main threat for hijacking attacks, where the model owner can be framed by the adversary to perform illegal or unethical tasks without knowing.
For instance, a completely different classification of some unethical tasks could be mounted in a benign classifier.
The second one is parasitic computing, where the model owner pays the model maintenance costs, while the adversary uses/offers it for their own application/service for free.
For instance, one NVIDIA V100 GPU chip costs $\$0.992-\$2.48$ per hour on the Google Cloud Platform.\footnote{\url{https://cloud.google.com/compute/all-pricing\#gpus}}
On the other hand, the model hijacking technique can also be adapted to compress models, i.e., training a single model for multiple tasks.

\subsection{Our Contributions}
\label{subsection: our contributions}

\mypara{Motivation}
The model hijacking attack mandates that the hijacking and original tasks have the same modality, which limits the applicable domains to computer vision (CV) related tasks.
However, ML has achieved great success in many domains, e.g., the multiple available translators such as DeepL and Google Translate, and the different face detectors on social media platforms.
Hence, relaxing this assumption will significantly increase the risks of the model hijacking attack, as the adversary can now target models with different modalities, i.e., more target models exist for the adversary to perform their attack.

\begin{figure*}[!t]
\centering
\includegraphics[width=1.3\columnwidth]{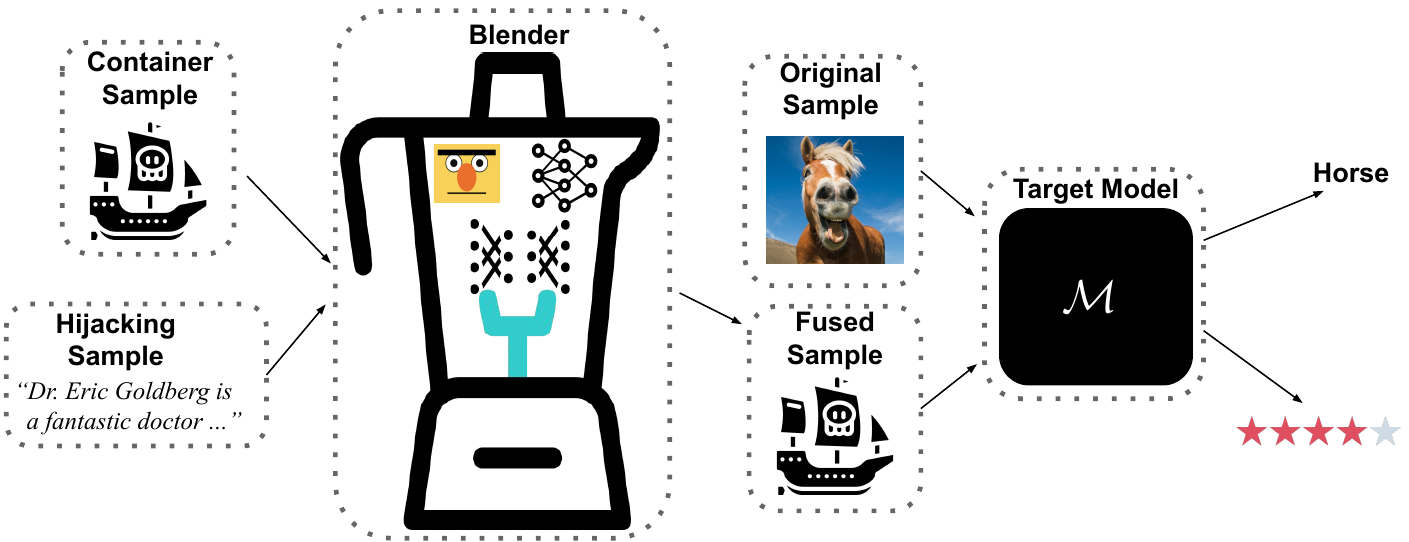}
\caption{An overview of the multimodal hijacking attack.
First, the Blender takes a sample from both the hijacking and container datasets.
It then mixes both of these inputs to have a fused image with the looks of the container one but with the features of the hijacking text input.
The model is able to perform the original classification task (classifying the image as a horse) and the hijacking one, i.e., classifying the fused image as 4-star (the label of the hijacking input).}
\label{figure:overview}
\end{figure*}

In this paper, therefore, we transform the model hijacking attack into a more general \emph{multimodal} setting, i.e., implementing a hijacking task in a completely different domain.
More concretely, the adversary can implement an NLP hijacking task by a CV target model, as illustrated in \autoref{figure:overview}.
For short, we refer to our attack as the \emph{modal hijacking attack}.
The goal of our modal hijacking attack can be summarized as two aspects:
(a) The effectiveness corresponds to the successful implementations of hijacking attacks in victim models;
(b) The stealthiness corresponds to victim models not realizing that they are hijacked.

Our modal hijacking attack follows the same assumption as the attacks requiring poisoning~\cite{JOBLNL18,SHNSSDG18,STLLXCS18,SBZ22}, i.e., the adversary is only able to poison the training dataset without any access to the target model's architecture or hyperparameters.
Our modal hijacking attack would increase the risks of accountability and parasitic computing.
Concretely, our attack could pose a more severe threat to accountability since multiple modalities are involved instead of a single one.
For instance, by model hijacking attacks, a benign classifier would be forced to claim some unethical statements are valid, where the scope of feasible hijacking tasks is extended.
Besides, the threat of parasitic computing can be more significant when NLP tasks are involved.
This is because the establishment of NLP models normally requires an enormous training budget.\footnote{\url{https://research.aimultiple.com/large-language-model-training/}}
And the risk of parasitic computing will get further increased when our Blender (introduced in the following soon) is reusable.

\mypara{Modal Hijacking Attack}
To perform the modal hijacking attack, the adversary needs to transform the NLP-based \emph{hijacking dataset} into the victim model's CV-based \emph{original dataset}.
To this end, we propose the Blender, a novel encoder-decoder-based model which integrates a language model, i.e., BERT~\cite{DCLT19} and multiple CNN models.
The Blender integrates two losses, i.e., visual and semantic losses, 
to fuse both the hijacking and original inputs.
Then the crafted sample has a similar visual appearance to the original inputs but maintains the semantic features of the hijacking one, as shown in \autoref{figure:overview}.
A successful modal hijacking attack should enable the target victim model to preserve its utility, i.e., has the same performance on the original CV task (``Stealthiness''), while performing the hijacking NLP task with high accuracy (``Effectiveness'').

Different from the previous work, our modal hijacking attack expands the scope from a single type of data to a multimodal setting.
The challenge that stems from the change is the needed transformation from a discrete domain (NLP) to a continuous one (CV), in which there exists a comprehension gap.
We believe this transformation is not trivial, as demonstrated in \autoref{subsection: hyperparameters/design decisions}.
Meanwhile, our modal hijacking attack is more general, i.e., the Blender can be applied with different hijacking and original datasets.
Besides, the reusability property provides a cheaper option for the adversary to hijack target models as shown later in \autoref{subsection: hyperparameters/design decisions}.
To the best of our knowledge, the modal hijacking attack is the first work to combine different data modalities, which increases the capability and flexibility for hijacking attacks.
Moreover, this work can encourage the exploration of the applicability of performing hijacking attacks for different modalities (which can also be with the benign aim of model compression).

\mypara{Evaluation}
To evaluate our modal hijacking attack, we use two NLP datasets \cite{ZZL15}, namely Yelp Review (Yelp) and Sogou News (Sogou), and three CV datasets, i.e., MNIST~\cite{MNIST}, CIFAR-10~\cite{CIFAR}, and STL-10~\cite{CNL11}.
We extensively evaluate the different setups for our modal hijacking attack.
Our results show that our modal hijacking attack can achieve strong performances with respect to both the attack success rate and the victim's model utility.
For instance, when victim models trained on MNIST, CIFAR-10, and STL-10 datasets are hijacked by the Yelp (Sogou) datasets, our modal hijacking attack achieves an attack success rate of $65\%$ ($94\%$), $68\%$ ($94\%$), and $65\%$ ($95\%$), respectively.
Meanwhile, the victim models' utility is not jeopardized, i.e., our modal hijacking achieves the utility of $99\%$ ($99\%$), $93\%$ ($93\%$), and $93\%$ ($92\%$), respectively, which is less than $2\%$ drop compared to clean models.
Moreover, we show the generalizability of our modal hijacking attack by evaluating it against different setups, e.g., different models to construct the Blender and target model.
Finally, we involve two possible defenses to evaluate the modal hijacking attack.

In addition, we discuss the advantages of our work from the aspects of practical usage, stealthiness, and computational efficiency in \autoref{section: discussion} for a better understanding.

\section{Background and Related Works}
\label{section: background and related works}

\mypara{Training Time Attacks}
The inclusion of new parties in the training pipeline for ML has induced a new attack vector.
Adversaries can utilize this to interfere with the training of the victim model.
These attacks are referred to as training time attacks.
One of the most famous training time attacks is data poisoning~\cite{STLLXCS18, SHNSSDG18, BG192, TTGL20, ZLDG20, SSTS20, CT21}.
This class of attacks allows the adversary to insert malicious samples into the victim model's training data.
One widespread target for the poisoning attacks is to jeopardize the model's utility, i.e., to make the training of the victim model fail, which is different from our modal hijacking attack.

Another related training time attack is the backdoor attack~\cite{YLZZ19, SSP20, ZMZBCJ20, CSBMSWZ21, SSBHZ20}.
In this attack, the adversary takes a further step and tries to associate a malicious behavior -- of the target model -- with a trigger, e.g., a white square on the corner of the input.
A successful backdoor attack results in a victim model that behaves benignly on clean inputs; while predicting a specific label when queried by backdoored data, i.e., inputs with triggers, but triggers are not as flexible as our modal hijacking attack.

\mypara{Testing Time Attack}
A similar attack is adversarial reprogramming~\cite{EGS19}.
In this attack, the adversary also tries to perform their own task using a victim model.
However, this is a testing time attack, i.e., the adversary only accesses the model after its training.
Unlike our hijacking attack, this attack requires assumptions on the target model such as white-box access.

\section{Modal Hijacking Attack}
\label{section: modal hijacking attack}

\subsection{Threat Model}
\label{subsection: threat model}

\mypara{Adversary's Goals}
The main goals of our attack could be summarized as:
(a) The effectiveness guarantees that victim models can successfully conduct our hijacking tasks, which will be evaluated by the metric of attack success rate;
(b) The stealthiness ensures that victim models will not realize the existence of hijacking attacks, which will be measured from multi-faceted aspects, e.g., the metric of utility, the visual appearances of fused images, and the poisoning rate.
The results and analyses in \autoref{subsection: hyperparameters/design decisions} and \autoref{section: discussion} show the superiority of our modal hijacking attack.

In this paper, we generalize hijacking attacks to a multi-modal setting, which represents a more practical scenario in the real world.
It might be hard for the adversary to find a victim model of the exact same task domain.
Thus, the previous model hijacking attack, which only focuses on the same modality, limits the applications of hijacking attacks.
However, the transformation to a multi-modal setting is challenging.
Specifically, NLP tasks are located in a discrete domain while CV tasks are in a continuous one.
In that case, how to understand the discrete information in a continuous form is not trivial.
To address this challenge, we propose the modal hijacking attack which enables the adversary to hijack the victim model of CV tasks by NLP tasks, enhancing the applicable scope and practical usage of hijacking attacks.

\mypara{Adversary's Knowledge}
We follow the same assumption as that widely used in poisoning~\cite{JOBLNL18,SHNSSDG18,STLLXCS18,TTGL20} and backdoor attacks~\cite{YLZZ19,SSP20,CSBMSWZ21,SWBMZ22}, i.e., we only assume the ability to poison the training dataset of the target model.
Besides the involvement of users who contribute training data and companies that provide computing resources, another possible approach under this assumption is intrusion techniques~\cite{LLLT13,KGVK19}.
However, when aiming at a malicious goal (e.g. frame-up or parasitic computing), no matter in which cases, fewer poisons indicate a more practical scenario and also more stealthiness, which will be further discussed in \autoref{subsection: hyperparameters/design decisions}.
On the other hand, when a model owner conducts model compression, the target model's training data is of course accessible.
In short, our modal hijacking attack does not require any extra information about the target model architecture or hyperparameters.

Moreover, we assume the adversary to have a container image dataset to fuse with the hijacking one.
This container dataset does not have to follow the same distribution as the original training dataset of the victim model.
However, the adversary can construct it relying on their preference for the visual appearance of the Blender's outputs, i.e., the fused dataset.
In addition, as victim models are used to perform the hijacking task, our modal hijacking attack assumes that the number of labels of the original dataset is at least equal to the hijacking dataset's one.

\subsection{Dataset Terminologies}
\label{section:datasets}

The modal hijacking attack uses four different datasets which are defined now for clarity:
First, the \emph{Original Dataset ($\originaldataset$)} is the victim model's training dataset for training the original task;
Second, the \emph{Hijacking Dataset ($\hijackingdataset$)} is the adversary's training dataset for training the hijacking task;
Third, the \emph{Container Dataset ($\containerdataset$)} is a set of images the adversary constructs/collects to fuse with the hijacking dataset samples;
Finally, the \emph{Fused Dataset ($\fuseddataset$)} is the container dataset after being fused with the hijacking one.

\subsection{Blender Design}
\label{subsection: blender design}

In general, the Blender aims to generate a fused dataset, which is used to hijack the victim model.
This is performed by fusing the hijacking NLP sentences with the container CV images.
We first present the design of our Blender, then how to deploy it.

\mypara{Design}
To fuse the text-hijacking dataset with an image container one, we first extract the hijacking dataset's features, following state-of-the-art works by using a language model~\cite{DCLT19,SQXH19}.
Then to construct the fused dataset, we first try the naive approach of building an adapter, which is a CNN, to resize the NLP features to the size of the victim model's input.
However, this approach does not perform well for some datasets as shown later in \autoref{subsection: hyperparameters/design decisions}.
Moreover, using this naive approach results in random-looking images as illustrated in \autoref{figure:adapterimages}, which can be easily detected.
To circumvent the limitations of the naive approach, we propose our encoder-decoder-based model, i.e., Blender.
More concretely, the Blender consists of an NLP feature extractor $\nlpextractor$, i.e., a language model, an adapter $\adapter$, two encoders {$\nlpencoder$} and {$\cvencoder$}, and a decoder {$\decoder$}.
Further, to improve the computational efficiency, we relax the requirement from using double encoders to a single one {$\singleencoder$}.

Another design decision we make is to use the complete embeddings of the hijacking sentence instead of only the last ``[cls]'' token.
As shown later in \autoref{subsection: hyperparameters/design decisions}, using all of the embeddings significantly improves the performance of the modal hijacking attack.

Our final design decision is to fine-tune the NLP feature extractor using the hijacking dataset before deployment.
This will allow the Blender to better understand the hijacking dataset, thus improving its ability to extract semantic information and integrate it into container images.
The performance gains of this step are evaluated in \autoref{subsection: hyperparameters/design decisions}.
It's worth noting that this step does not require any additional assumptions as the adversary, who owns the hijacking dataset, can control the NLP feature extractor independent of the target victim model.
Additionally, the cost of fine-tuning is not a major concern for the adversary, as it can be mitigated by reusing the Blender as demonstrated in \autoref{subsection: hyperparameters/design decisions}.

\mypara{Operation}
We first explain how double-encoder Blender operates.
The Blender uses the NLP feature extractor $\nlpextractor$ to extract the features/embeddings of the text hijacking input $\hijackingsample \in \hijackingdataset$.
Next, these features are input to the adapter $\adapter$ to preprocess before being input to the first encoder {$\nlpencoder$}.
In parallel, the container image $\containersample \in \containerdataset$ is input to the second encoder {$\cvencoder$}.
Next, the outputs of both encoders are concatenated together and input to the decoder {$\decoder$}.
Finally, the decoder constructs the output fused image $\fusedsample$, which has the visual appearance of the container image $\containersample$, while having the features of the text one $\hijackingsample$.
More formally,
\[\decoder\Big(\nlpencoder\Big(\adapter\big(\nlpextractor(\hijackingsample)\big)\Big) \Big|\Big| \cvencoder(\containersample)\Big) = \fusedsample,\]
where $\Big|\Big|$ is the concatenation operator.

In the single-encoder setting, the container image $\containersample$ is directly concatenated with the adapter result $\adapter(\cdot)$, which is then sent to the encoder $\singleencoder$.
This process is formulated as
\[\decoder\Bigg(\singleencoder\Big(\adapter\big(\nlpextractor(\hijackingsample)\big)\Big|\Big| \containersample\Big) \Bigg) = \fusedsample,\]

\mypara{Training}
To train the Blender, we use two losses, namely the visual and semantic losses.

\emph{Visual Loss:}
The visual loss ($\visualloss$) is responsible for forcing the fused image to have a similar look to the container one.
To accomplish this, we utilize the mean squared error (MSE) to measure the pixel-wise difference between the fused and container inputs, i.e., $\visualloss = ||\fusedsample - \containersample||_2^2$.

\emph{Semantic Loss:}
The semantic loss $\semanticloss$ is designed to fuse the NLP features in the container image.
Similar to the visual loss, we use MSE for the semantic loss too.
However, the MSE here is calculated between the features extracted from the text input and the ones extracted from the fused image.
Feature extraction here is performed with different feature extractors according to the input type, i.e., the text/image feature extractor ($\nlpextractor/\cvextractor$) is used to extract the features of $\hijackingsample$/$\fusedsample$.
Since the MSE expects the same sizes for both inputs, we further process the adapter's output with a linear layer ($\linear$) to adjust its size to match the CV ones.
More formally,
$$\semanticloss = ||\linear(\adapter(\nlpextractor(\hijackingsample))) - \cvextractor(\fusedsample)||_2^2.$$

The Blender is trained as the following 4 steps:

\textbf{(1)} The adversary first constructs their container dataset $\containerdataset$.
The only requirement for this dataset is to be an image dataset.
Ideally, this dataset has a similar visual appearance as the original dataset $\originaldataset$, to make the modal hijacking attack more stealthy.
However, the adversary can construct this dataset as they desire.

\textbf{(2)} Second, every sentence in the hijacking dataset is randomly mapped to an image in the container dataset.
This mapping does not have to be one-to-one, i.e., a single container image can be associated with multiple sentences.

\textbf{(3)} Next, (sentence, image) pairs are processed by our Blender as previously presented; both semantic and visual losses are calculated.

\textbf{(4)} Finally, both losses are added and the Blender is updated accordingly, i.e., $\parameters = \mathrm{argmin}_{\parameters}(\visualloss+\semanticloss),$ where $\parameters$ is the parameters of the Blender and the linear layer $\linear$.

It is important to mention that the CV feature extractor $\cvextractor$ and the linear layer $\linear$ are only needed when training the Blender, then they can be discarded.

\subsection{Attack Deployment}
\label{subsection: attack deployment}

Our modal hijacking attack is executed in two phases.

\mypara{Poisoning the Victim Model}
The adversary starts by training the Blender as previously presented in \autoref{subsection: blender design}.
Next, they use the Blender to create the fused dataset, i.e., by fusing the container and hijacking datasets.
They then perform a label mapping between the original and hijacking datasets to decide the labels of the fused dataset, i.e., by unidirectionally one-to-one mapping the hijacking samples' labels to the fused dataset's ones.
Finally, the fused dataset with its labels is used to poison the victim model.

\mypara{Inference Time}
After training the victim model, the adversary executes the modal hijacking attack on a target hijacking input $\hijackingsample$ as presented in \autoref{figure:overview}.
As the figure shows, the adversary first samples a container image and then uses the trained Blender to fuse it with the hijacking input and create the fused image.
The adversary then queries the fused image to the victim model and receives the output label.
Finally, they map the received label back to its corresponding one in the hijacking dataset.

\subsection{Advantages}
\label{subsection: advantages}

There are 3 main advantages of our multimodal hijacking attack over the previous work~\cite{SBZ22}:

\vspace{0.1cm}
\noindent
- Hijacking attacks are generalized from homogeneous to heterogeneous modalities, which improves practical usage.

\vspace{0.1cm}
\noindent
- A single encoder in our Blender can support two completely different modalities in our attack, which largely improves the computational efficiency.

\vspace{0.1cm}
\noindent
- Our poisons (i.e., fused images) look much more natural, where the hijacking content is not exposed.
This advantage enhances our attack's stealthiness.

\section{Evaluation}
\label{section: evaluation}

\subsection{Dataset Description}
\label{subsection: dataset description}

\mypara{CV Datasets}
We use three commonly-used benchmark datasets in our evaluation, i.e., MNIST, CIFAR-10, and STL-10.
\emph{MNIST} is a handwriting digits datasets, which contains 70,000 $28\times28$ gray-scale images;
\emph{CIFAR-10} is a real-world objects dataset, which contains 60,000 $32\times32$ color images;
Finally, \emph{STL-10} is also a real-world objects dataset, which has some common (e.g., airplane, cat, and dog) classes with CIFAR-10.
We use the labeled subset of STL-10 that consists of 13,000 $96\times96$ color images.

\mypara{NLP Datasets}
We use two well-established NLP datasets, i.e., Yelp Review (Yelp) and Sogou News (Sogou).
\emph{Yelp} is a dataset of English reviews with labels corresponding to scores (between 1 and 5).
It includes 650,000 training and 50,000 testing samples.
\emph{Sogou} is a dataset of news articles with labels associated with five categories, i.e., sports, finance, entertainment, automobile, and technology.
It includes 90,000 training and 12,000 testing samples.

\subsection{Model Architectures}
\label{subsection: model architectures}

\mypara{Blender}
To recap, our Blender is composed of an NLP feature extractor, an adapter, two encoders/a single encoder, and a decoder.
The architectures of these components are presented as below: 

\emph{NLP feature extractor.}
We use the Bidirectional Encoder Representations from Transformers (BERT) language model~\cite{DCLT19}, and fine-tune it.
Then we discard the last layer, and use the embeddings of each token as our NLP features (as previously described in~\autoref{subsection: blender design}).
We also try different language models and present the results in \autoref{subsection: hyperparameters/design decisions}.

\emph{Adapter.}
The adapter is composed of an average pooling layer, and 4 convolutional ones.

\emph{Encoder(s).}
We use the same architecture for both double- and single-encoder settings.
Each encoder consists of four convolutional layers with batch normalization and $\mathrm{ReLU}$ activation function.

\emph{Decoder.}
The decoder consists of four convolutional transpose layers.
The first three use batch normalization and $\mathrm{ReLU}$ activation function, while the fourth one only uses a $\mathrm{tanh}$ activation function.

\mypara{CV Feature Extractor}
We adopt VGG11~\cite{SZ15} as our CV feature extractor and use its output as the features.
We evaluate the attack performances when using other CV feature extractors in \autoref{subsection: hyperparameters/design decisions}.

\mypara{Victim/Target Model}
We use the ResNet18~\cite{HZRS16} model for our evaluation and show the generalizability of our modal hijacking attack with different target models in \autoref{subsection: hyperparameters/design decisions}.

\subsection{Evaluation Metrics}
\label{subsection: evaluation metrics}

\mypara{Utility}
We use utility to measure the performance of the victim model's original task.
To this end, we use a clean testing dataset to evaluate its performance on the clean and the hijacked victim models, where the clean testing dataset is the testing dataset of the original CV model, i.e., from the same distribution as the original dataset but not used when training.
The closer the performances of these two models on the clean test dataset are, the better the utility is.

\mypara{Attack Success Rate}
We use attack success rate (ASR) to measure the effectiveness of the modal hijacking attack, i.e., the performance of the hijacking task on the victim model.
We first create a hijacking test dataset by fusing a clean testing dataset (for the hijacking task) with the container dataset using the Blender.
Next, we compute the accuracy of the victim model on that hijacking test dataset.
Moreover, we train an NLP classification model for the hijacking task to compare its performance with the victim model on the hijacking task, where the data and labels of the hijacking dataset are not changed.
This model represents the case where the adversary wants to perform the hijacking task without the modal hijacking attack, i.e., the upper bound of the attack.
The closer the performance between our modal hijacking attack and this model is, the better the modal hijacking attack is.

\subsection{Results}
\label{subsection: results}

\begin{figure}[!t]
\centering
\subfigure[Attack Success Rate.]{
\label{figure:experiment_asr}
\includegraphics[width=0.41\textwidth]{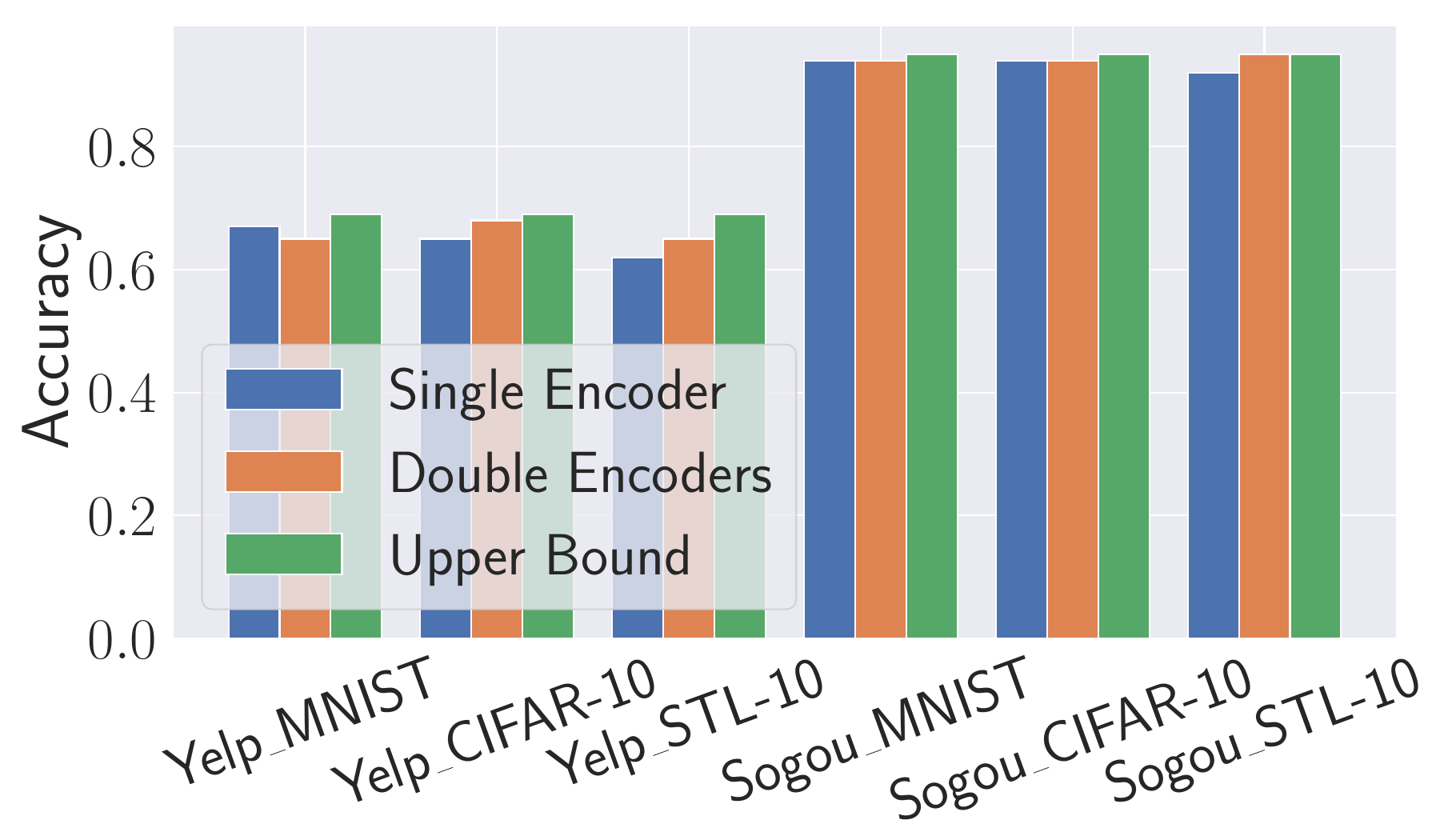}}
\subfigure[Utility.]{
\label{figure:experiment_utility}
\includegraphics[width=0.41\textwidth]{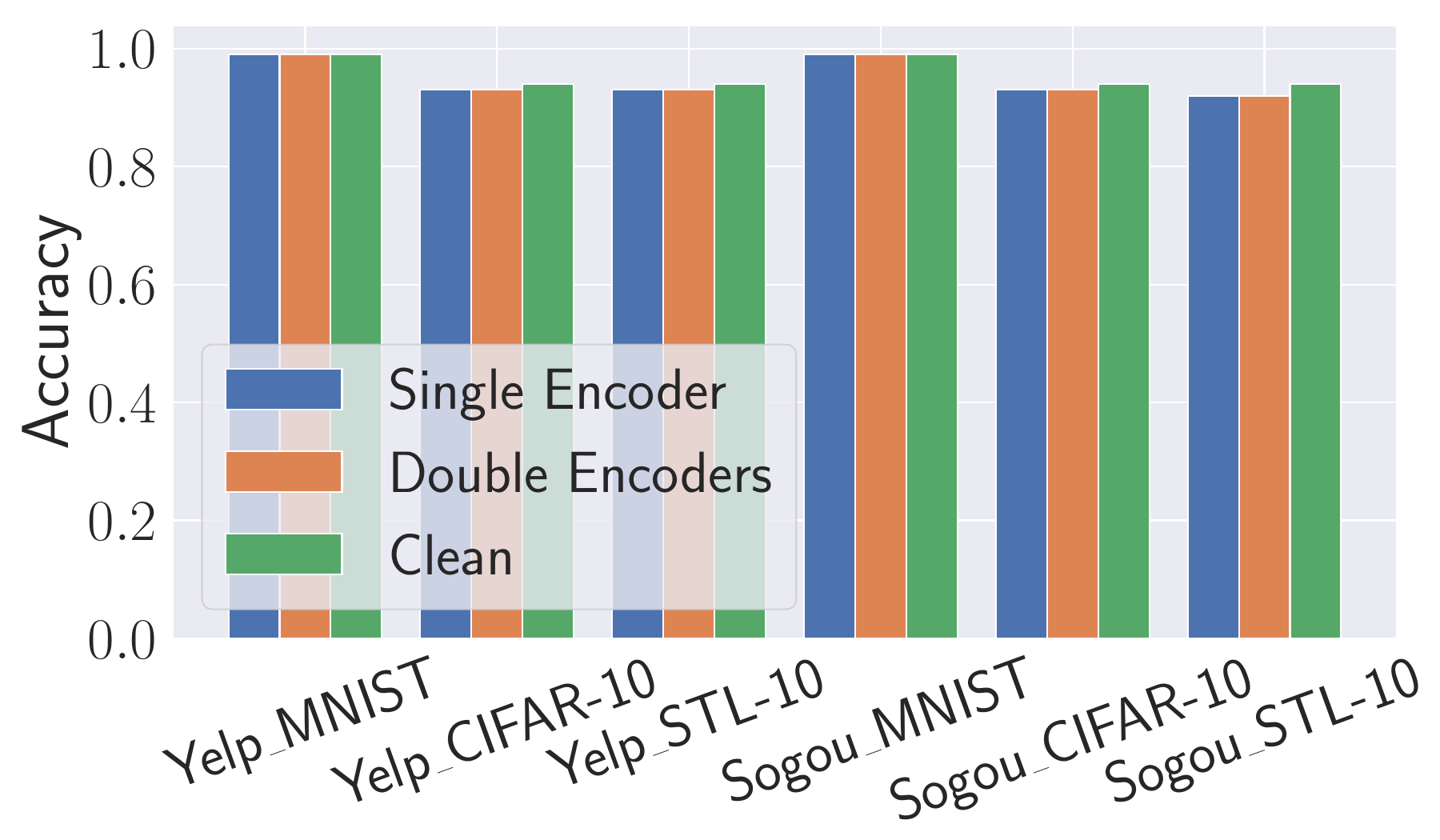}}
\caption{Our multimodal hijacking attack performance, where x\_y notation is the hijacking\_original dataset pair.}
\label{figure:experiments}
\end{figure}

\begin{figure}[!t]
\centering
\subfigure[Attack Success Rate.]{
\label{figure:tinyimagenet_asr}
\includegraphics[width=0.41\textwidth]{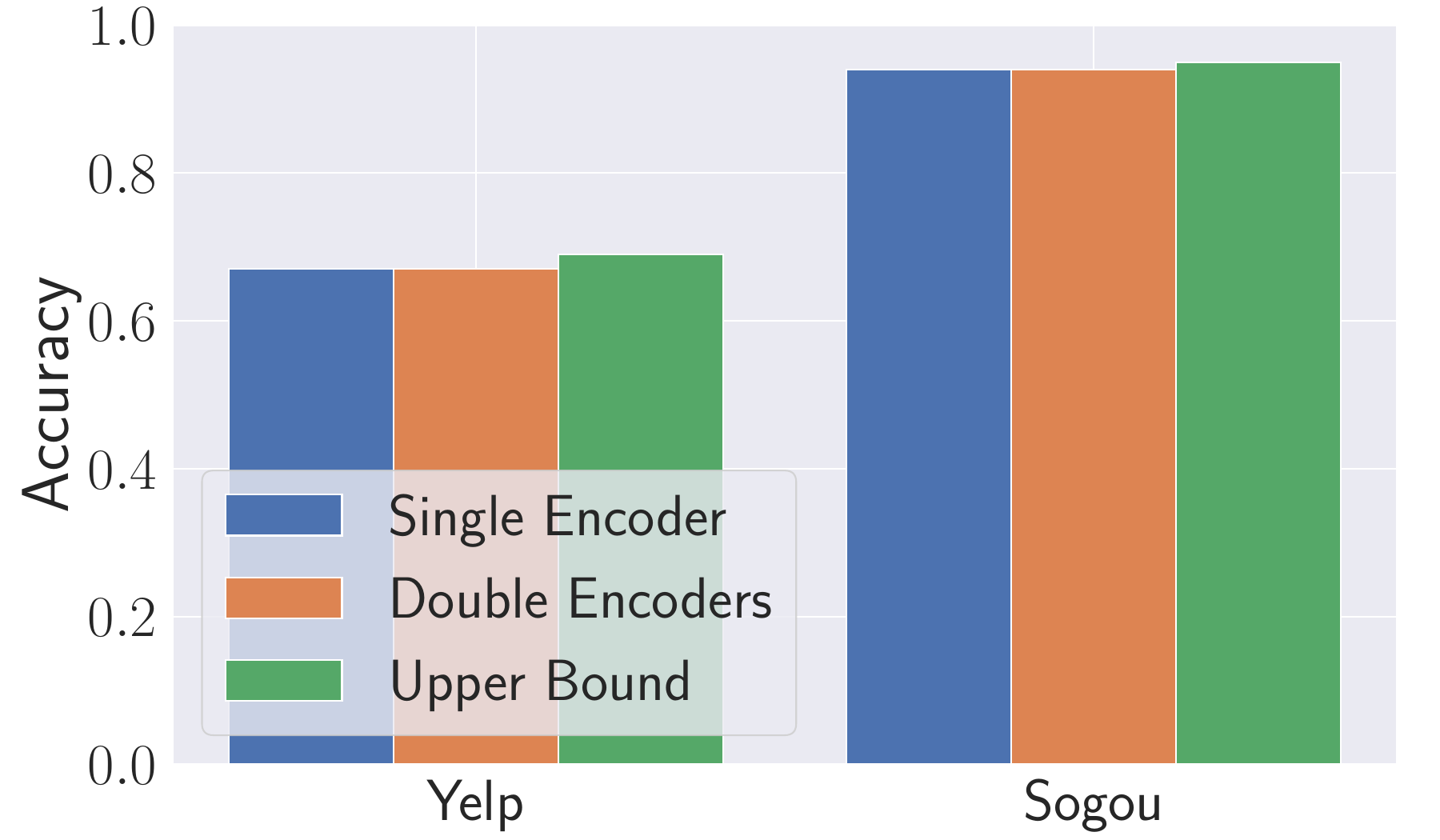}}
\subfigure[Utility.]{
\label{figure:tinyimagenet_utility}
\includegraphics[width=0.41\textwidth]{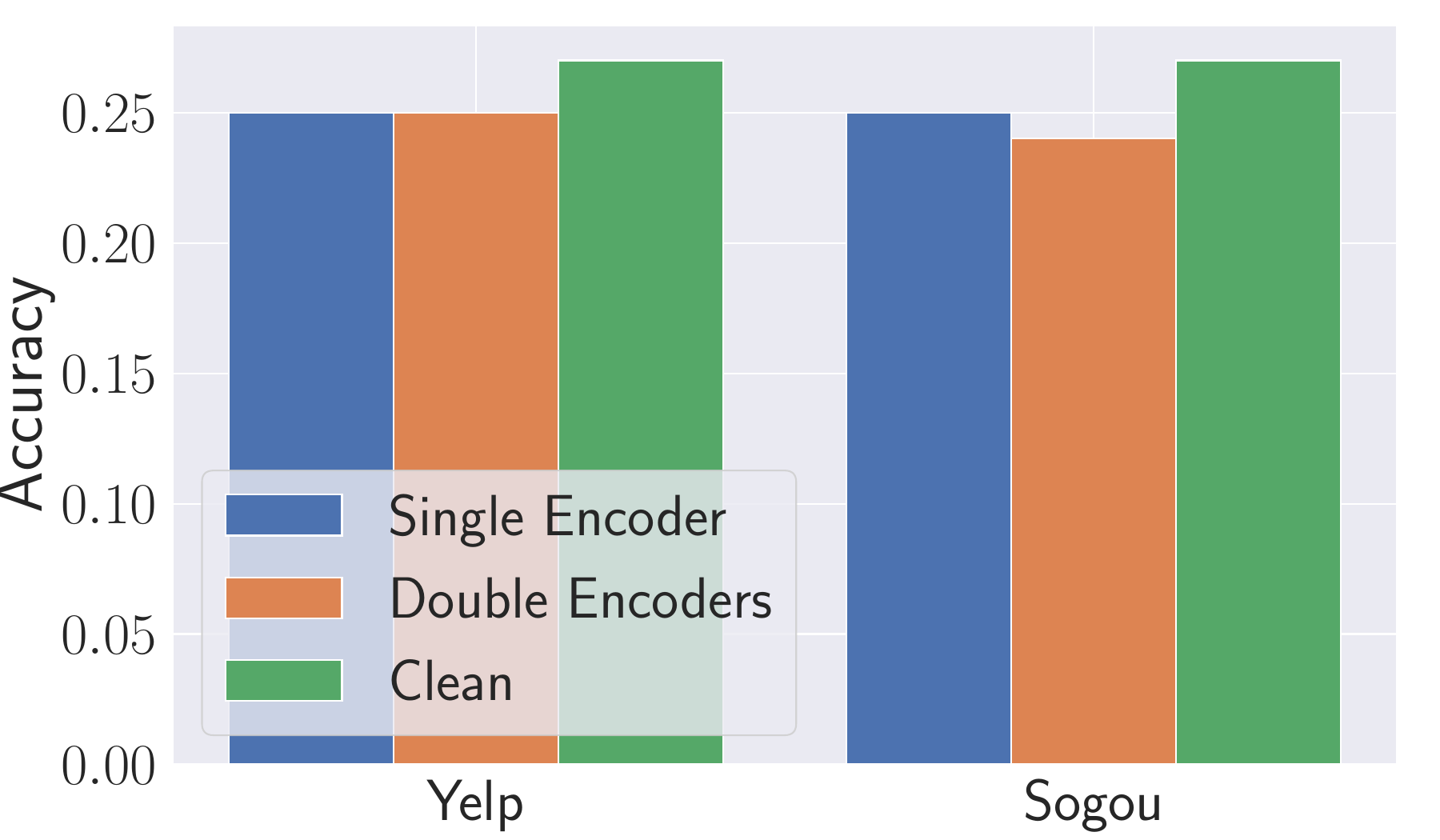}}
\caption{Our multimodal hijacking attack performance on different hijacking datasets, where Tiny ImageNet is our original dataset and mobilenetv2 is the target model.
The low utility of the clean models (27\%) is due to a large number of labels (1,000), and the limited number of samples in the dataset.}
\label{figure:tinyimagenet}
\end{figure}

\begin{table*}[!t]
    \caption{The details of the fused and corresponding container images across different hijacking and container datasets. For the fused images, labels are assigned based on the NLP sentences. For example, if the ground truth of a given NLP sentence is ``5 stars'', the label ``Digit 4'' would be assigned to all fused MNIST-like images that contain the NLP feature of this sentence. The complete table and the scaled-up version of the images are depicted in \autoref{table:completevision appendix} and \autoref{figure:completevision appendix} (\autoref{appendix: visual results})}
    \resizebox{\textwidth}{!}{
    \centering
    \begin{tabular}{l | c | c c c c | c c}
        \toprule
        \multirow{2}{*}{\textbf{\makecell{Hijacking\\dataset}}} & \multirow{2}{*}{\textbf{\makecell{Container\\dataset}}} & \multicolumn{4}{c|}{\textbf{Fused}} & \multicolumn{2}{c}{\textbf{Container}} \\
        & & Image & Image label & NLP sentence & \makecell{NLP\\ground truth} & Image & Image label  \\
        \midrule
        & MNIST & 
            \begin{minipage}[b]{0.3\columnwidth}
    		\centering
    		\raisebox{-.5\height}{\includegraphics[width=\linewidth]{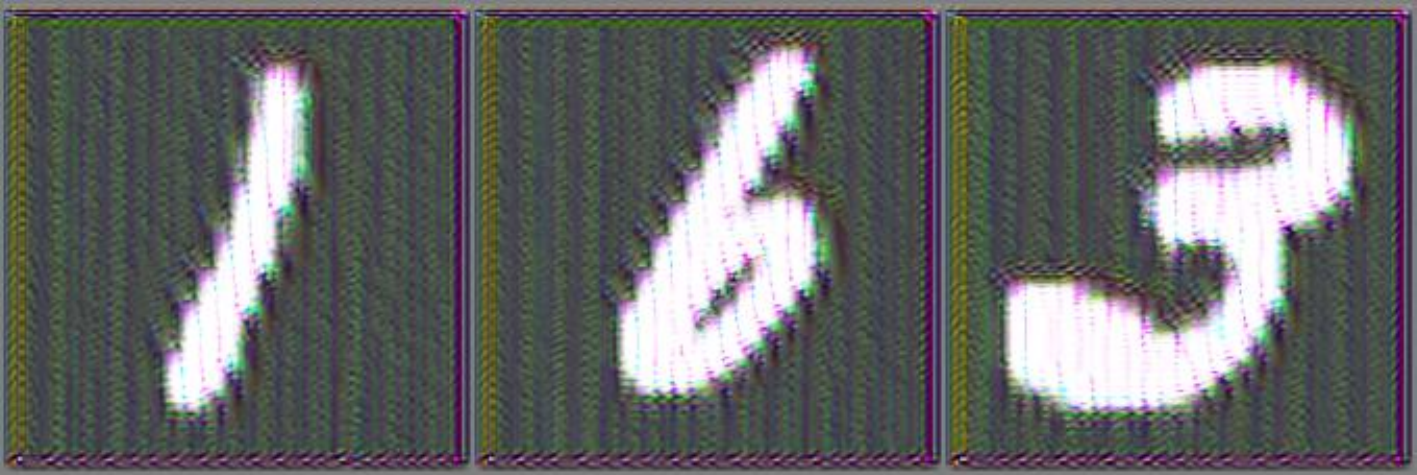}}
    	\end{minipage} & Digit 4 & \makecell{``Some of the best chow\\around--love this place.\\The bread and salads\\and soups are great.''} & 5 stars & 
            \begin{minipage}[b]{0.3\columnwidth}
    		\centering
    		\raisebox{-.5\height}{\includegraphics[width=\linewidth]{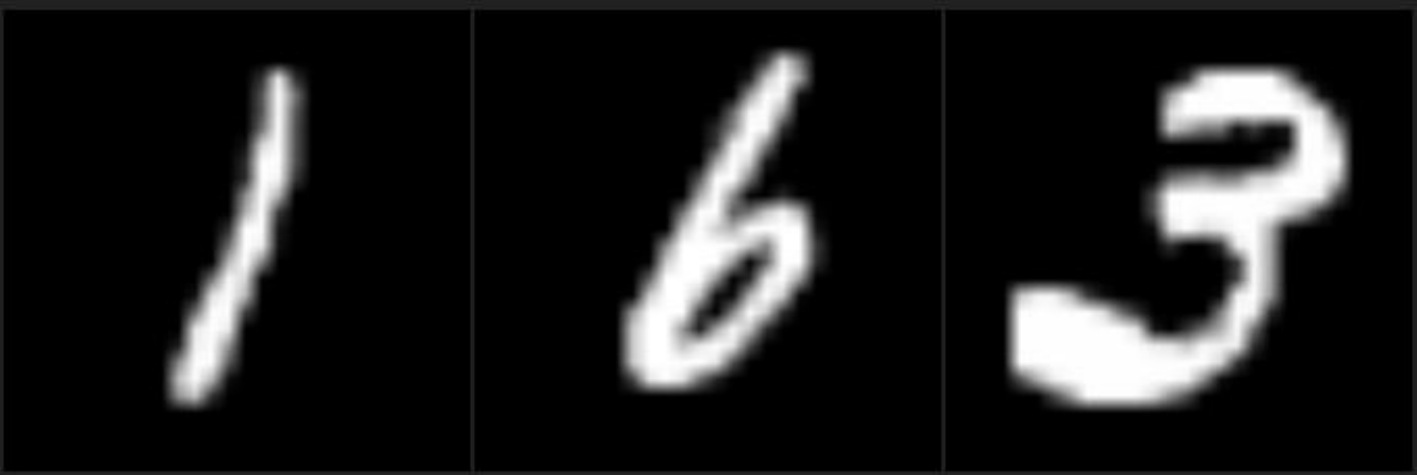}}
    	\end{minipage} & Digit 1, 6, 3 \\
        \multirow{3}{*}{Yelp} & Cifar-10 & 
            \begin{minipage}[b]{0.3\columnwidth}
    		\centering
    		\raisebox{-.5\height}{\includegraphics[width=\linewidth]{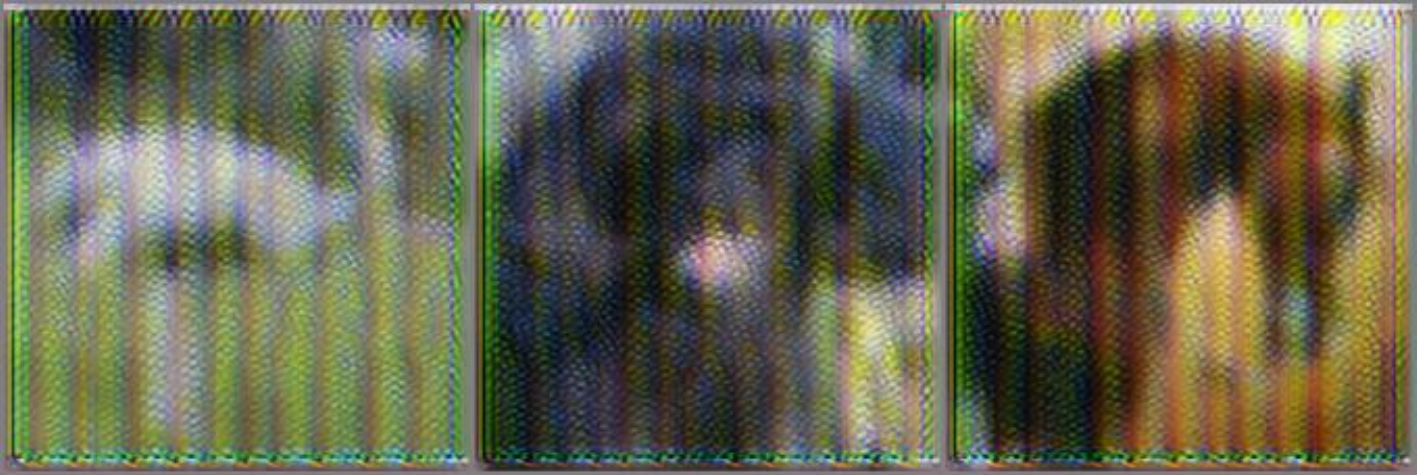}}
    	\end{minipage} & Airplane & \makecell{``The worst dental office\\I ever been. No one\\can beat it!!! You should\\avoid it at any time.''} & 1 star & 
            \begin{minipage}[b]{0.3\columnwidth}
    		\centering
    		\raisebox{-.5\height}{\includegraphics[width=\linewidth]{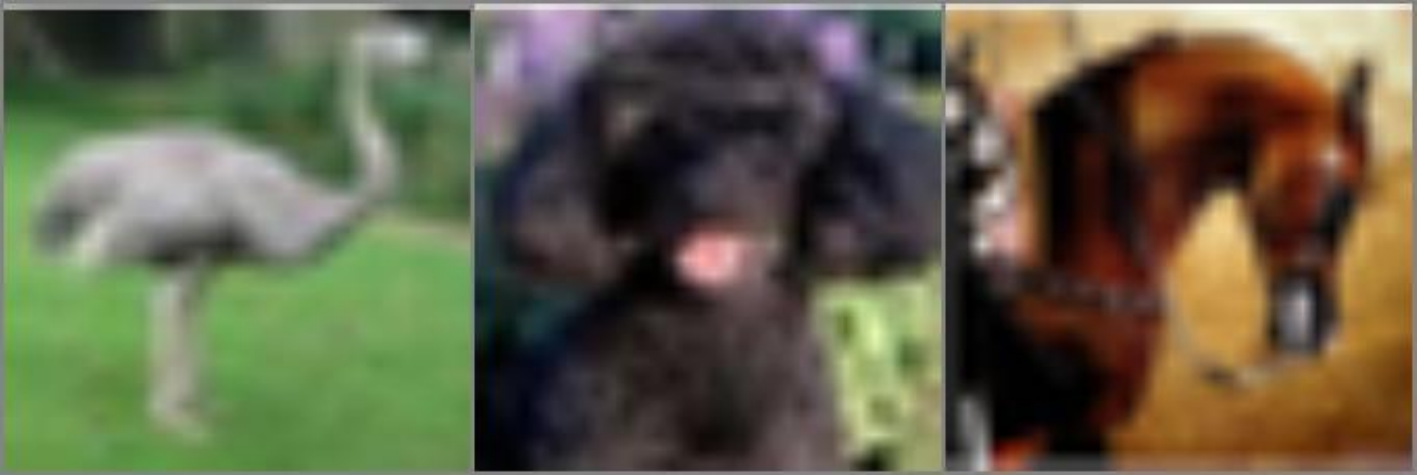}}
    	\end{minipage} & Bird, Dog, Horse \\
        & STL-10 & 
            \begin{minipage}[b]{0.3\columnwidth}
    		\centering
    		\raisebox{-.5\height}{\includegraphics[width=\linewidth]{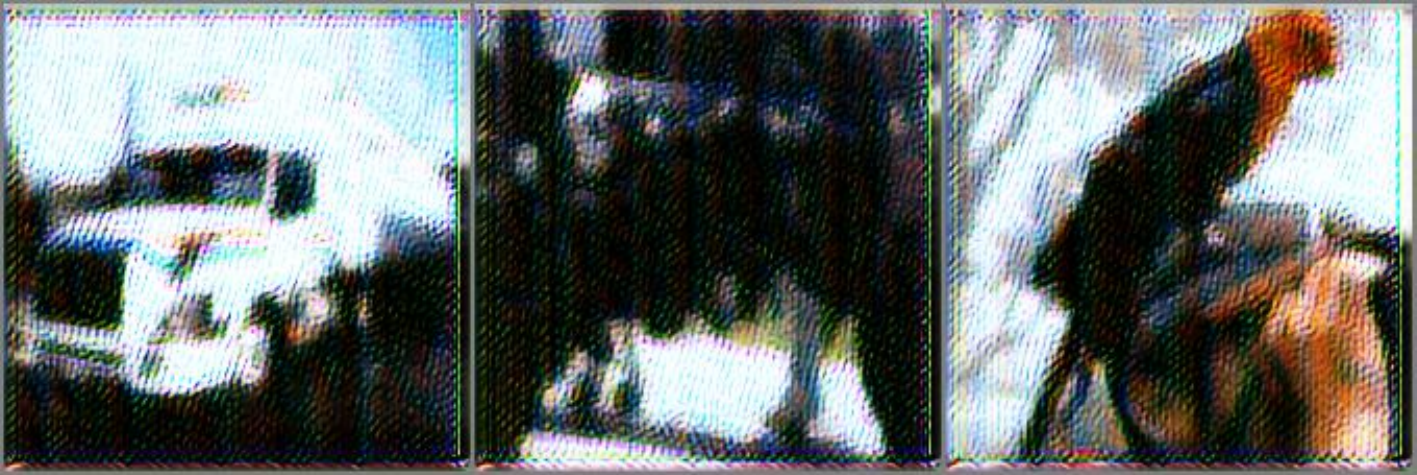}}
    	\end{minipage} & Bird & \makecell{``Far away from real\\Chinese food. Doesn't\\even taste good as American\\style Chinese food.''} & 2 stars & 
            \begin{minipage}[b]{0.3\columnwidth}
    		\centering
    		\raisebox{-.5\height}{\includegraphics[width=\linewidth]{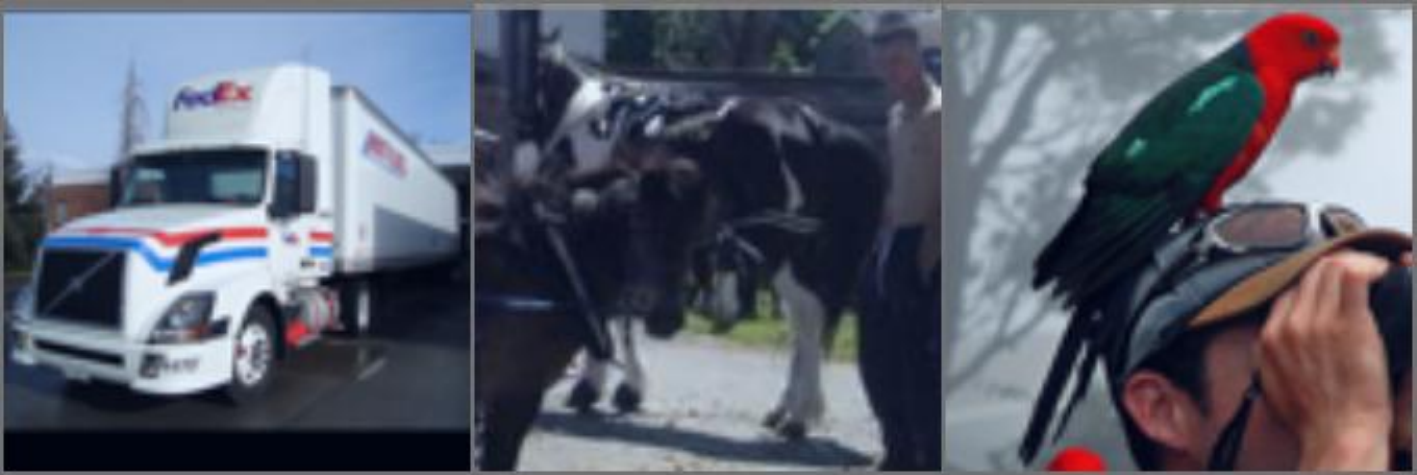}}
    	\end{minipage} & Truck, Horse, Bird \\
        \bottomrule
    \end{tabular}
    }
    \label{table:completevision}
\end{table*}

To recap, the adversary first needs to construct a container dataset to train the blender, as previously mentioned in \autoref{section: modal hijacking attack}.
We randomly sample 100 images from each target dataset to construct their corresponding container datasets.
These 100 images are then removed from the target datasets, i.e., $\containerdataset \cap\originaldataset = \Phi$.
Next, we sample 5,000 samples from each hijacking dataset, with 1,000 instances per label.
For each hijacking-original dataset pair, we randomly map the corresponding 5,000 hijacking samples with the 100 container images and train the Blender as mentioned in \autoref{subsection: blender design}.
Afterward, we use the trained Blender to fuse the 5,000 hijacking samples and poison the victim model.
The fused samples are combined with the complete training dataset for each victim model, i.e., 60,000, 50,000, and 5,000 samples for the MNIST, CIFAR-10, and STL-10 datasets, respectively.

To evaluate the performance of our modal hijacking attack in terms of both utility and attack success rate, we train clean models for all original and hijacking tasks. 
For the original tasks, we use the complete original dataset when training the models.
And we use the whole clean testing dataset to evaluate the performances.
As mentioned in \autoref{subsection: evaluation metrics}, the closer the performances -- with respect to the clean testing dataset -- of the hijacked and clean models are, the better the modal hijacking attack is.
For the hijacking tasks, we calculate the upper bound of the performance by using the complete training datasets, not just the 5,000 sentences used to hijack the victim model.
We then sample a testing dataset from the hijacking's task test dataset and use it to evaluate these models.
Finally, we fuse this dataset and use the fused version of the dataset to evaluate the attack success rate of the victim model.
The closer the attack success rate to the upper bound performance is, the better the modal hijacking attack is.

\mypara{Quantitative Evaluation}
We plot the attack success rate (ASR) in \autoref{figure:experiment_asr}.
As the figure shows, our modal hijacking attack achieves strong performance independent of the hijacking and the original datasets.
For instance, our attack achieves $94\%$, $94\%$, and $95\%$, when using the Sogou dataset to hijack MNIST, CIFAR-10, and STL-10 victim models by our Blender of double encoders, respectively, which is only $1\%$ worse than then upper bound models.
Similarly, for the Yelp dataset, our attack's ASR is only $4\%$, $1\%$, and $4\%$ less than the upper bound for the MNIST, CIFAR-10, and STL-10 victim models, respectively.
This clearly demonstrates the effectiveness of our modal hijacking attack.
Especially taking into consideration that we only use 5,000 hijacking samples to hijack the victim models unlike the entire dataset when training the upper bound ones, we next evaluate our modal hijacking attack's utility.
We plot the performances of the victim models and the ones trained with clean datasets in \autoref{figure:experiment_utility}.
As the figure shows, our modal hijack attack achieves comparable performance on the clean testing dataset with the clean models.
For instance, the victim model achieves the utility of $99\%$ ($99\%$), $93\%$ ($93\%$), and $93\%$ ($92\%$) on the MNIST, CIFAR-10, and STL-10 models when being hijacked by the Yelp (Sogou) dataset in the double-encoder case, respectively.
This shows the negligible drop in model utility for our modal hijacking attack.
In addition, we evaluate the attack performance on the Tiny ImageNet~\cite{imagenet} as the original dataset.
The results show that our attack is generalizable to higher-quality images, as shown in \autoref{figure:tinyimagenet}.
Furthermore, when we restrict the Blender using a single encoder, our method achieves comparable attack performance.
Besides, in the following exploration of different hyperparameters/decision designs (\autoref{subsection: hyperparameters/design decisions}), using a single encoder consistently shows similar power as double encoders.
These results provide a potential option to gain strong attack performance with less cost of computation resources.

\mypara{Qualitative Evaluation}
Concretely, \autoref{table:completevision} shows randomly sampled fused samples together with their container images, as well as corresponding details.
As the figure shows, the output of our Blender is very similar to the original dataset, with few visible artifacts.
As shown later (\autoref{subsection: defense}), these artifacts do not make the fused images easier to detect.
Besides, compared to the fused images of \cite{SBZ22}, where the hijacking digits are visible, it is hard to recognize the hijacking content in our attack.
This advantage further increases the stealthiness of our modal hijacking attack.
Moreover, increasing the training epochs of the Blender can reduce the artifacts as shown in \autoref{figure:q5vision}.
To compare the performances of the modal hijacking attack with the naive approach, i.e., only using the adapter but not the Blender, we plot the outputs of the adapter of the Yelp samples in \autoref{figure:adapterimages}.
Comparing both figures, the output of our Blender is clearly more similar to the original dataset, hence, showing the stealthiness of our modal hijacking attack.

\subsection{Hyperparameters/Design Decisions}
\label{subsection: hyperparameters/design decisions}

\begin{figure}[!t]
\centering
\subfigure[Attack Success Rate.]{
\label{figure:q1asr}
\includegraphics[width=0.41\textwidth]{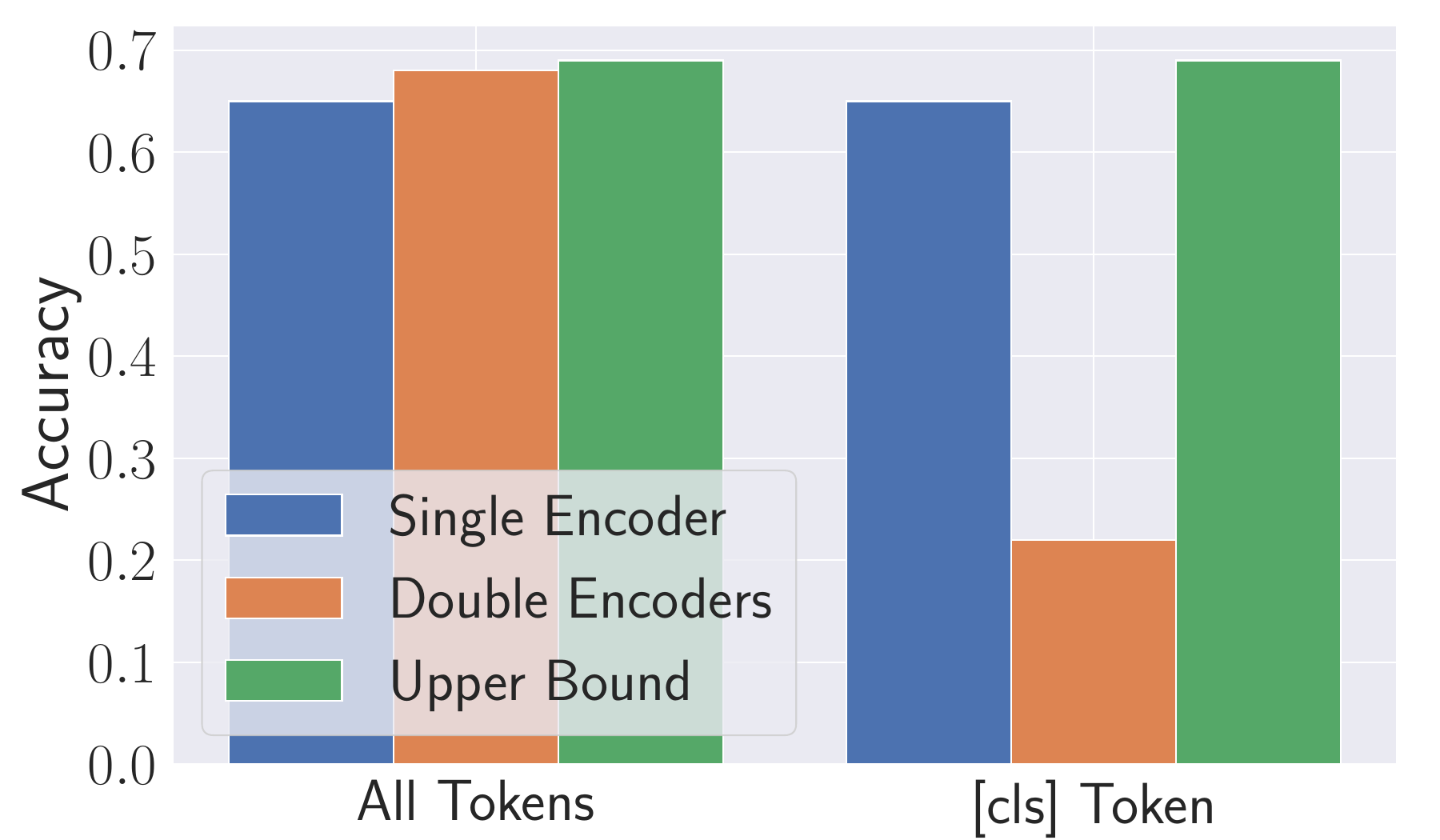}}
\subfigure[Utility.]{
\label{figure:q1utility}
\includegraphics[width=0.41\textwidth]{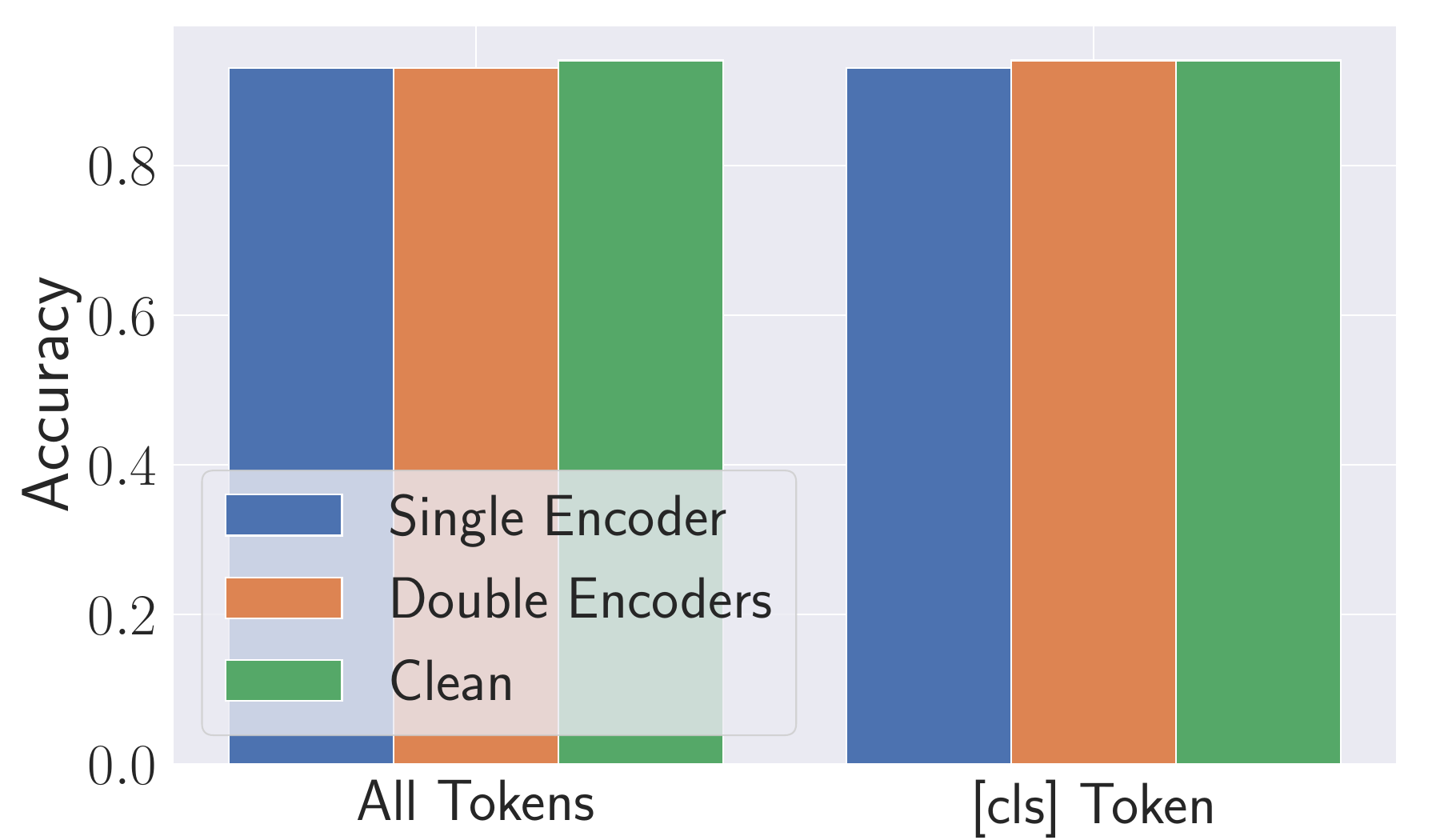}}
\caption{The comparison of our multimodal hijacking attack performances between using complete embeddings of hijacking sentence and the last -- ``[cls]'' -- token.
The hijacking dataset is Yelp and the original dataset is CIFAR-10.}
\label{figure:q1performance}
\end{figure}

\begin{figure}[!t]
\centering
\subfigure[Attack Success Rate.]{
\label{figure:q2asr}
\includegraphics[width=0.41\textwidth]{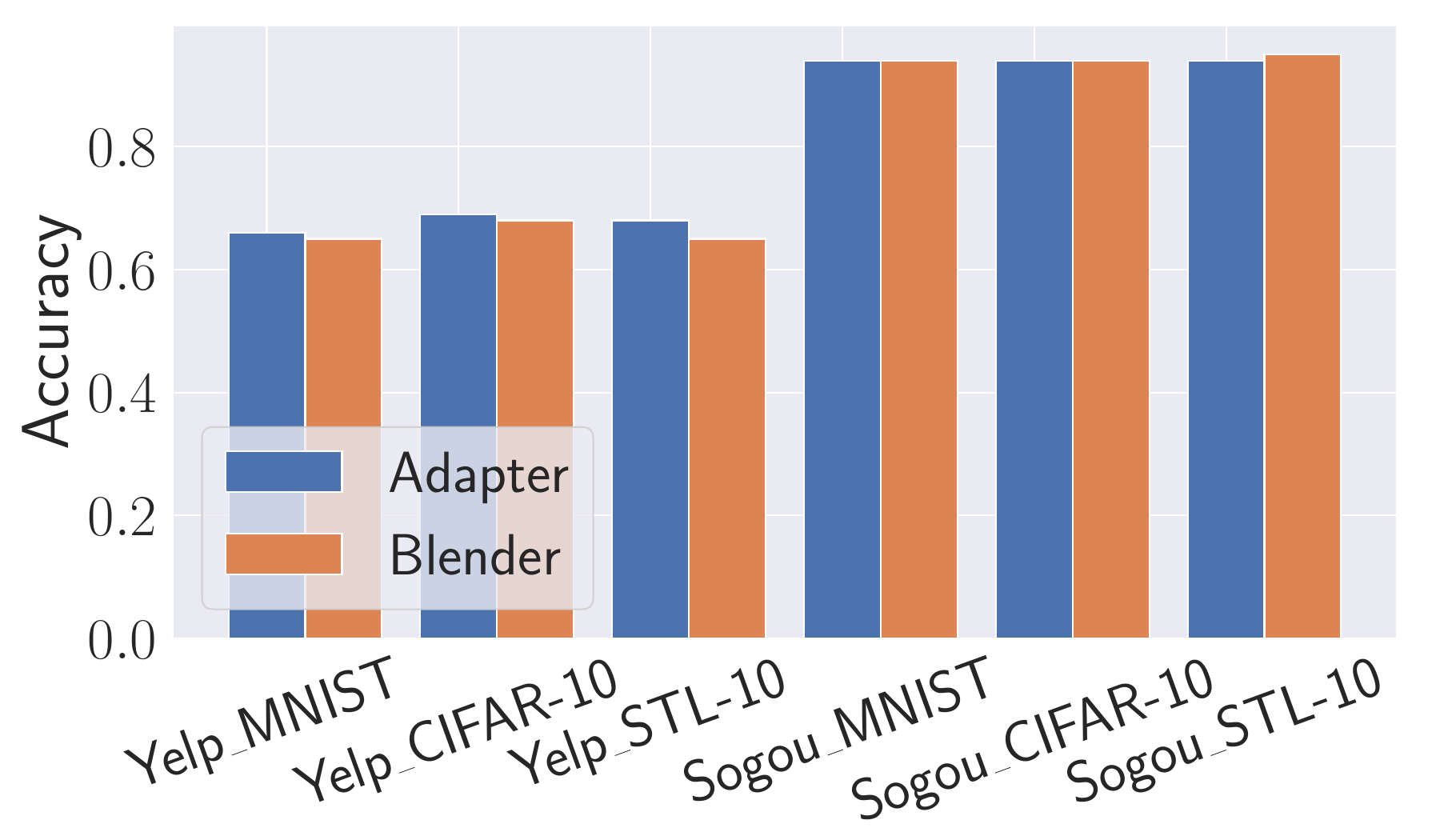}}
\subfigure[Utility.]{
\label{figure:q2utility}
\includegraphics[width=0.41\textwidth]{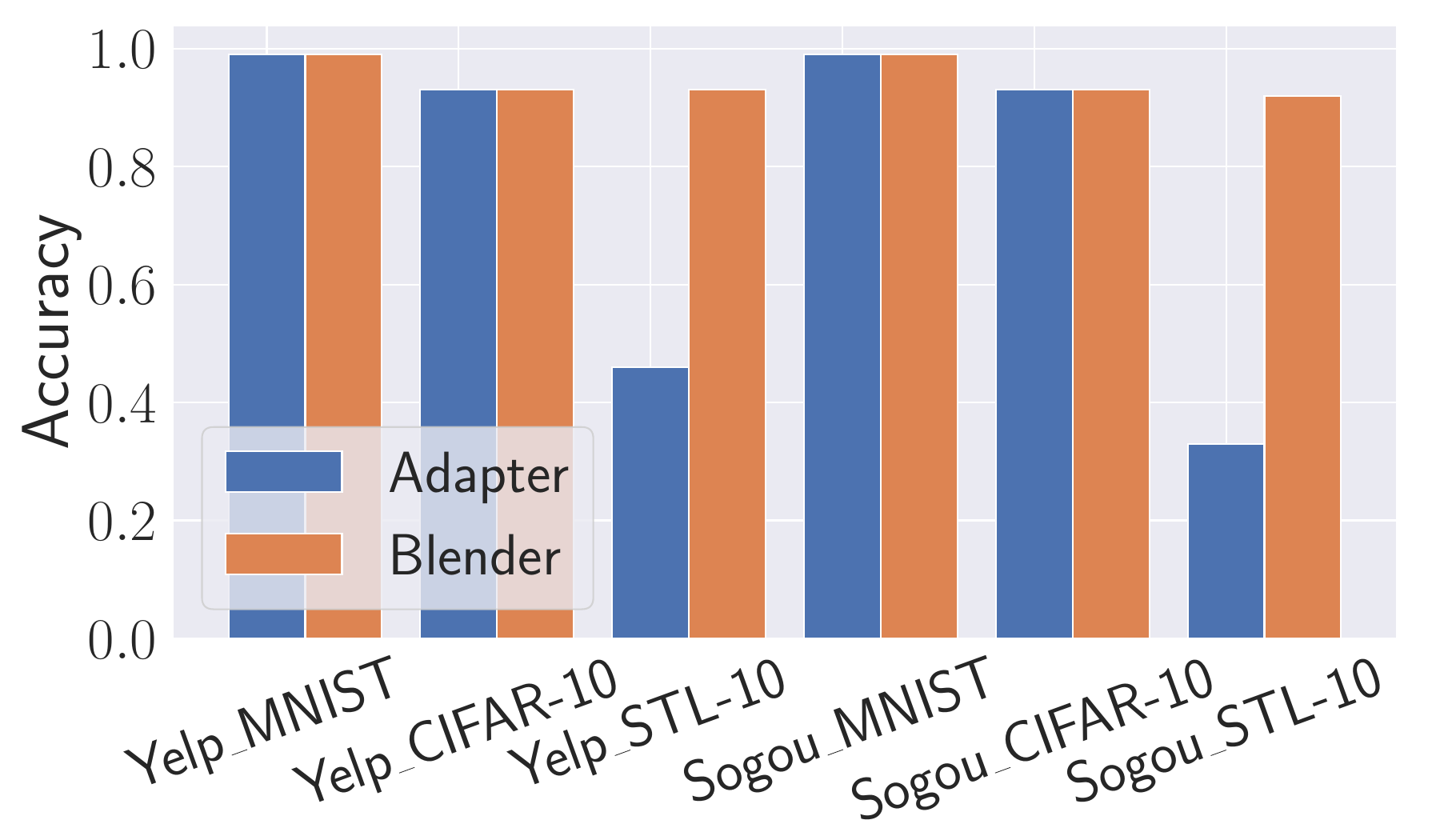}}
\caption{The comparison of our multimodal hijacking attack performances between using the adapter and our Blender, where x\_y notation is the hijacking\_original dataset pair.}
\label{figure:q2performance}
\end{figure}

\begin{figure}[!t]
\centering
\subfigure[Y \#1]{
\label{figure:yelp_adapter0}
\includegraphics[width=0.06\textwidth]{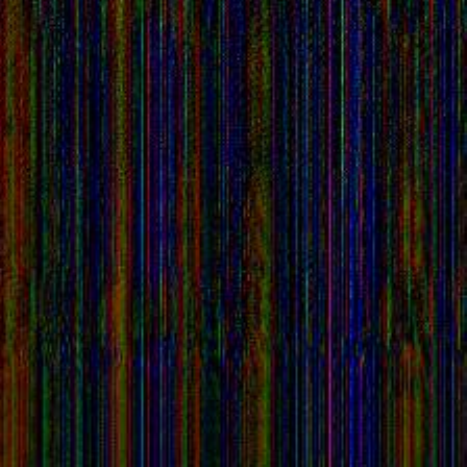}}
\subfigure[Y \#2]{
\label{figure:yelp_adapter1}
\includegraphics[width=0.06\textwidth]{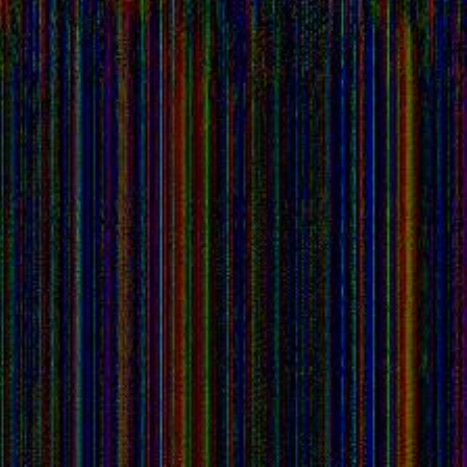}}
\subfigure[Y \#3]{
\label{figure:yelp_adapter2}
\includegraphics[width=0.06\textwidth]{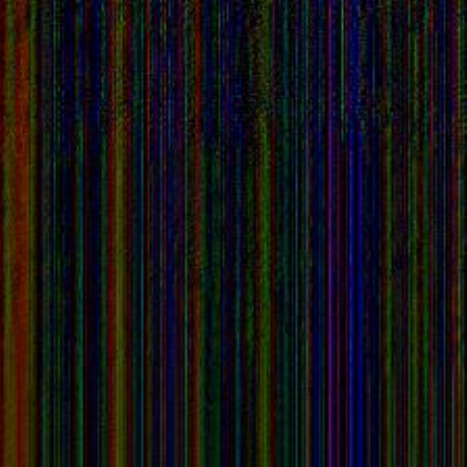}}
\subfigure[S \#1]{
\label{figure:sogou_adapter0}
\includegraphics[width=0.06\textwidth]{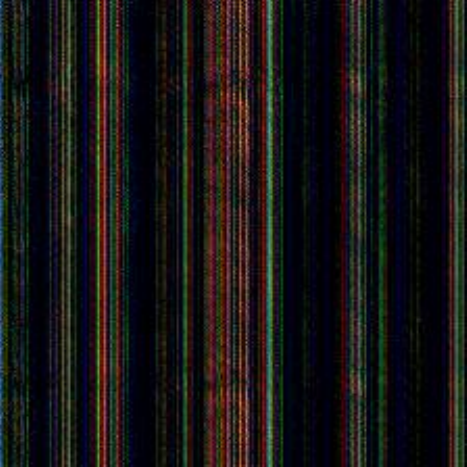}}
\subfigure[S \#2]{
\label{figure:sogou_adapter1}
\includegraphics[width=0.06\textwidth]{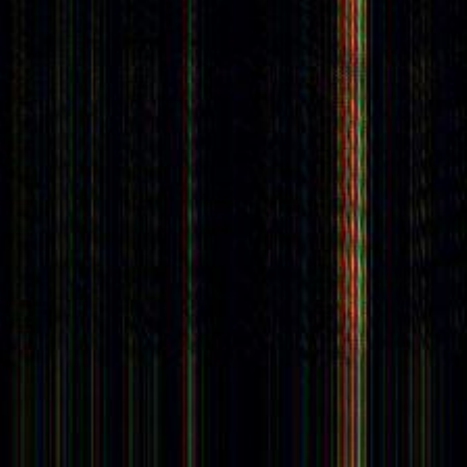}}
\subfigure[S \#3]{
\label{figure:sogou_adapter2}
\includegraphics[width=0.06\textwidth]{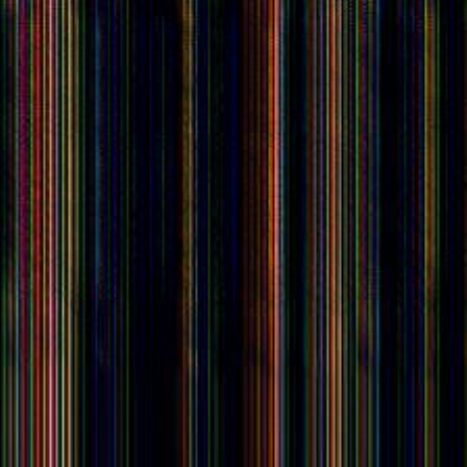}}
\caption{The visual result examples of the adapter's output using the hijacking sentences of Yelp (Y) and Sogou (S).}
\label{figure:adapterimages}
\end{figure}

\begin{figure}[!t]
\centering
\subfigure[Attack Success Rate.]{
\label{figure:q3asr}
\includegraphics[width=0.41\textwidth]{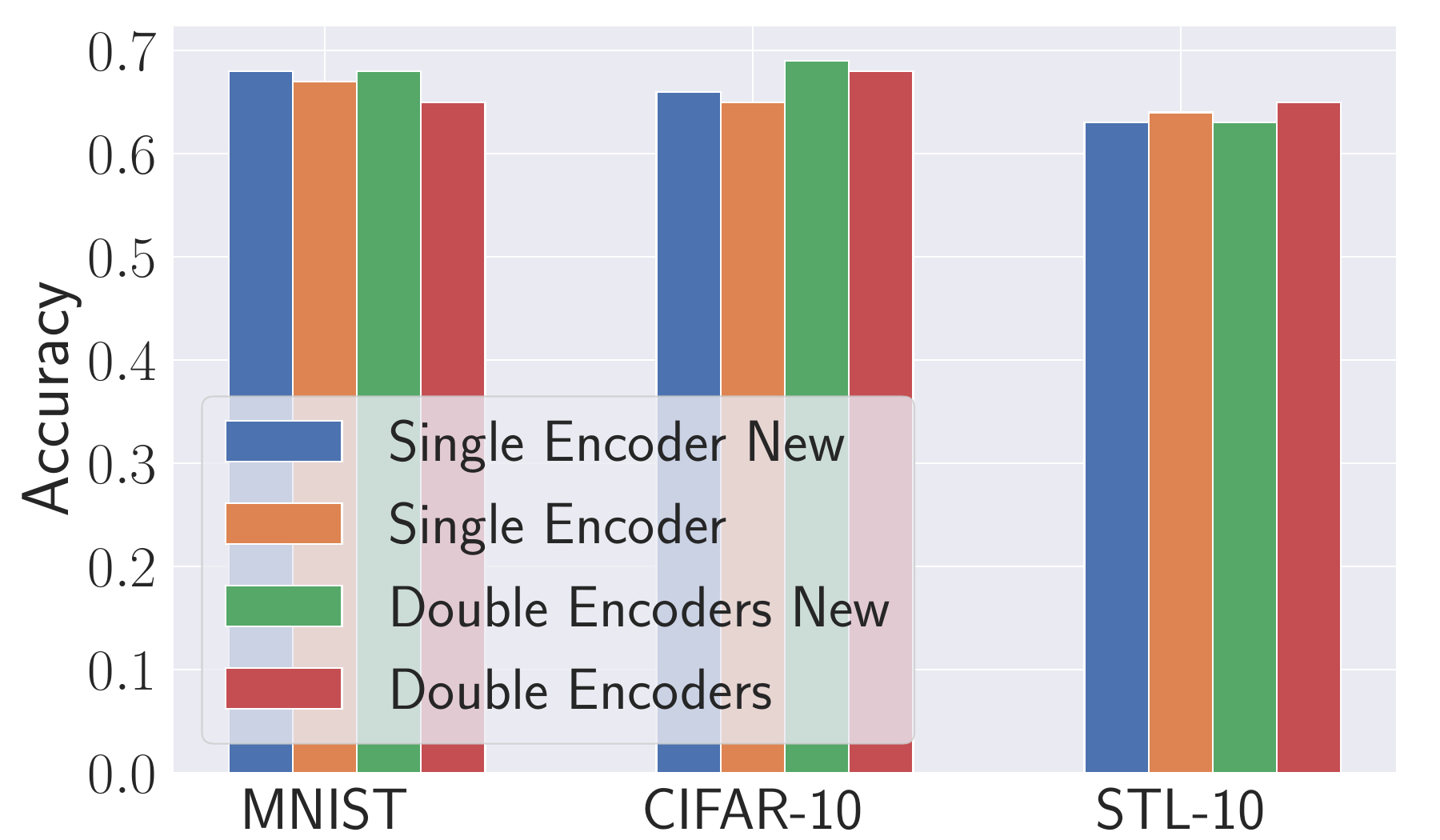}}
\subfigure[Utility.]{
\label{figure:q3utility}
\includegraphics[width=0.41\textwidth]{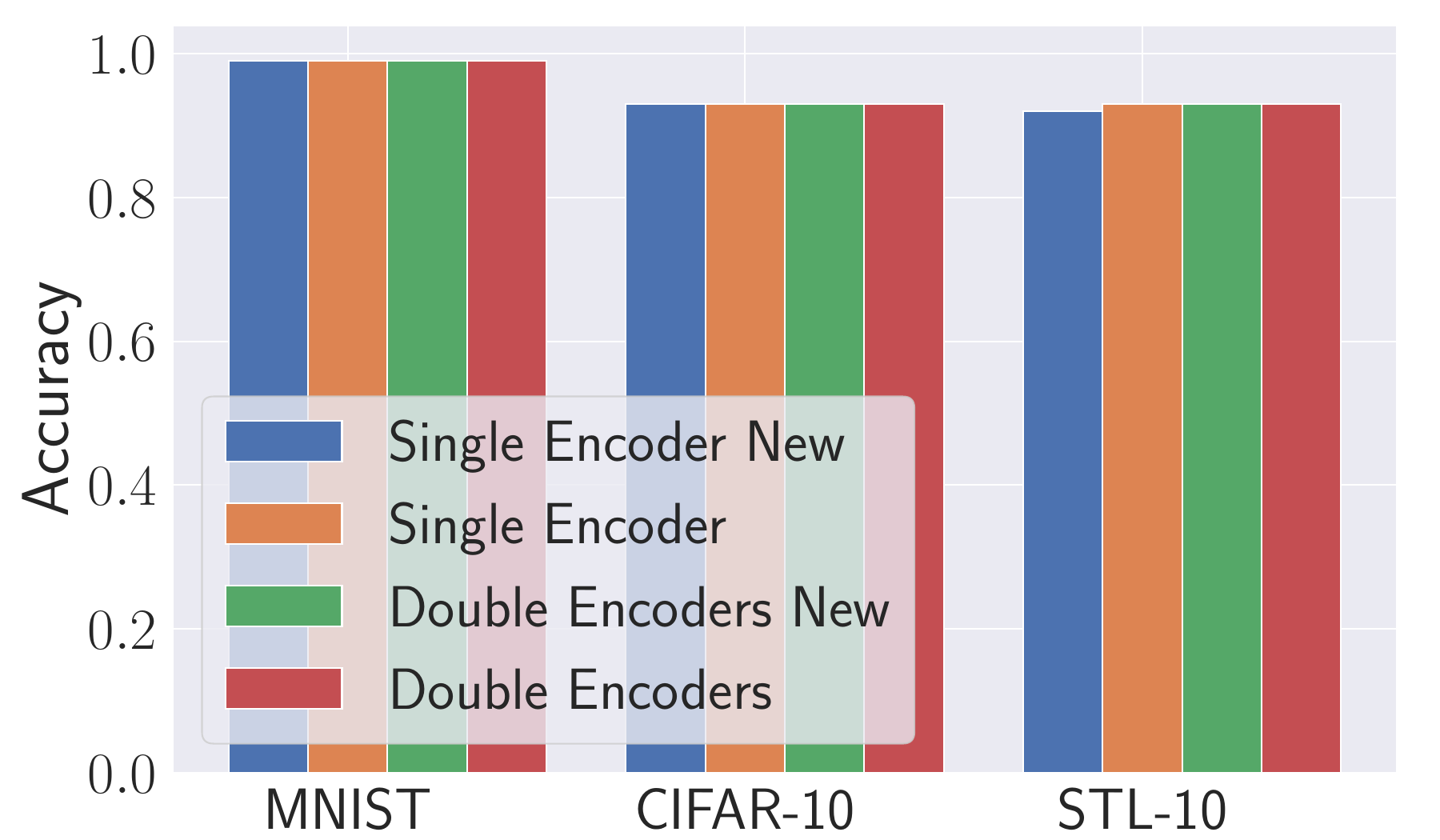}}
\caption{Our multimodal hijacking attack performance of BART and MobileNetv2 as the -- NLP -- and -- CV -- feature extractors, of which the setting is denoted as ``New'', where x\_y notation is the hijacking\_original dataset pair.}
\label{figure:q3performance}
\end{figure}

\begin{figure}[!t]
\centering
\subfigure[Attack Success Rate.]{
\label{figure:q4asr}
\includegraphics[width=0.41\textwidth]{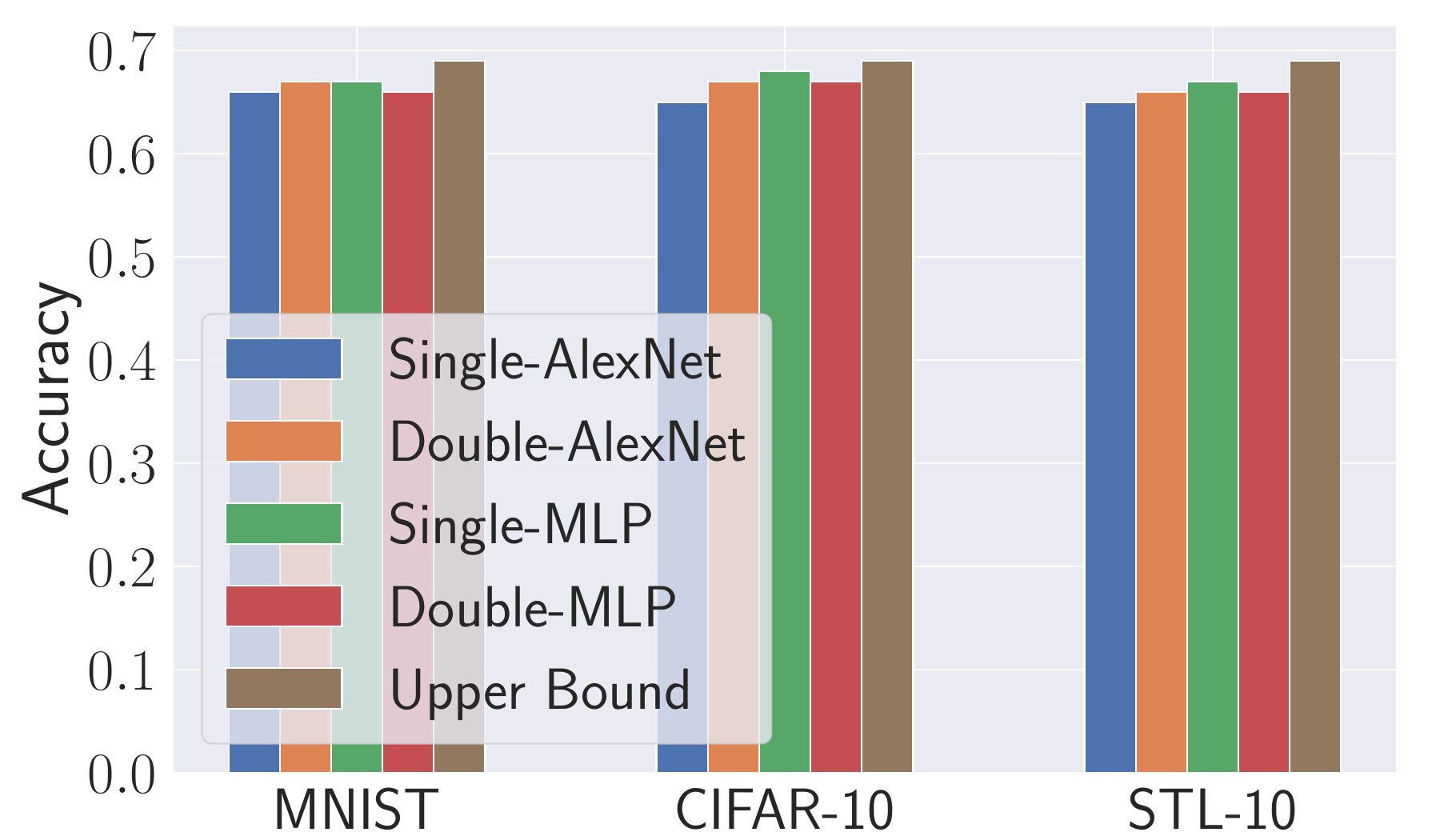}}
\subfigure[Utility.]{
\label{figure:q4utility}
\includegraphics[width=0.41\textwidth]{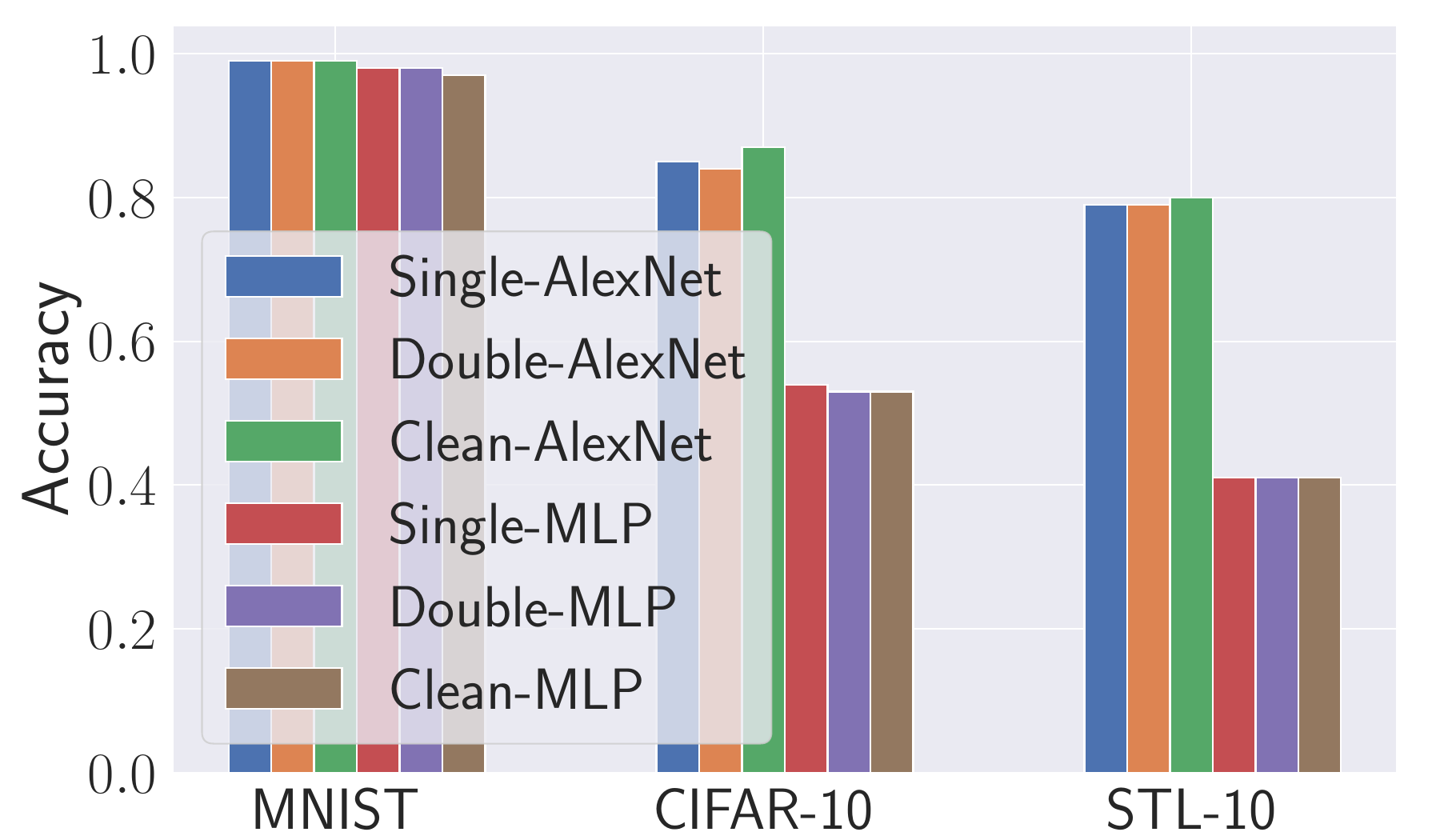}}
\caption{Our multimodal hijacking attack performance when the victim model is AlexNet on different hijacking datasets and the Yelp original dataset.}
\label{figure:q4performance}
\end{figure}

\begin{figure}[!t]
\centering
\subfigure[The Attack Performance.]{
\label{figure:q5accuracy}
\includegraphics[width=0.41\textwidth]{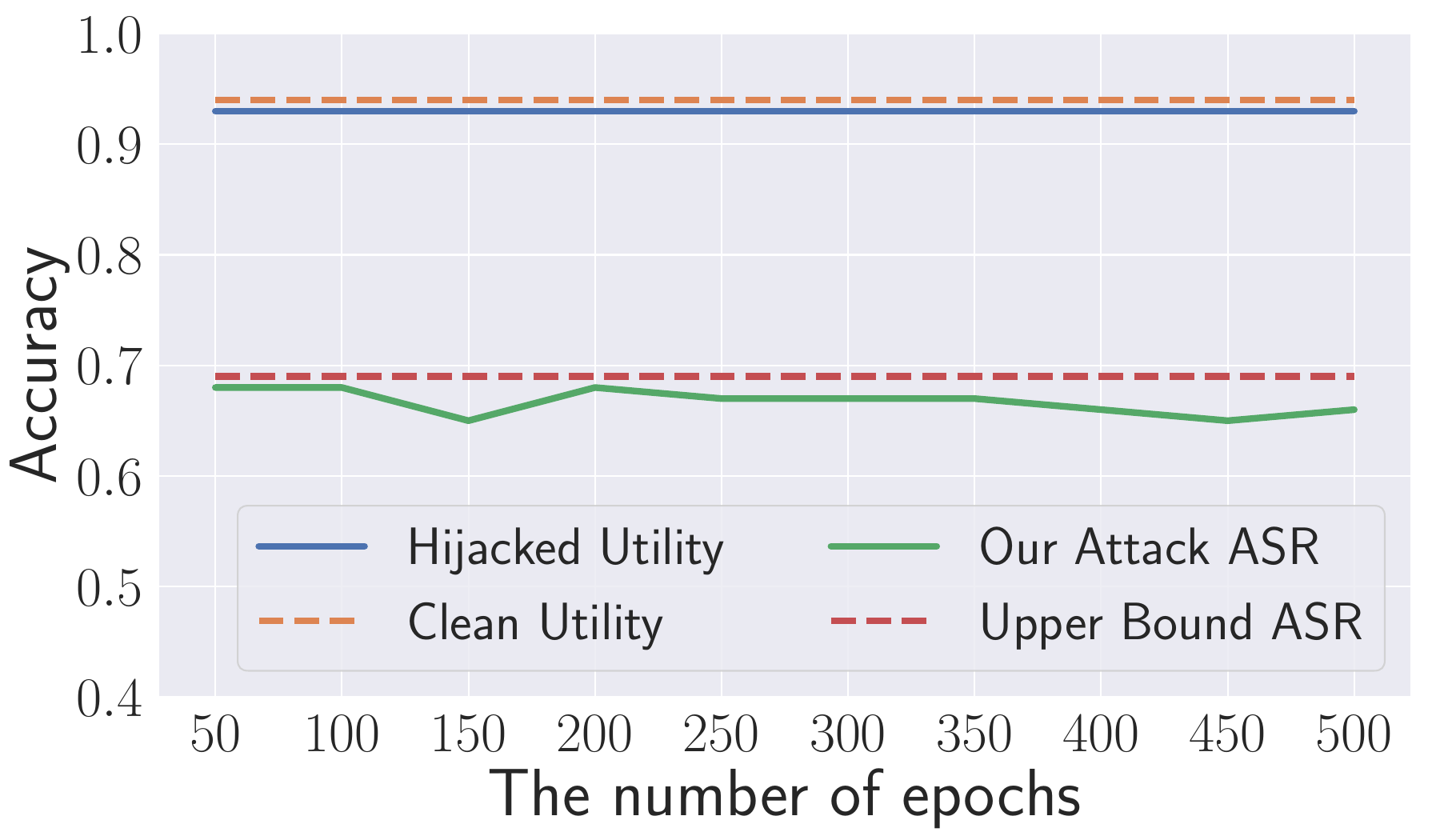}}
\subfigure[The Visual Results.]{
\label{figure:q5vision}
\includegraphics[width=0.44\textwidth]{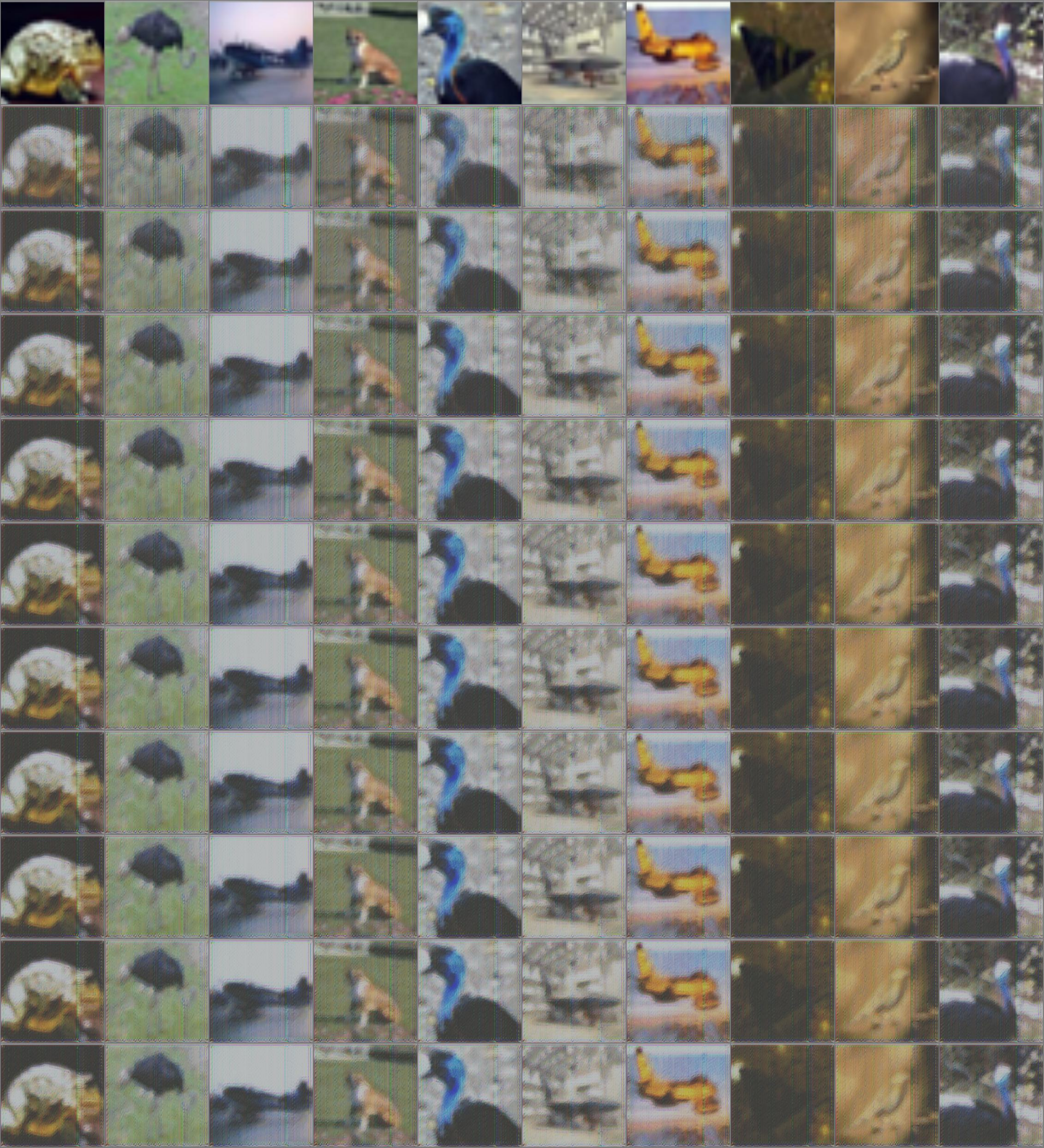}}
\caption{Our multimodal hijacking attack performance and the visual results of the Blender's output.
The hijacking and original datasets are Yelp and CIFAR-10.
The Blender trained with the different numbers of epochs, from 50 to 500 with steps of 50.
In (a), the x-axis denotes the number of epochs which our Blender is trained with.
And in (b), the first row presents the visual results of container images, and the following rows indicate the visual results of the output of our Blender trained with different numbers of epochs.}
\label{figure:q5performance}
\end{figure}

\begin{figure}[!t]
\centering
\subfigure[Attack Success Rate.]{
\label{figure:q6asr}
\includegraphics[width=0.41\textwidth]{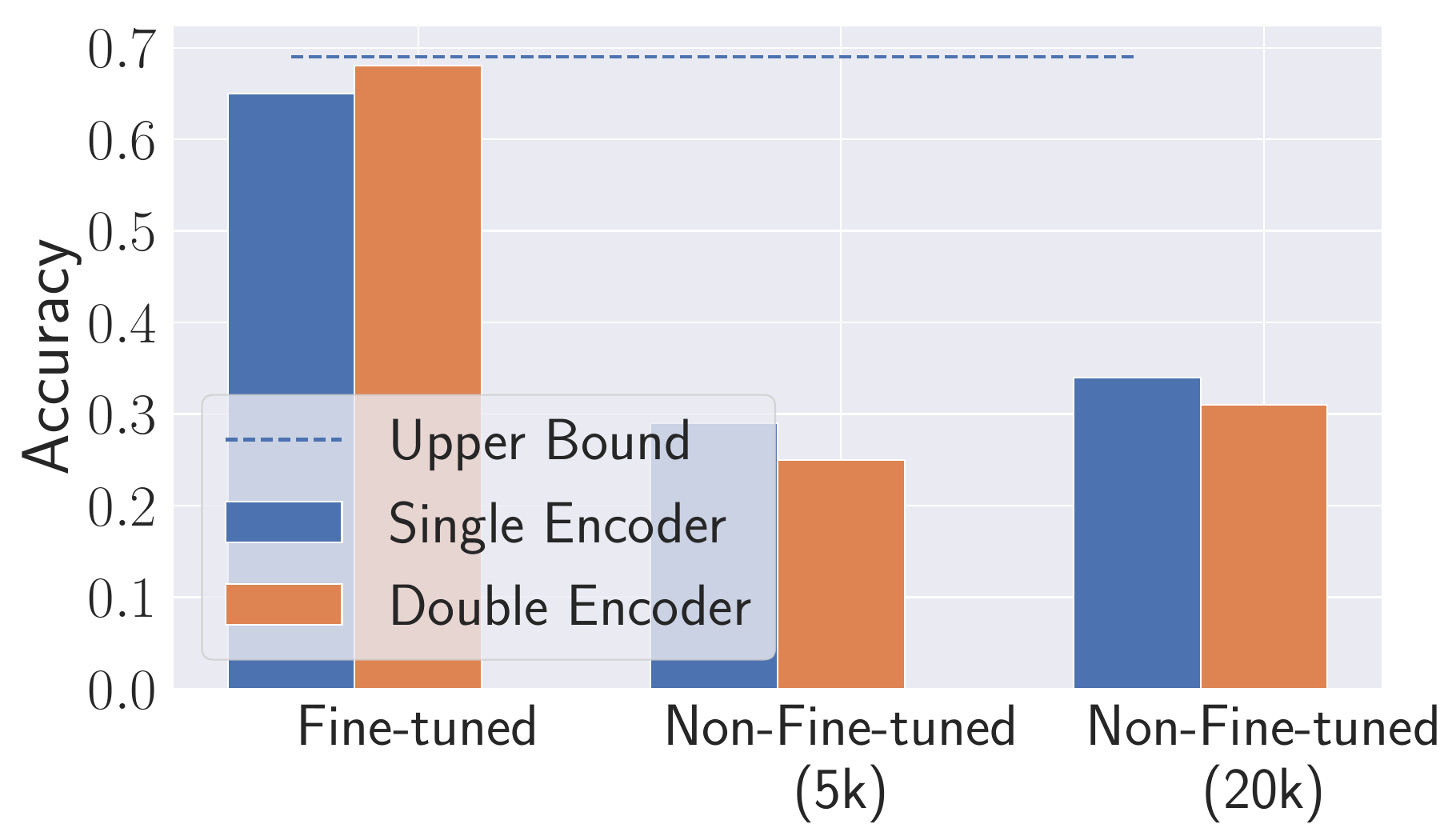}}
\subfigure[Utility.]{
\label{figure:q6utility}
\includegraphics[width=0.41\textwidth]{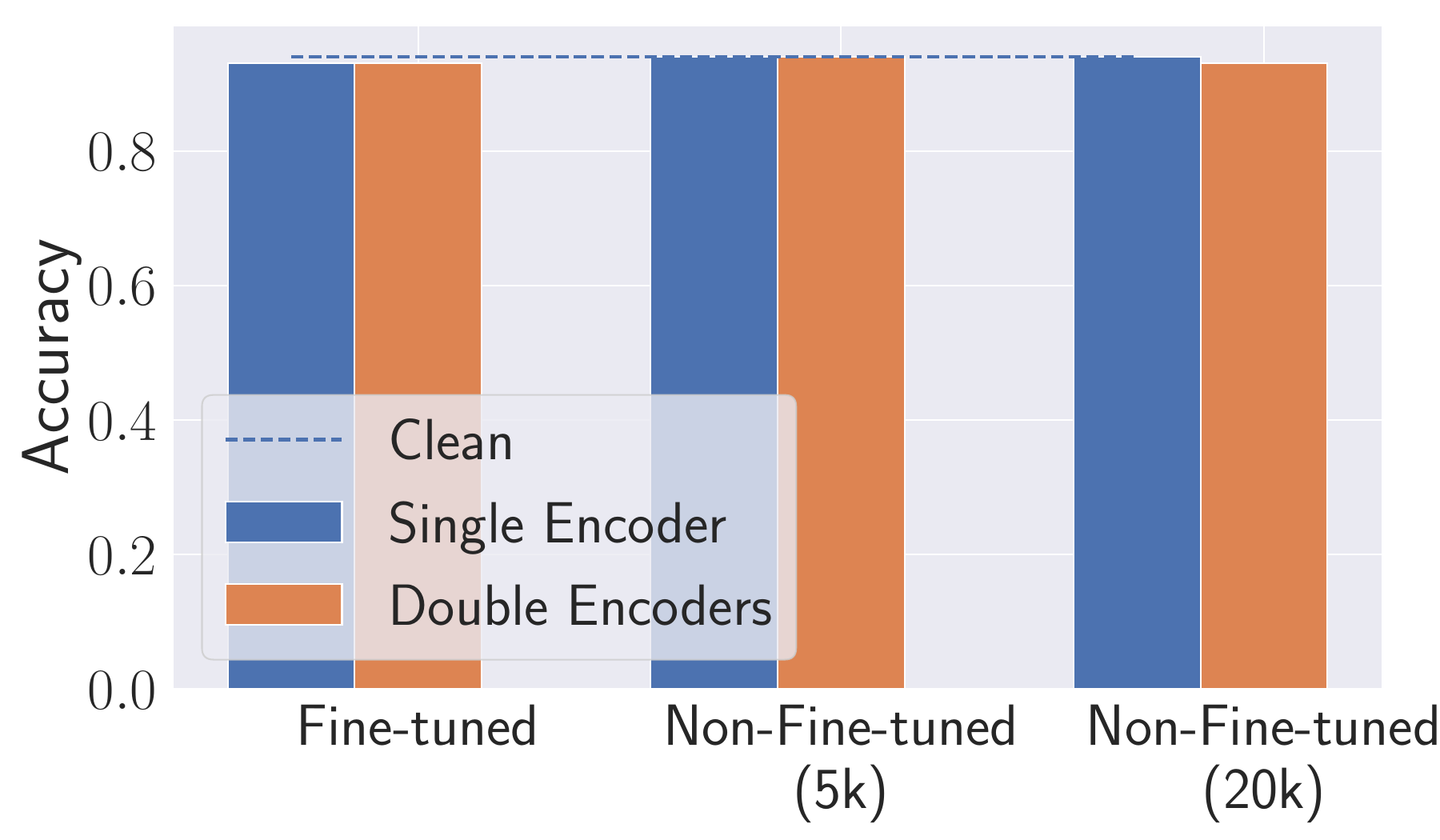}}
\caption{Our multimodal hijacking attack performances of fine-tuning and not fine-tuning BERT, as well as using a non-fine-tuned BERT with more poisons.
The hijacking dataset is Yelp and the original dataset is CIFAR-10.}
\label{figure:q6performance}
\end{figure}

\begin{figure}[!t]
\centering
\subfigure[Attack Success Rate.]{
\label{figure:q7asr}
\includegraphics[width=0.41\textwidth]{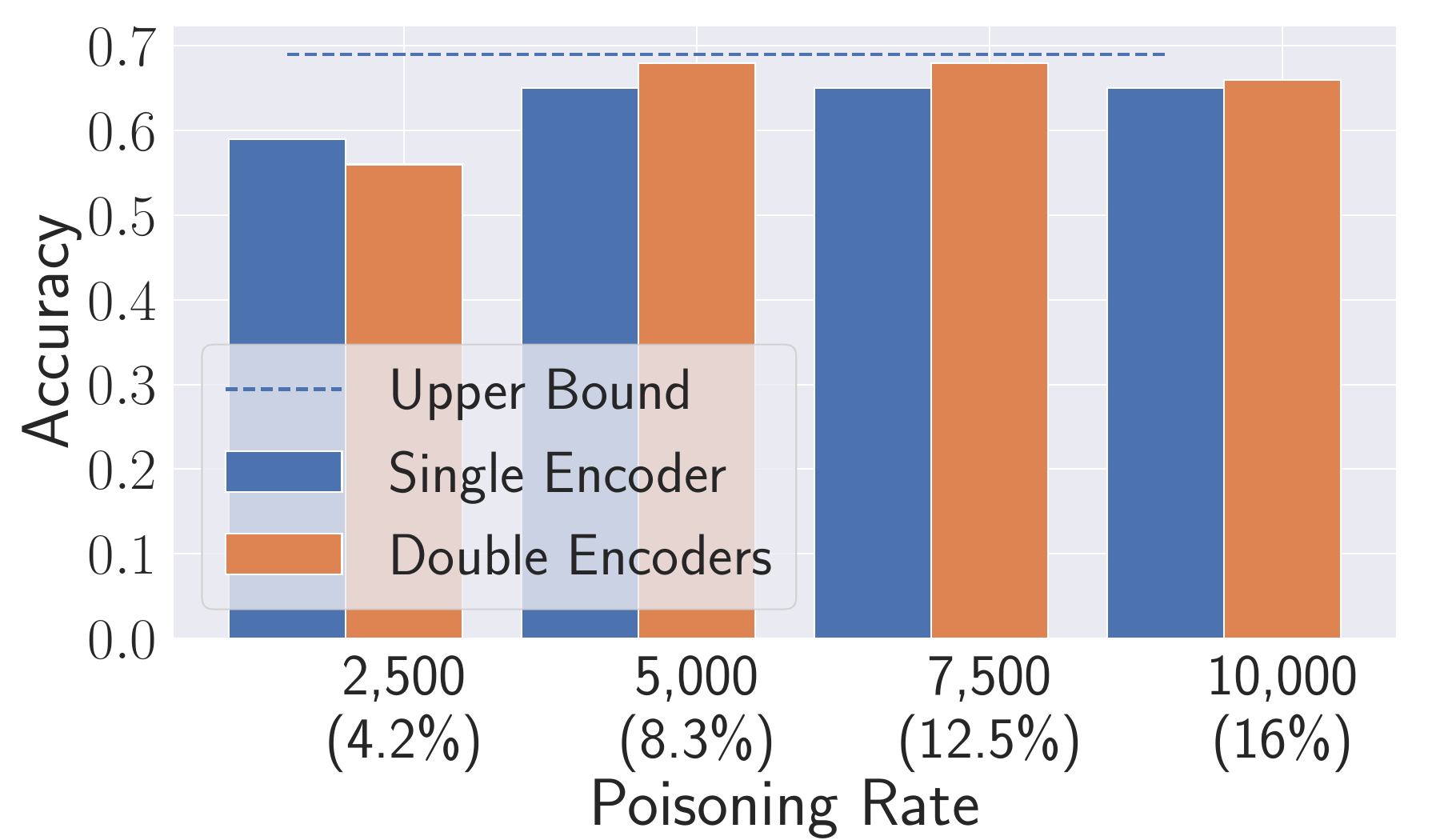}}
\subfigure[Utility.]{
\label{figure:q7utility}
\includegraphics[width=0.41\textwidth]{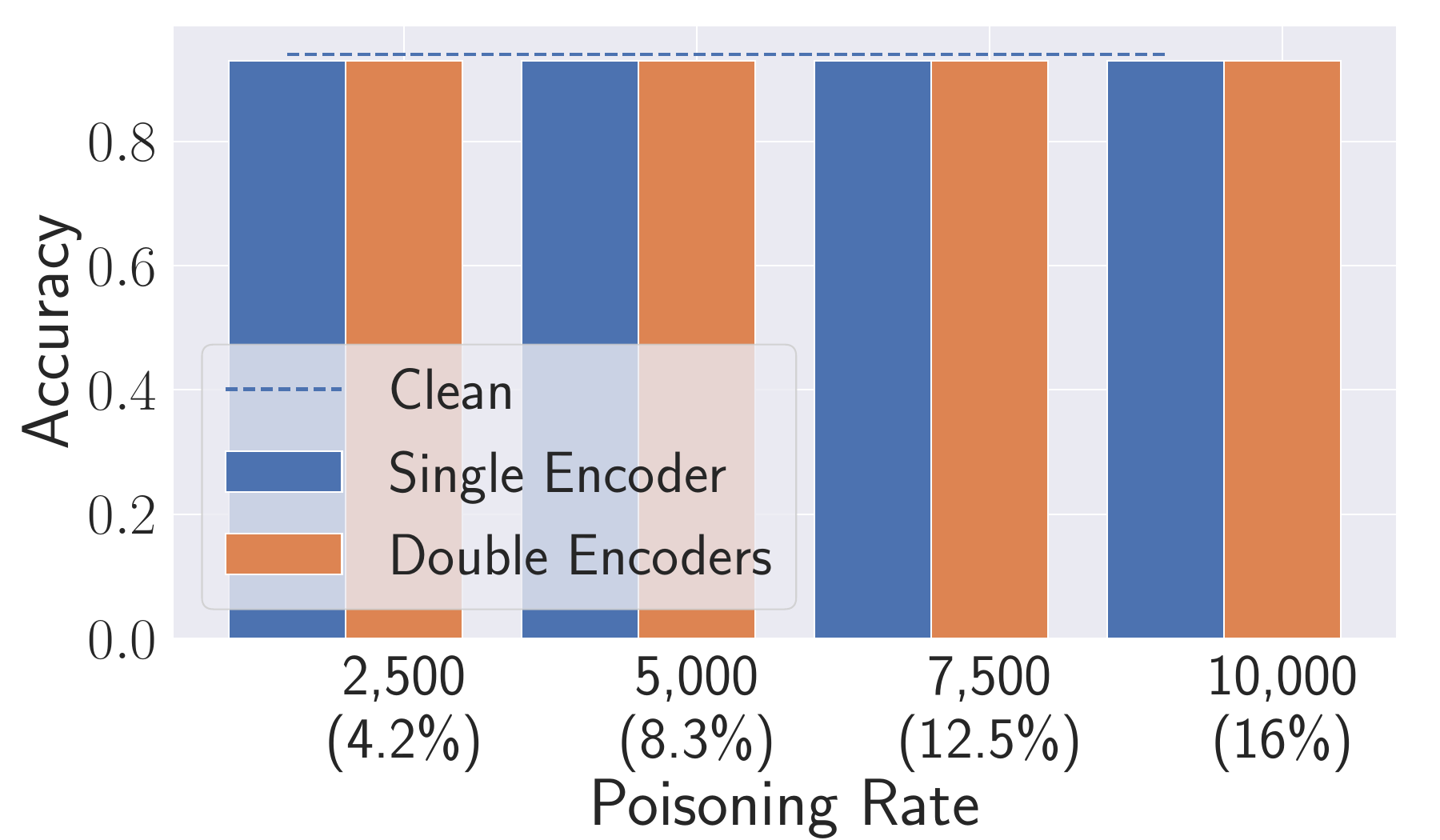}}
\caption{The comparison of our multimodal hijacking attack performances between using different numbers of poisons, from 2,500 to 10,000 with steps of 2,500.
The hijacking dataset is Yelp and the original dataset is CIFAR-10.}
\label{figure:q7performance}
\end{figure}

\begin{figure}[!t]
\centering
\subfigure[Attack Success Rate.]{
\label{figure:q8asr}
\includegraphics[width=0.41\textwidth]{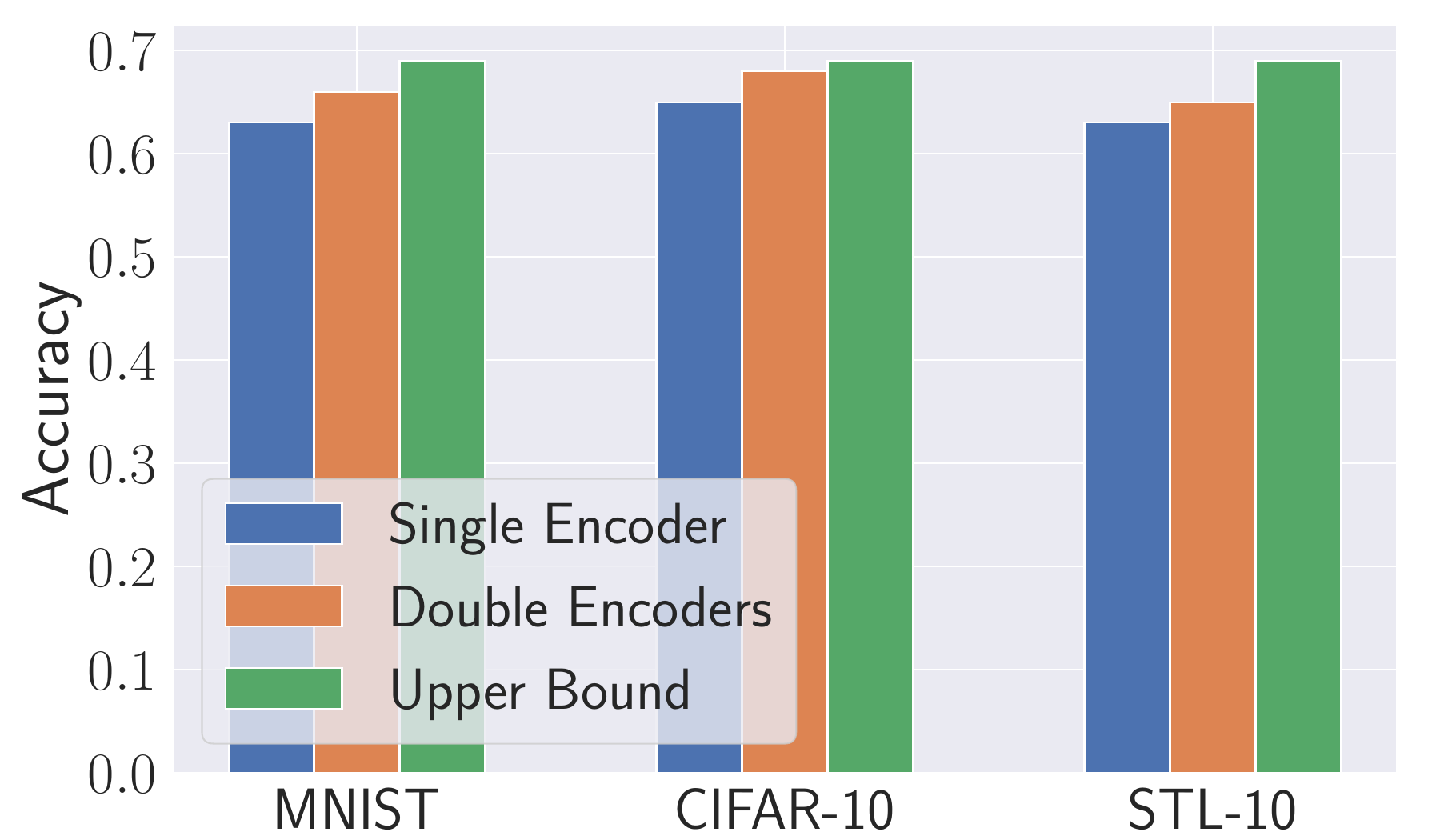}}
\subfigure[Utility.]{
\label{figure:q8utility}
\includegraphics[width=0.41\textwidth]{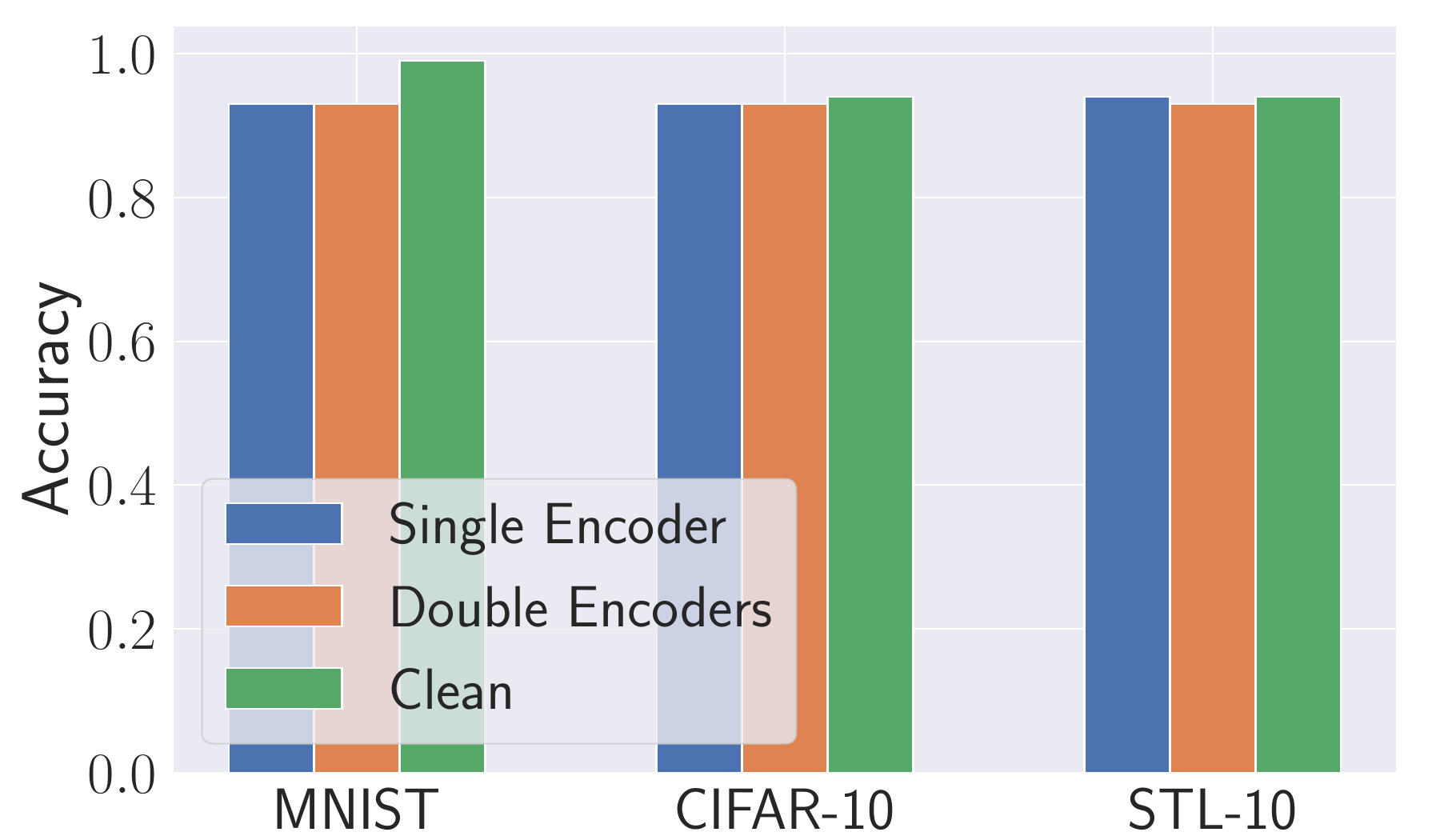}}
\caption{Our multimodal hijacking attack performance when the same Blender is used to fuse the hijacking dataset with different container datasets on different container datasets.
The Blender is trained on the hijacking dataset of Yelp and the container dataset of CIFAR-10.}
\label{figure:q8performance}
\end{figure}

\begin{figure}[!t]
\centering
\subfigure[Attack Success Rate.]{
\label{figure:transferability_1vs1_asr}
\includegraphics[width=0.41\textwidth]{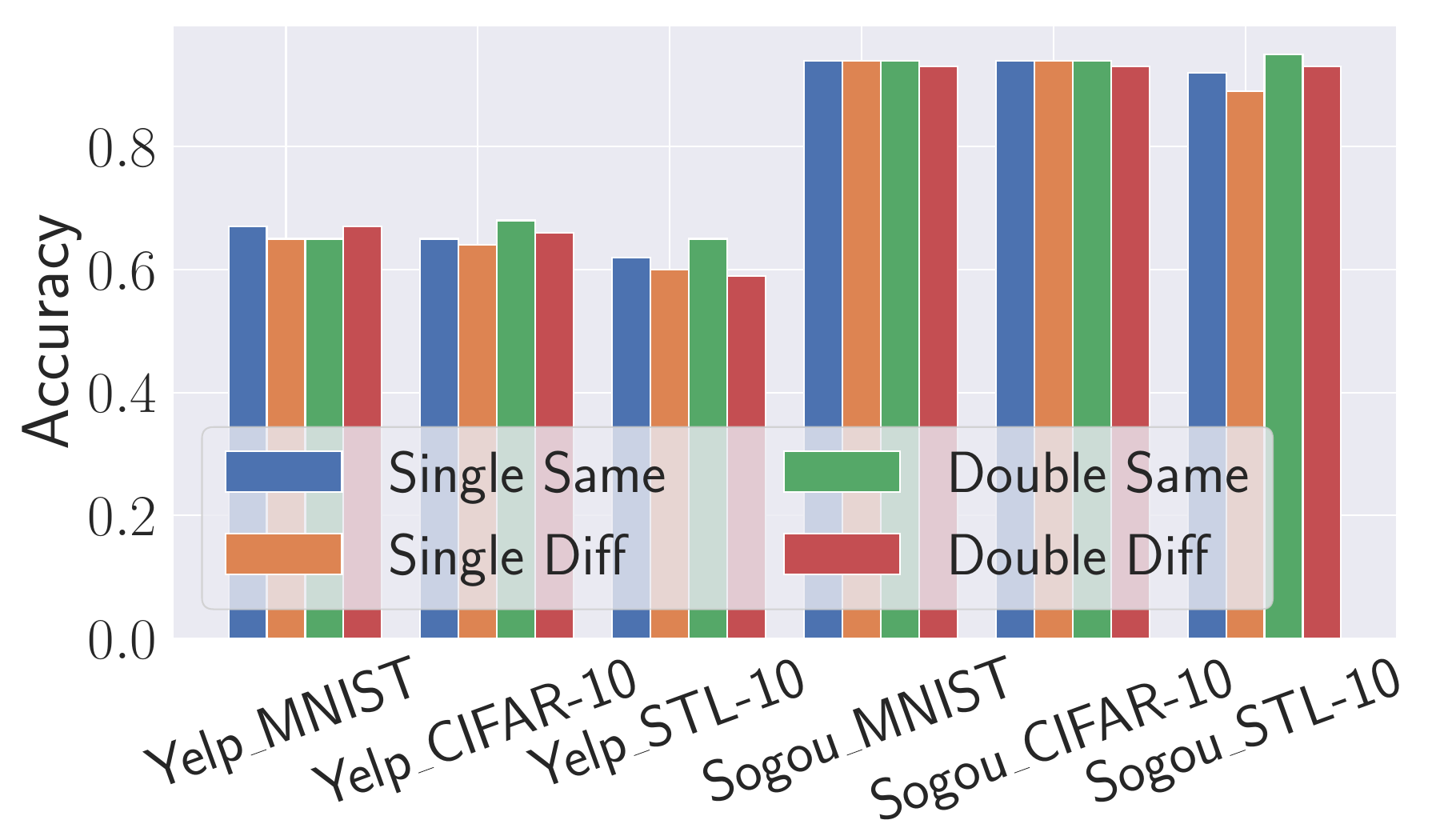}}
\subfigure[Utility.]{
\label{figure:transferability_1vs1_utility}
\includegraphics[width=0.41\textwidth]{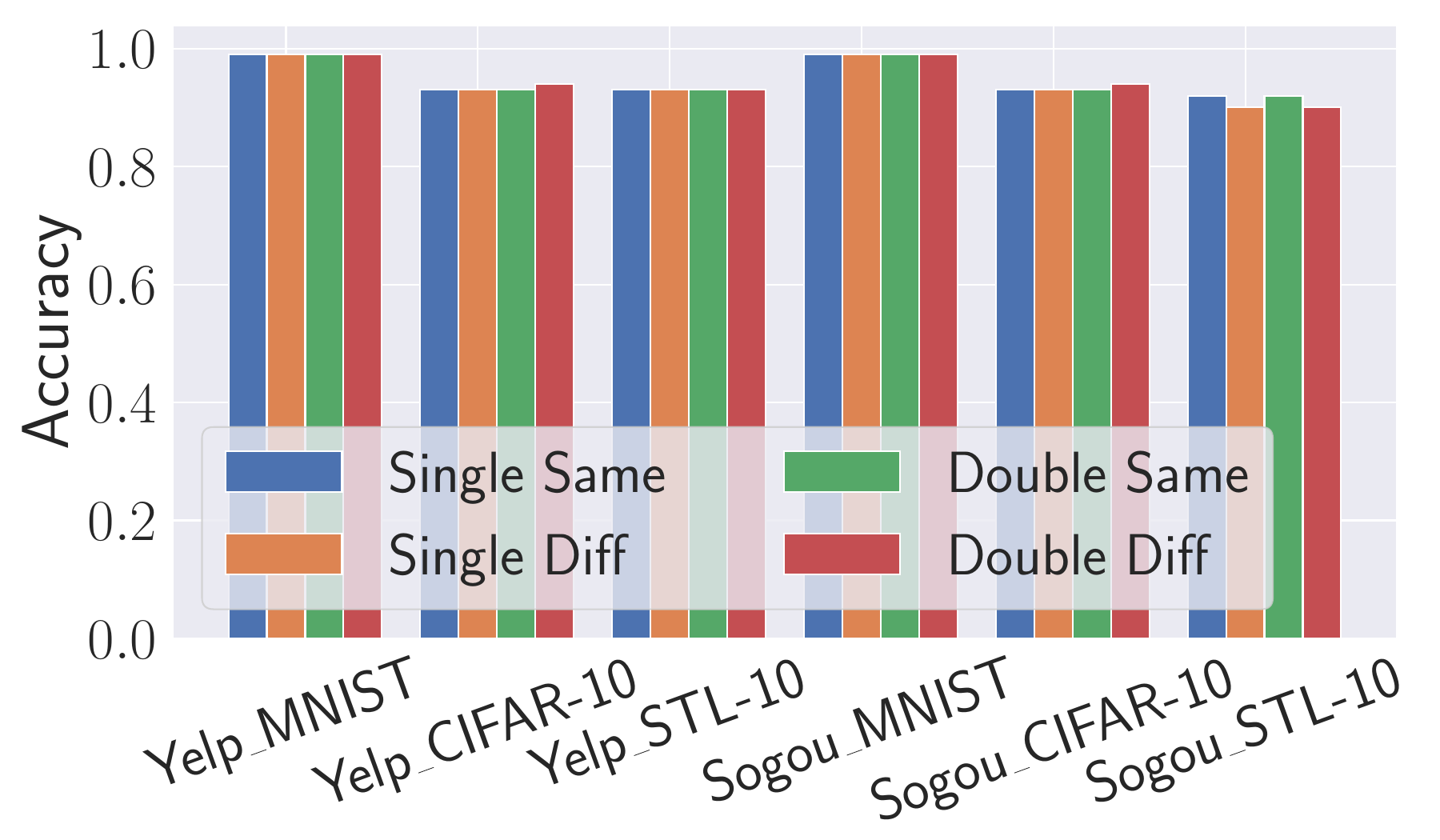}}
\caption{Our multimodal hijacking attack performance on different container datasets when the Blender is trained on Yelp(Sogou) but the attack is conducted on the hijacking dataset of Sogou (Yelp).
The ``Same'' indicates the same training and hijacking datasets, while the ``Diff'' means different.}
\label{figure:transferability_1vs1}
\end{figure}

\begin{figure}[!t]
\centering
\subfigure[Attack Success Rate.]{
\label{figure:transferability_2vs1_asr}
\includegraphics[width=0.41\textwidth]{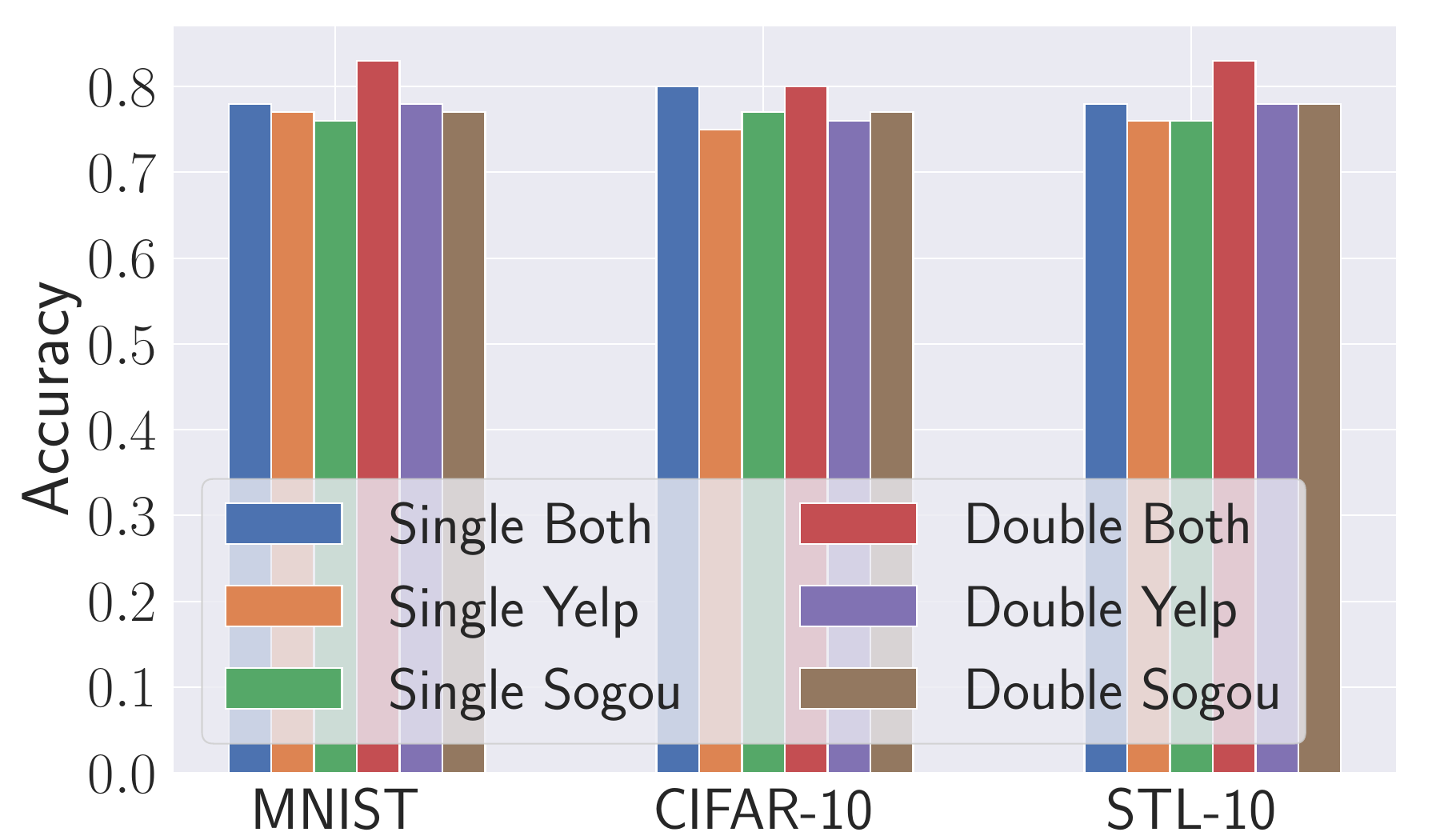}}
\subfigure[Utility.]{
\label{figure:transferability_2vs1_utility}
\includegraphics[width=0.41\textwidth]{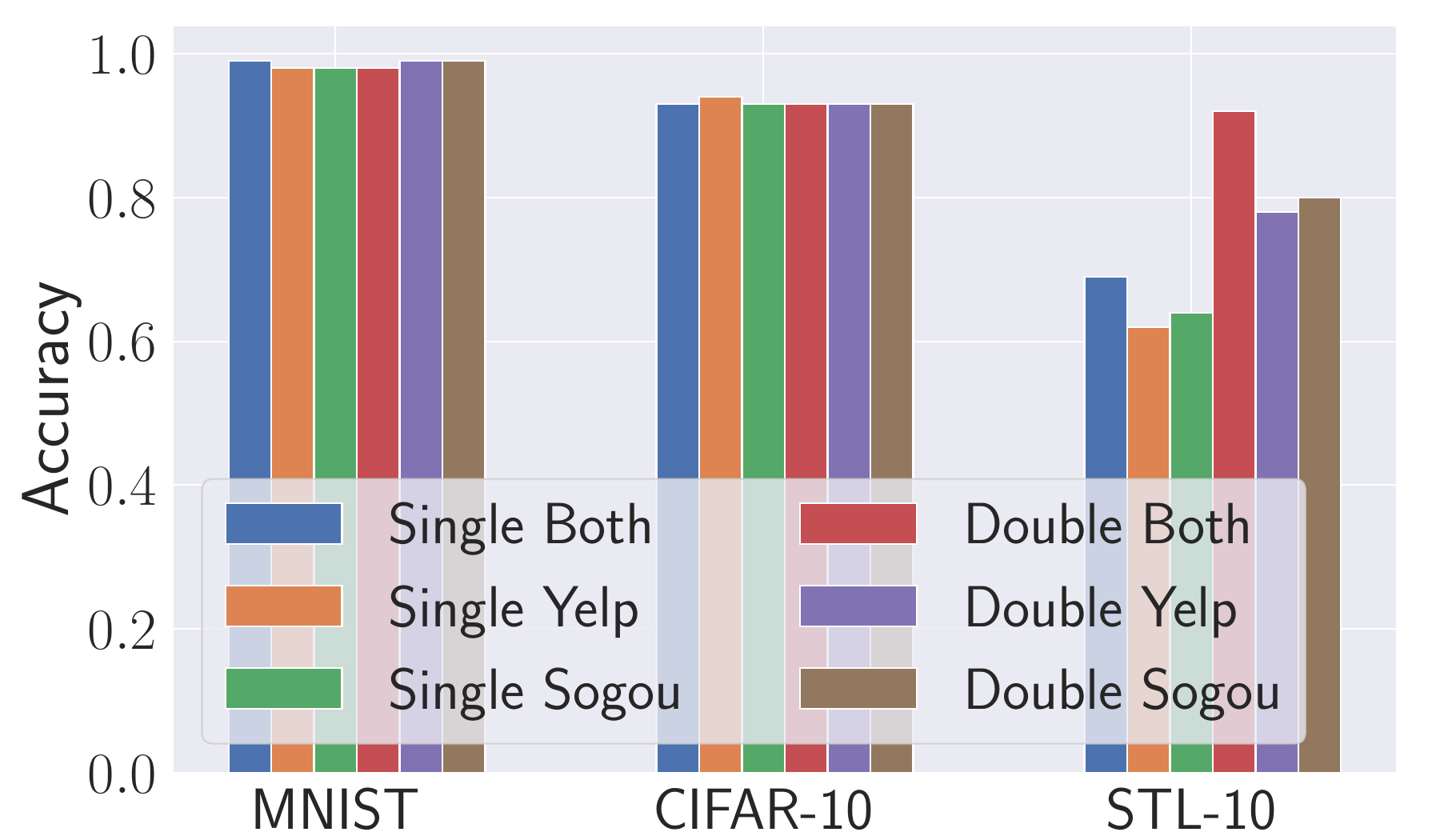}}
\caption{Our multimodal hijacking attack performance on different container datasets.
The ``Both'' indicates that the Blender is trained on both Yelp and Sogou.
The ``Yelp'' indicates that the Blender is trained on Yelp.
The ``Sogou'' indicates that the Blender is trained on Sogou.
The attack is conducted on a third hijacking dataset (CoLA).}
\label{figure:transferability_2vs1}
\end{figure}

\begin{figure}[!t]
\centering
\subfigure[Airplane.]{
\label{figure:notclear_airplane}
\includegraphics[width=0.08\textwidth]{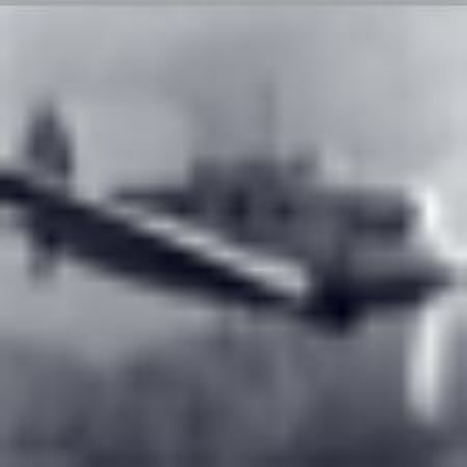}}
\subfigure[Car.]{
\label{figure:notclear_automobile}
\includegraphics[width=0.08\textwidth]{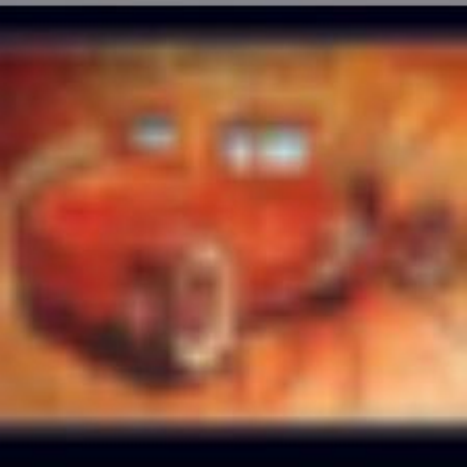}}
\subfigure[bird.]{
\label{figure:notclear_bird}
\includegraphics[width=0.08\textwidth]{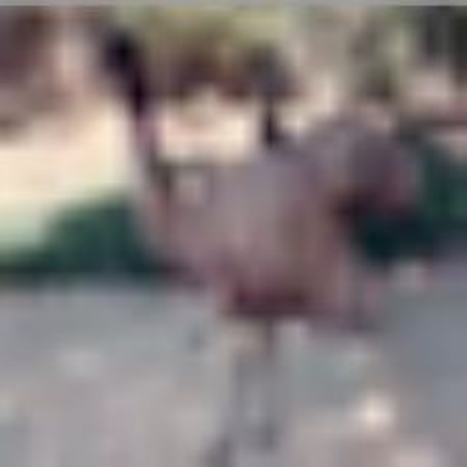}}
\subfigure[Cat.]{
\label{figure:notclear_cat}
\includegraphics[width=0.08\textwidth]{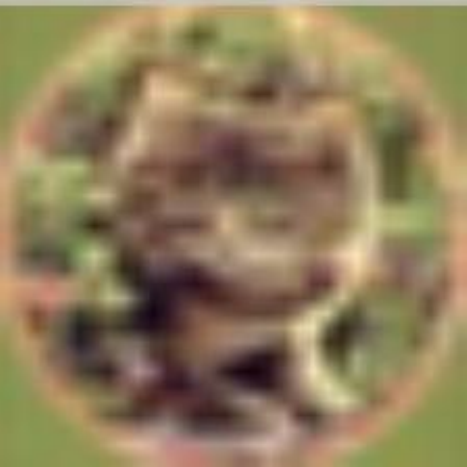}}
\subfigure[Deer.]{
\label{figure:notclear_deer}
\includegraphics[width=0.08\textwidth]{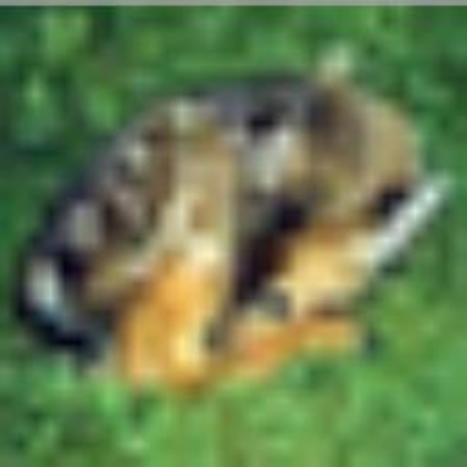}}

\subfigure[Dog.]{
\label{figure:notclear_dog}
\includegraphics[width=0.08\textwidth]{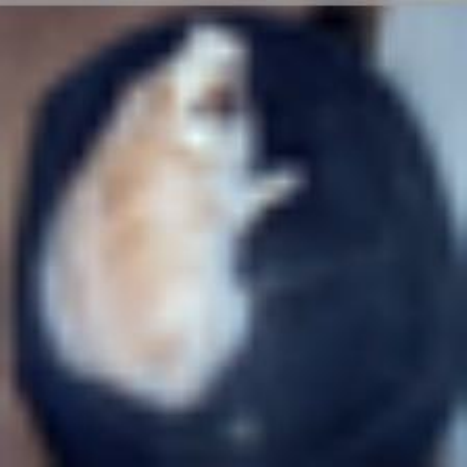}}
\subfigure[Frog.]{
\label{figure:notclear_frog}
\includegraphics[width=0.08\textwidth]{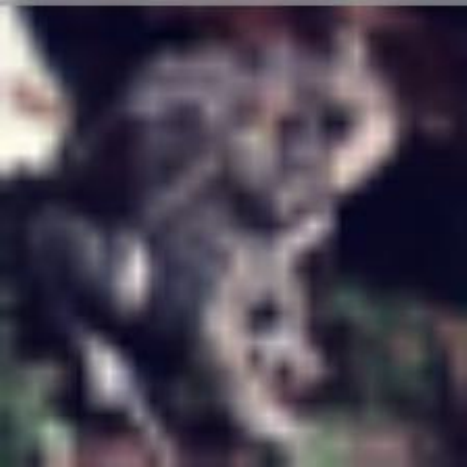}}
\subfigure[Horse.]{
\label{figure:notclear_horse}
\includegraphics[width=0.08\textwidth]{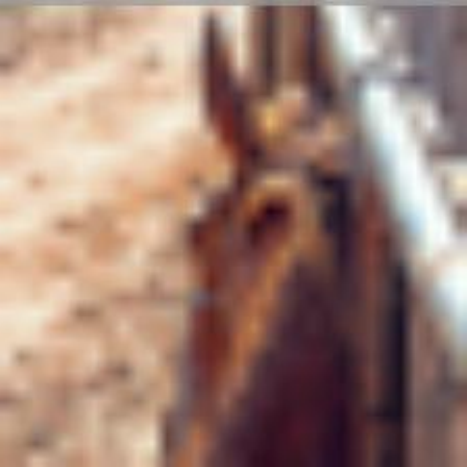}}
\subfigure[Ship.]{
\label{figure:notclear_ship}
\includegraphics[width=0.08\textwidth]{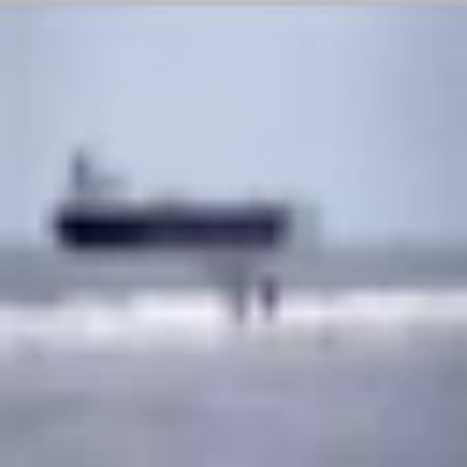}}
\subfigure[Truck.]{
\label{figure:notclear_truck}
\includegraphics[width=0.08\textwidth]{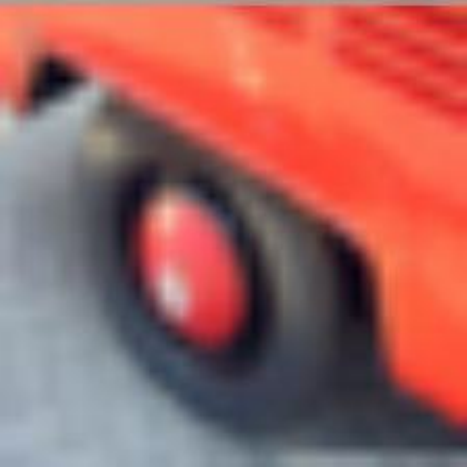}}
\caption{The visual results of examples for each class of CIFAR-10 that are misclassified with the most confidence.}
\label{figure:q9vision}
\end{figure}

\mypara{Using Different NLP Features}
To recap, our attack utilizes the embeddings on all tokens instead of only using the``[cls]'' one~\cite{DCLT19}.
We now evaluate the performance of our attack when using only the ``[cls]'' token's embedding.
To this end, we use the Yelp and CIFAR-10 hijacking and original datasets, respectively.
As depicted in \autoref{figure:q1performance}, with double encoders, using only the ``[cls]'' token's embedding reduces the attack success rate from $68\%$ to $22\%$.
However, surprisingly, using a single encoder could even slightly improve the attack success rate.
Besides, using the ``[cls]'' token's embedding does not change the model's utility (similar to our decision of encoding all tokens as \autoref{figure:q1performance} shows), as both approaches result in fused images – with features – that are distinct from the original images.
Hence, the target hijacked model is able to learn the original task without any loss of utility.

\mypara{Naive Attack}
We now evaluate the naive approach for the modal hijacking attack, i.e., directly using the output of the adapter instead of the Blender.
In this part, we take the Blender with double encoders as an example.
To this end, we evaluate the naive approach for all of the hijacking and original datasets.
The results show that the naive approach can achieve almost the same performance as our attack for some datasets, e.g., MNIST and CIFAR-10.
However, the victim models' utility drops significantly on STL-10.
For instance, the utility of the victim models drops separately to $46\%$ and $33\%$ when using the Yelp and Sogou datasets.
We present the full results in \autoref{figure:q2performance}.
Moreover, we plot the resulting images from the naive approach in \autoref{figure:adapterimages}.
As the figures show, the resulting images clearly look random and consequently easy to detect, which violates the goal of stealthiness.
These results show the advantage of using the Blender to perform the attack over the naive approach.

\mypara{Different Feature Extractors}
We now show the generalizability of our modal hijacking attack by using different feature extractors.
More concretely, instead of using BERT and VGG11 to train the Blender, we use BART~\cite{LLGGMLSZ20} and MobileNetv2~\cite{SHZZC18} as the NLP and CV feature extractors, respectively.
The results show that changing the feature extractors yields similar performance regardless of the encoder setting.
For instance, with double encoders, it achieves $68\%$, $69\%$, and $63\%$ ASR with a negligible drop in utility, when using the Yelp dataset to hijack MNIST, CIFAR-10, and STL-10, respectively.
We present the full results in \autoref{figure:q3performance}.
This result shows that our attack can use different feature extractors depending on the adversary's preference.

\mypara{Different Victim Models}
We now show the generalizability of our modal hijacking attack against different target victim models.
To this end, we use the Yelp dataset as the hijacking task for all of the CV tasks while using the AlexNet~\cite{KSH12} as the victim model's architecture.
As expected, our modal hijacking attack still achieves strong performance against the AlexNet-based models in both cases of using double encoders and a single one.
For example, with double encoders, it achieves $67\%$, $67\%$, and $66\%$ ASR with the utility drop within $3\%$ when hijacking MNIST, CIFAR-10, and STL-10 classification models, respectively.
We present the full results in \autoref{figure:q4performance}.
These results show that our modal hijacking attack is indeed independent of the victim's model architecture.
We believe this independence is due to the robust feature extraction of our attack.
More concretely, using state-of-the-art language models to extract features from the hijacking datasets results in robust features that can be picked up by different target model architectures when fused with container images.

Intuitively, we hypothesize that the effectiveness of our hijacking attack relies on the redundant model capacity.
That is, the victim model could simultaneously support its original task (i.e., stealthiness) and the hijacking task (i.e., effectiveness).
To this end, we explore the case of using an MLP model as the victim model, as the MLP model can only achieve the utility of $53\%$, which is much lower than CNNs.
Such a utility gap indicates that MLP models do not own redundant capacity.
In that case, if the hypothesis is accepted, our hijacking attack against MLP models should achieve poor performance.
However, in \autoref{figure:q4performance}, it is surprising to find that our attack performs strongly in both stealthiness and effectiveness.
These results indicate that low-capacity victim models are also vulnerable, which further increases the risks of our attack.

\mypara{Number of Training Epochs for Blender}
We evaluate the effect of varying the number of the Blender training epochs using the Yelp-CIFAR-10 setting, where we take the Blender with double encoders as an example.
We train Blenders using from 50 to 500 epochs with steps of 50, then evaluate our attack performance on them.
Our results show that the utility and ASR do not change much with the number of epochs (\autoref{figure:q5accuracy}).
However, the quality of the fused images does get better with a higher number of epochs.
The visual quality saturates at approximately 200 epochs.
Hence, the adversary can use 50 epochs if they care less about the visual appearance of the fused images; otherwise 200 epochs would be a good compromise.
We show a set of randomly sampled fused images for the different epochs in \autoref{figure:q5vision}.

\mypara{Effect of Fine-tuning BERT}
We now evaluate the effect of fine-tuning the NLP feature extractor ($\nlpextractor$).
To this end, we use the Yelp dataset to hijack a CIFAR-10 classification model with a non-fine-tuned BERT.
Our results show that the utility of the victim model does not change; however, the ASR is significantly impacted, i.e., the ASR significantly drops.
This shows the need to fine-tune $\nlpextractor$ on the hijacking dataset.
Besides, even the non-fine-tuned one with more poisons is not comparable to a fine-tuned NLP feature extractor.
Another interesting finding is that our single-encoder Blender is more resilient to the non-fine-tuning case.
We present the full results in \autoref{figure:q6performance}.

\mypara{The Poisoning Rate Effect}
We now, to better understand the adversary's capacity, evaluate the influence of the poisoning rate on our attack.
We use the Yelp-CIFAR-10 setting (with 60,000 clean images) and set the number of poisoning -- fused -- samples from 2,500 (4.2\%) to 10,000 (16\%) with steps of 2,500 to hijack different models.
The results show that using 2,500 samples is too few to hijack the model, i.e., the ASR is less than $60\%$.
However, increasing the points beyond 5,000 (8.3\%) does not improve the ASR.
Hence, we use 5,000 hijacked samples for our evaluations in \autoref{section: evaluation}.

In application, the amount of data to poison the dataset here depends on the adversary.
On the one hand, as our attack implements a new task, it just depends on how accurate the adversary wants their hijacking task to be.
On the other hand, even if the poisons could be injected by involving new parities or intrusion techniques~\cite{LLLT13,KGVK19}, less modification of the dataset is intuitively less possible to be revealed, which indicates a more practical setting and more stealthiness.
It is also important to note that even if the accuracy is not state-of-the-art, the ability of the model to perform an unethical task (even with low performance) can make the model owner accountable for a violation.
Even, in that case, one successful attack case might be enough to support it.
Besides, state-of-the-art models usually have millions of inputs, e.g., ImageNet has more than {1.2~M} images.
Hence, the amount of poisoning data can be negligible depending on the application.
Further, we find that the Blender with double encoders are more sensitive to the poisoning rate.
We present the full results in \autoref{figure:q7performance}.

\mypara{Reusability of the Blender}
As the Blender can be expensive to train, we evaluate reusing a trained Blender to hijack different settings. 
We try four different setups:
The first hijacks models using the same container dataset but different victim models;
The second increases the flexibility and uses a different container dataset;
The third and the fourth further increase the efficiency of the attack by using a pre-trained Blender to camouflage different hijacking datasets.
For the first case, we already presented its results when evaluating our attack against different target victim models, i.e., AlexNet and MLP.
For the second case, we use the Blender trained using the Yelp hijacking and CIFAR-10 container datasets to hijack MNIST and STL-10 models.
In other words, we use the CIFAR-10 trained Blender to separately fuse images from the MNIST and STL-10 datasets.
Our results show that the modal hijacking attack still achieves strong performance.
More concretely, it achieves the same utility compared to using their original Blender (\autoref{subsection: results}) with a non-significant ASR drop for the MNIST and STL-10 hijacked models, respectively.
We present the full results in \autoref{figure:q8performance}.
For the third case, we use the Blender trained on Yelp (Sogou) but conduct the attack with the hijacking dataset of Sogou (Yelp).
In that setting, our modal hijacking can gain almost the same performance.
For instance, when the adversary trains a double-encoder Blender on the Yelp (Sogou) dataset to hijack a CIFAR-10 model using the Sogou (Yelp) hijacking dataset, our attack achieves an ASR of $93\%$ ($66\%$) and utility of $94\%$ ($94\%$), which is only $0\%-2\%$ lower compared to training a specific Blender for each hijacking dataset.
Meanwhile, the single-encoder cases show similar results.
We present the full results in \autoref{figure:transferability_1vs1}.
For the final case, we try a different setup where two hijacking datasets (Yelp and Sogou) are used to build a Blender.
Then this Blender is used to attack a third dataset (CoLA~\cite{WSMHLB18}, which is a binary dataset).
This approach shows that it is indeed better to use two datasets instead of one when building the Blender.
More concretely, using the double-encoder Blender trained jointly on both datasets (Yelp and Sogou) improves the performance to 80\% ASR and 93\% utility, which is 4\% (3\%) stronger in ASR than when using a Blender trained on a single dataset, e.g., Yelp (Sogou).
We present the full results in \autoref{figure:transferability_2vs1}.
These results demonstrate that a single Blender is reusable to different setups, which significantly reduces the cost and increases the flexibility of our modal hijacking attack.

\mypara{Container Dataset Creation}
Finally, we propose a way of constructing the container dataset.
So far, we use a randomly selected container dataset.
One problem is that the labels might not align with the fused samples.
Hence, a manual inspection can raise some flags.
To this end, we now propose a more stealthy way to construct the container dataset.
We first train a CV classifier on the container dataset.
Next, we sort the images that are misclassified with the most confidence.
Finally, we manually select the container dataset out of these images. \autoref{figure:q9vision} shows a subset of these images.
As the figure shows, it is hard to manually assign a label to these images, which makes the fused samples more stealthy.

\subsection{Defense}
\label{subsection: defense}

\begin{figure}[!t]
\centering
\subfigure[Attack Success Rate.]{
\label{figure:defense_filter_asr}
\includegraphics[width=0.41\textwidth]{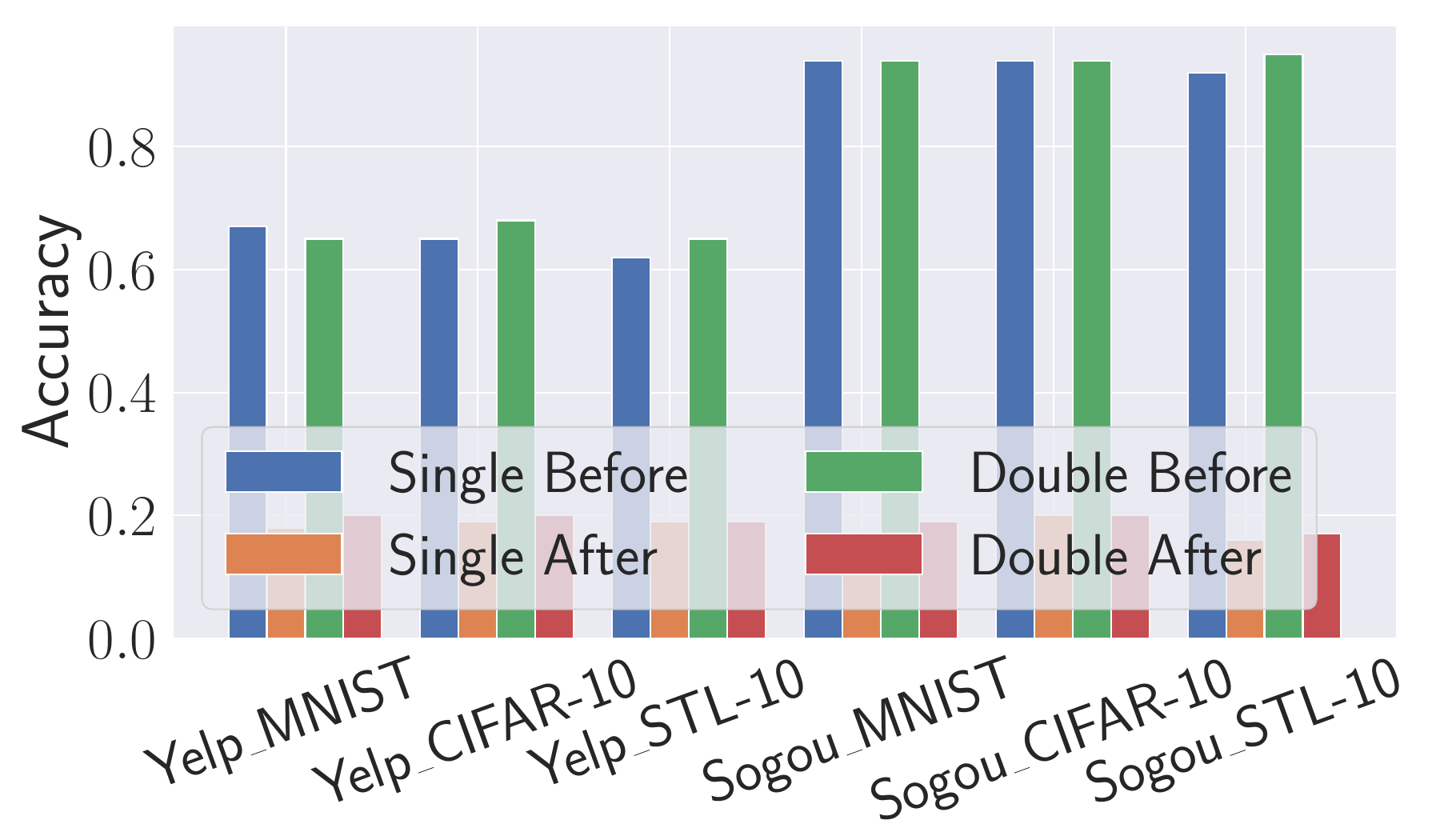}}
\subfigure[Utility.]{
\label{figure:defense_filter_utility}
\includegraphics[width=0.41\textwidth]{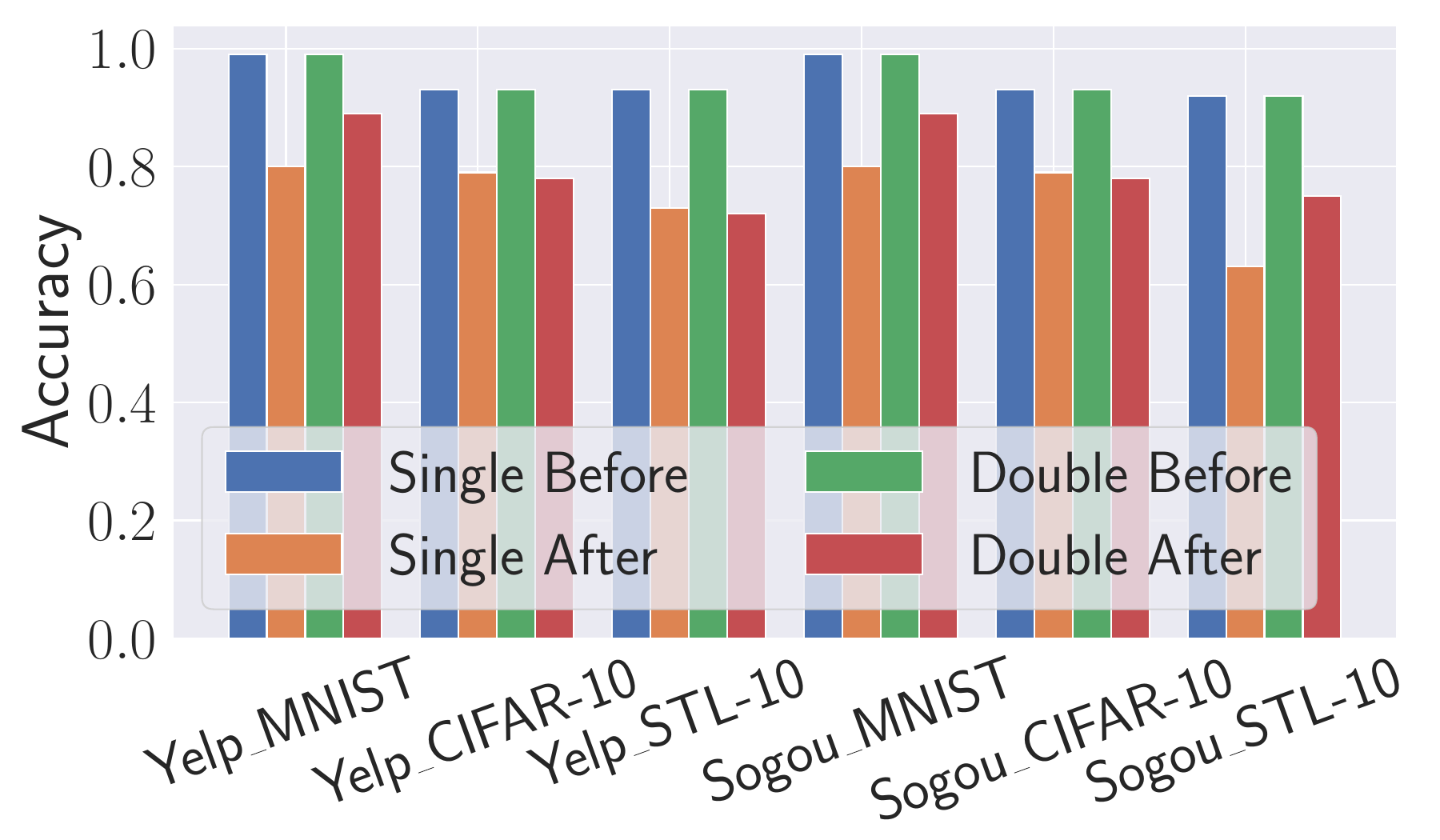}}
\caption{The defense performance against our modal hijacking attack by filtering poisons~\cite{SKL17}, where the hijacking tasks are Yelp and Sogou and the original tasks are MNIST, CIFAR-10 and STL-10.}
\label{figure:defense_filter}
\end{figure}

\begin{figure}[!t]
\centering
\subfigure[Attack Success Rate.]{
\label{figure:defense_epic_asr}
\includegraphics[width=0.41\textwidth]{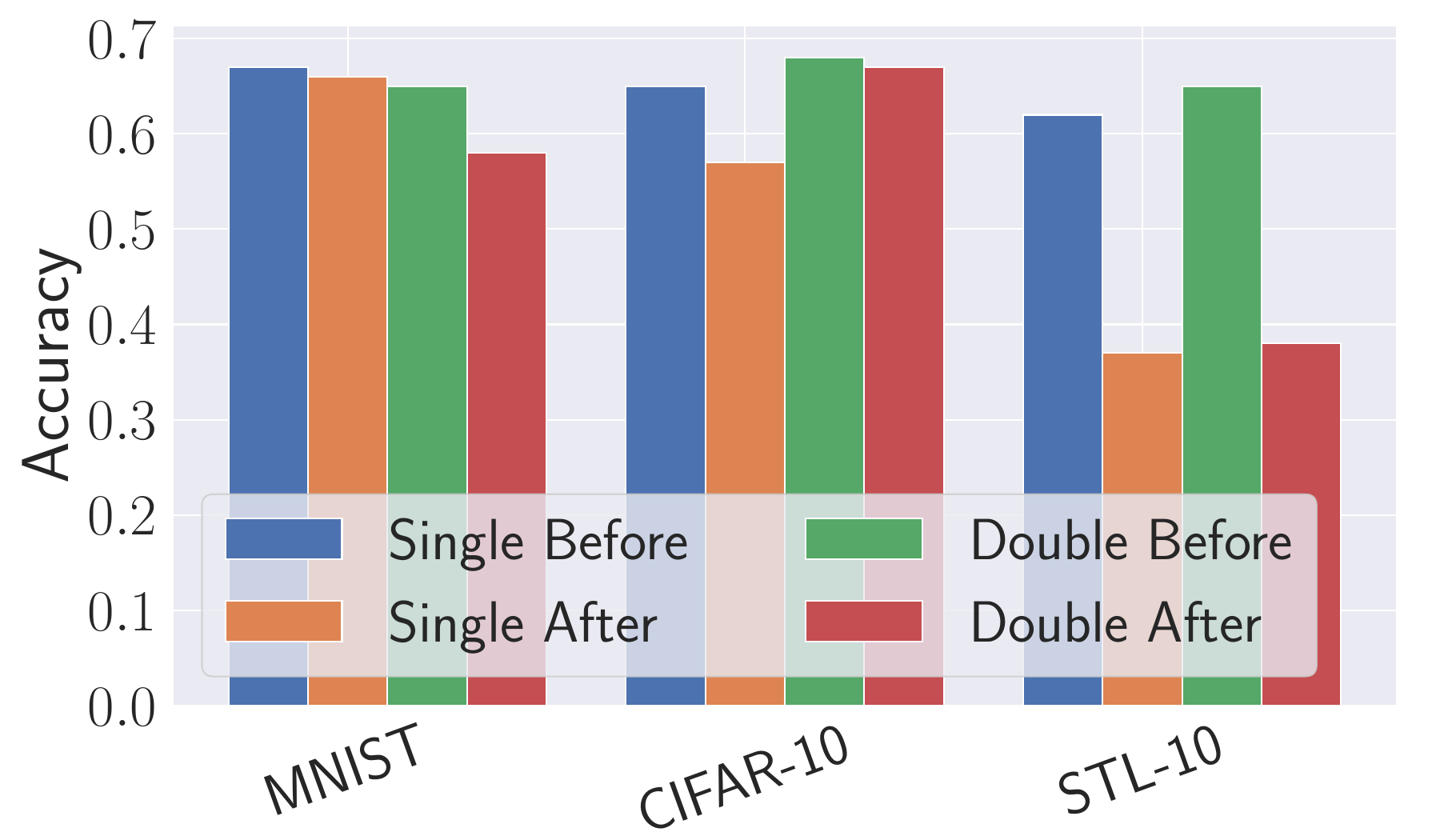}}
\subfigure[Utility.]{
\label{figure:defense_epic_utility}
\includegraphics[width=0.41\textwidth]{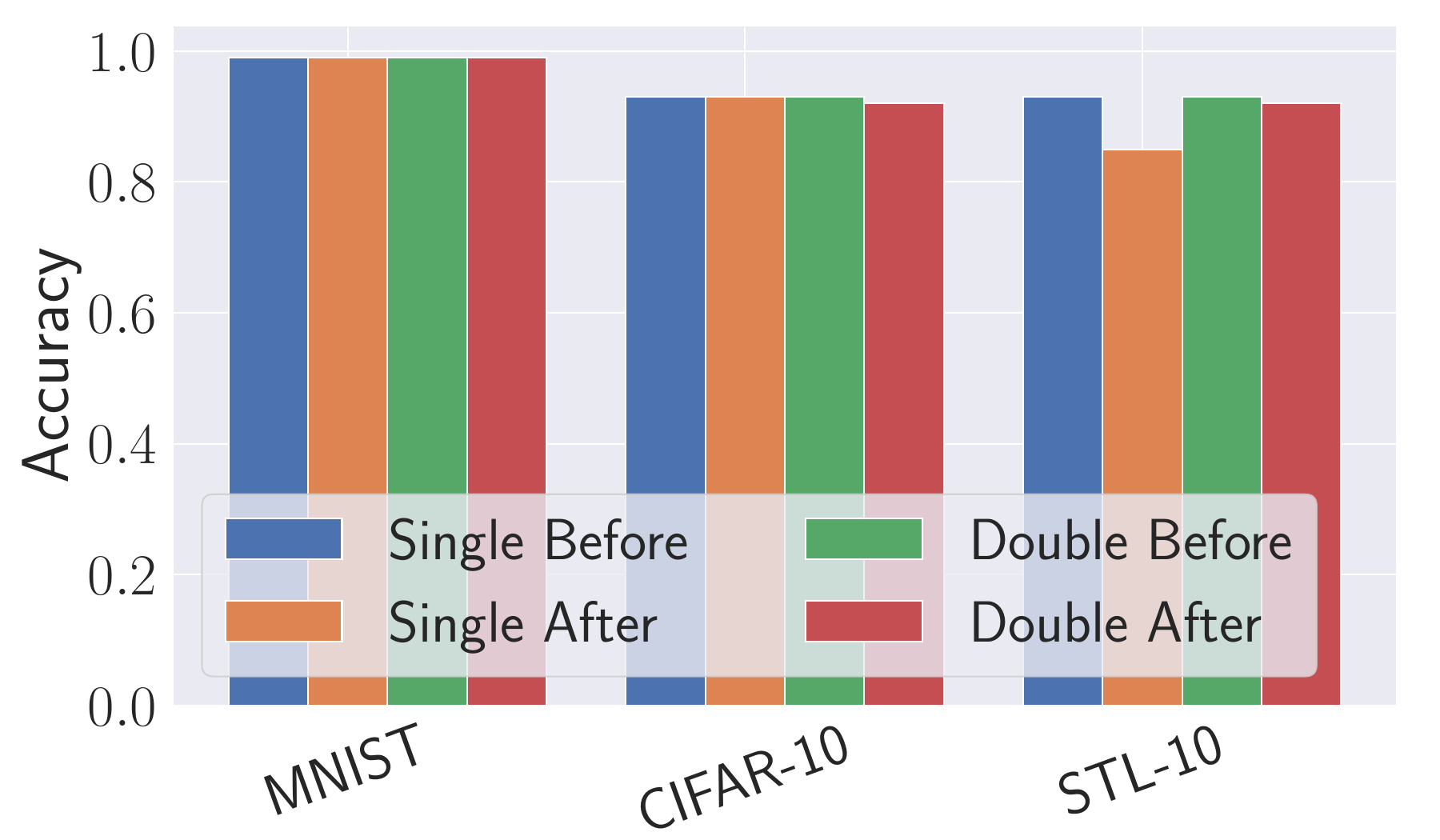}}
\caption{The defense performance against our modal hijacking attack by EPIC~\cite{YLM22}, where using the three original datasets MNIST, CIFAR-10 and STL-10, Resnet18 as the target model, and Yelp as the hijacking dataset.}
\label{figure:defense_epic}
\end{figure}

We evaluate using two established defenses against data poisoning attacks to mitigate our modal hijacking attack~\cite{SKL17,YLM22}.
The first defense, intuitively, clusters a clean dataset and computes the centroid of each label.
Then for any given input, the distance between it and its corresponding centroid -- depending on its label -- is calculated.
Inputs with distances exceeding the threshold are then discarded.
We evaluate this defense technique on the hijacking tasks of Yelp and Sogou and the original tasks of MNIST, CIFAR-10 and STL-10.
The results consistently show that the attack performance drops to almost random guesses, however, it induces an average drop of the utility of around 15\%, where using double encoders is more robust.
Moreover, this type of defense requires access to clean data which is sometimes hard to get for many applications in practice.
The second defense EPIC aims to find effective poisons and drop them in low-density gradient regions during training.
The results show that the defense indeed reduces the attack success rate, however, our hijacking attack is still effective in most cases.
We present the complete results in \autoref{figure:defense_filter}.

\section{Discussion}
\label{section: discussion}

\subsection{Challenges in Multiple-Modality Scenario}
\label{subsection:challengeandfuturework}

As aforementioned, the main challenge in our work is that NLP and CV tasks are separately located in discrete and continuous feature spaces.
This gap boosts the design of our Adapter and of course our Blender, which generalizes our straightforward but effective framework of hijacking attacks from the homogeneous to the heterogeneous modalities.
Such an improvement largely benefits the practical usage of hijacking attacks.
In practice, this multimodal setting is prevalent since deep learning is now deployed in various applications, e.g., facial recognition, self-driving, and text-to-image generation.
Intuitively, our technique can be adapted for different scenarios where the hijacking task has feature extractors, e.g., language models for the NLP tasks.
However, different modalities have different difficulties, i.e., how easy/hard the original data can be modified, which indicates a non-trivial challenge of adaptively implementing hijacking attacks.

Utilizing the representative tasks in NLP and CV domains, our work serves as the pioneer in exploring hijacking attacks in the multimodal scenario.
We hope that inspired by our work, more complicated scenarios could be applicable in future works.
For instance, hijacking an NLP model with a CV task is an interesting next step but would be significantly difficult/different, as text inputs are not continuous and hence cannot be changed as the images do.

\subsection{Stealthiness of Fused Images}
\label{subsection: stealthiness of fused images}

Compared to the fused images of previous work~\cite{SBZ22} (where MNIST digits are visible in the fused CIFAR-10 images), ours look much more natural as shown in \autoref{table:completevision}, indicating more stealthiness.
Nevertheless, few noises could still be observed.
We plan to reduce these artifacts by adding a GAN-like discriminator~\cite{GPMXWOCB14} to the training of the Blender in future work.
This discriminator is trained to differentiate between the container and fused samples.
The Blender is penalized for any container sample that the discriminator can identify, hence improving the appearance of the fused samples.

\subsection{Computational Costs of Training Blender}
\label{subsection: computational costs of training blender}

Another issue to consider is the computational cost of training the Blender.
Even though this is not the main scope of our work, minimizing the cost of hijacking attacks could improve the adversary's ability in practice.
To this end, we show in \autoref{subsection: hyperparameters/design decisions} that our Blender can be trained once and then utilized for multiple hijacking attacks, even for different hijacking tasks, making it more efficient.
Besides, our extensive evaluations show the feasibility of only using a single encoder in our Blender instead of double ones, which largely saves computational costs.

\section{Conclusion}
\label{section: conclusion}

Model hijacking attack is a new threat that exploits the inclusion of new parties in the ML training pipeline.
In this attack, the adversary can hijack CV-based models to implement their own image classification task.
However, as ML has improved in multiple domains besides CV, the hijacking and original tasks might be from different data modalities.
In the paper, we propose a more general hijacking attack in the \emph{multimodal} scenario, where the adversary can hijack CV models using NLP tasks, which increases the practical usage of hijacking attacks.
To this end, we propose an autoencoder-based model that mixes language models with CNNs, namely the Blender, to perform our \emph{modal hijacking attack}.
Using the Blender, the adversary can hijack image classification models using text/NLP hijacking tasks.
We extensively evaluate our attack using five different datasets, including three image classification datasets and two text ones.
Our results show that the modal hijacking attack achieves strong performances (i.e., effectiveness) with a negligible drop in the model's utility (i.e., stealthiness).
Besides, the relaxed encoder requirement and improved visual appearances of poisons indicate the advantages of computational efficiency and stealthiness over the previous model hijacking attack.

We aim by this work to first raise awareness of possible accountability risks in some of the realistic machine learning training pipelines.
Then, to motivate the community to work on different mitigation techniques to address this risk, we already present a couple of possibilities.
Finally, our modal hijacking technique can also be used for compressing target models, hence reducing the training or maintenance costs.


\bibliographystyle{plain}
\bibliography{cite}

\clearpage
\appendix

\onecolumn
\section{Visual Results}
\label{appendix: visual results}

\begin{table}[H]
    \caption{The complete version of \autoref{table:completevision} which depicts the details of the fused and corresponding container images across different hijacking and container datasets. For the fused images, labels are assigned based on the NLP sentences. For example, if the ground truth of a given NLP sentence is ``5 stars'', the label ``Digit 4'' would be assigned to all fused MNIST-like images that contain the NLP feature of this sentence.}
    \resizebox{\textwidth}{!}{
    \centering
    \begin{tabular}{l | c | c c c c | c c}
        \toprule
        \multirow{2}{*}{\textbf{\makecell{Hijacking\\dataset}}} & \multirow{2}{*}{\textbf{\makecell{Container\\dataset}}} & \multicolumn{4}{c|}{\textbf{Fused}} & \multicolumn{2}{c}{\textbf{Container}} \\
        & & Image & Image label & NLP sentence & \makecell{NLP\\ground truth} & Image & Image label  \\
        \midrule
        & MNIST & 
            \begin{minipage}[b]{0.3\columnwidth}
    		\centering
    		\raisebox{-.5\height}{\includegraphics[width=\linewidth]{pics/yelp_mnist_fused.pdf}}
    	\end{minipage} & Digit 4 & \makecell{``Some of the best chow\\around--love this place.\\The bread and salads\\and soups are great.''} & 5 stars & 
            \begin{minipage}[b]{0.3\columnwidth}
    		\centering
    		\raisebox{-.5\height}{\includegraphics[width=\linewidth]{pics/yelp_mnist_container.pdf}}
    	\end{minipage} & Digit 1, 6, 3 \\
        \multirow{4}{*}{Yelp} & Cifar-10 & 
            \begin{minipage}[b]{0.3\columnwidth}
    		\centering
    		\raisebox{-.5\height}{\includegraphics[width=\linewidth]{pics/yelp_cifar10_fused.pdf}}
    	\end{minipage} & Airplane & \makecell{``The worst dental office\\I ever been. No one\\can beat it!!! You should\\avoid it at any time.''} & 1 star & 
            \begin{minipage}[b]{0.3\columnwidth}
    		\centering
    		\raisebox{-.5\height}{\includegraphics[width=\linewidth]{pics/yelp_cifar10_container.pdf}}
    	\end{minipage} & Bird, Dog, Horse \\
        & STL-10 & 
            \begin{minipage}[b]{0.3\columnwidth}
    		\centering
    		\raisebox{-.5\height}{\includegraphics[width=\linewidth]{pics/yelp_stl10_fused.pdf}}
    	\end{minipage} & Bird & \makecell{``Far away from real\\Chinese food. Doesn't\\even taste good as American\\style Chinese food.''} & 2 stars & 
            \begin{minipage}[b]{0.3\columnwidth}
    		\centering
    		\raisebox{-.5\height}{\includegraphics[width=\linewidth]{pics/yelp_stl10_container.pdf}}
    	\end{minipage} & Truck, Horse, Bird \\
        & \makecell{Tiny \\ ImageNet} & 
            \begin{minipage}[b]{0.3\columnwidth}
    		\centering
    		\raisebox{-.5\height}{\includegraphics[width=\linewidth]{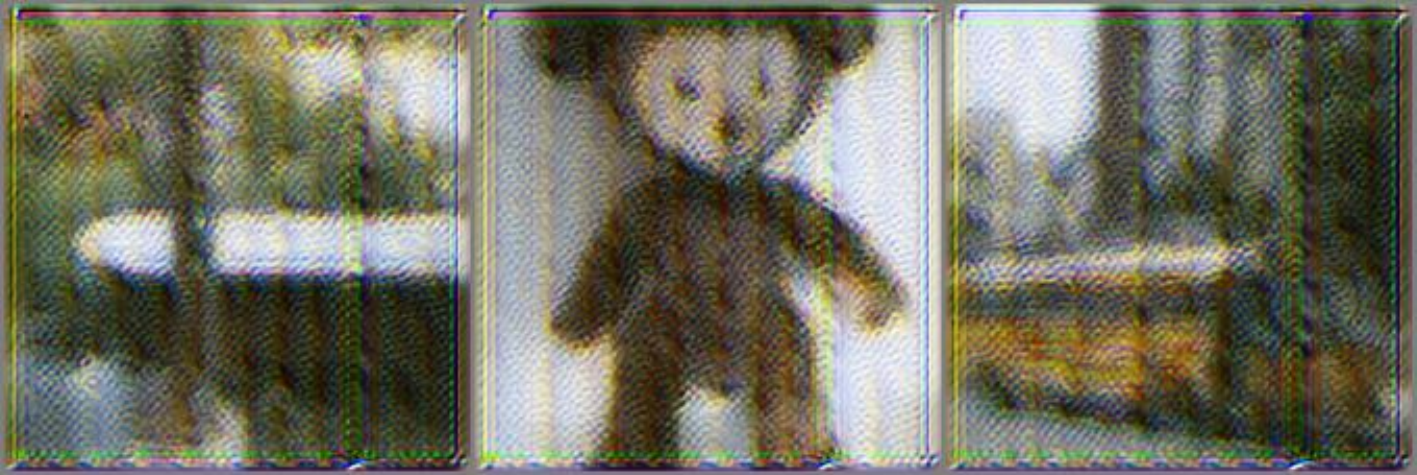}}
    	\end{minipage} & Rocker & \makecell{``Wast there last Friday.\\Seats right in front if\\the stage. The show was\\good. The headliner...''} & 4 stars & 
            \begin{minipage}[b]{0.3\columnwidth}
    		\centering
    		\raisebox{-.5\height}{\includegraphics[width=\linewidth]{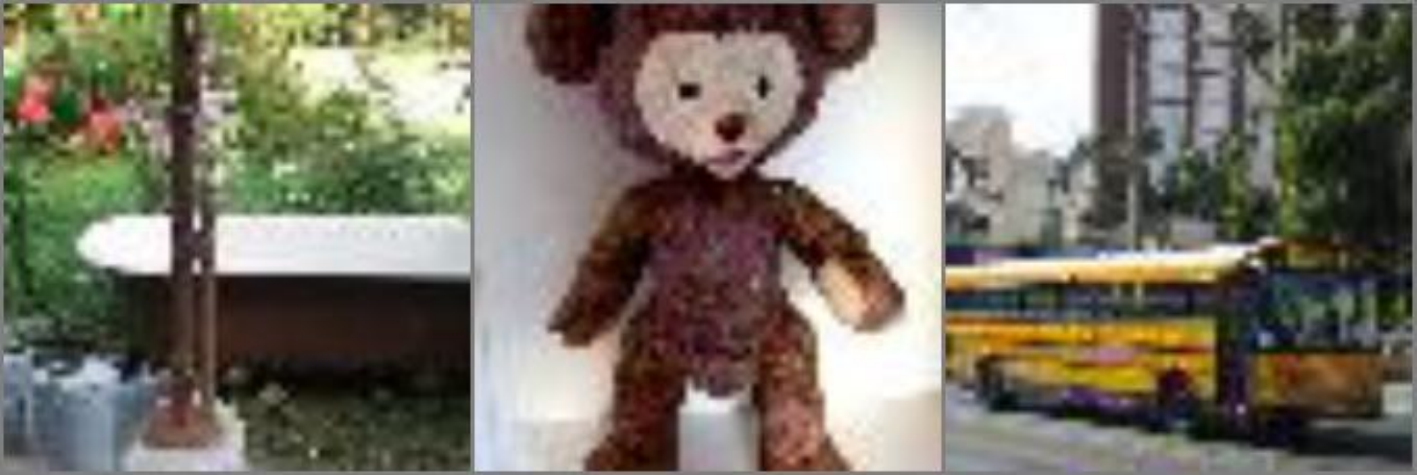}}
    	\end{minipage} & \makecell{Bathtub, Teddy,\\School bus} \\
        \midrule
        & MNIST & 
            \begin{minipage}[b]{0.3\columnwidth}
    		\centering
    		\raisebox{-.5\height}{\includegraphics[width=\linewidth]{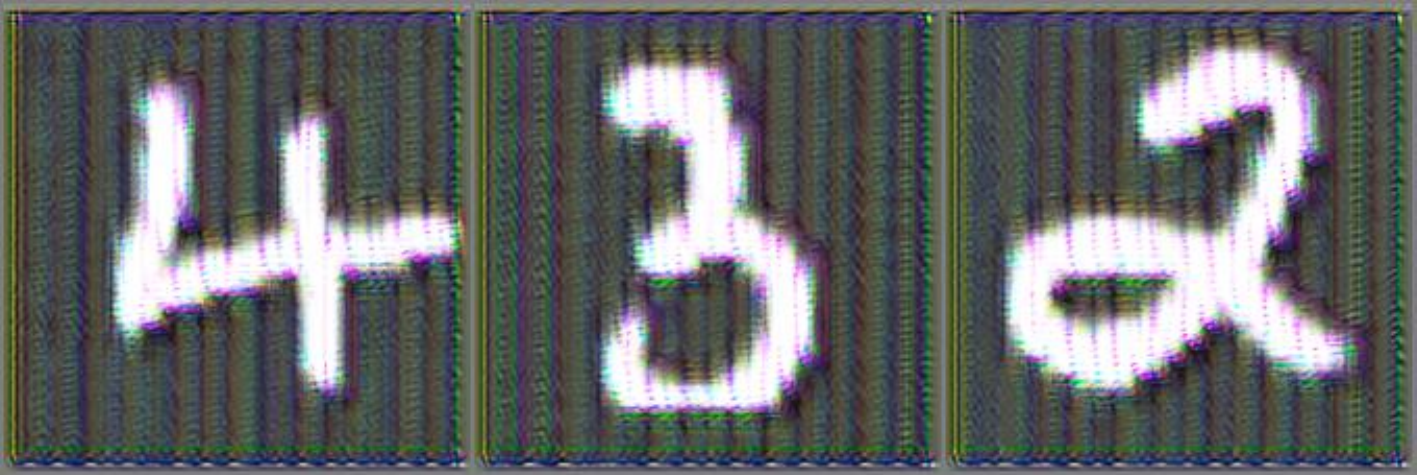}}
    	\end{minipage} & Digit 3 & \makecell{``zho1ng hua2 ju4n\\jie2 FRV ya4o shi''} & Automobile & 
            \begin{minipage}[b]{0.3\columnwidth}
    		\centering
    		\raisebox{-.5\height}{\includegraphics[width=\linewidth]{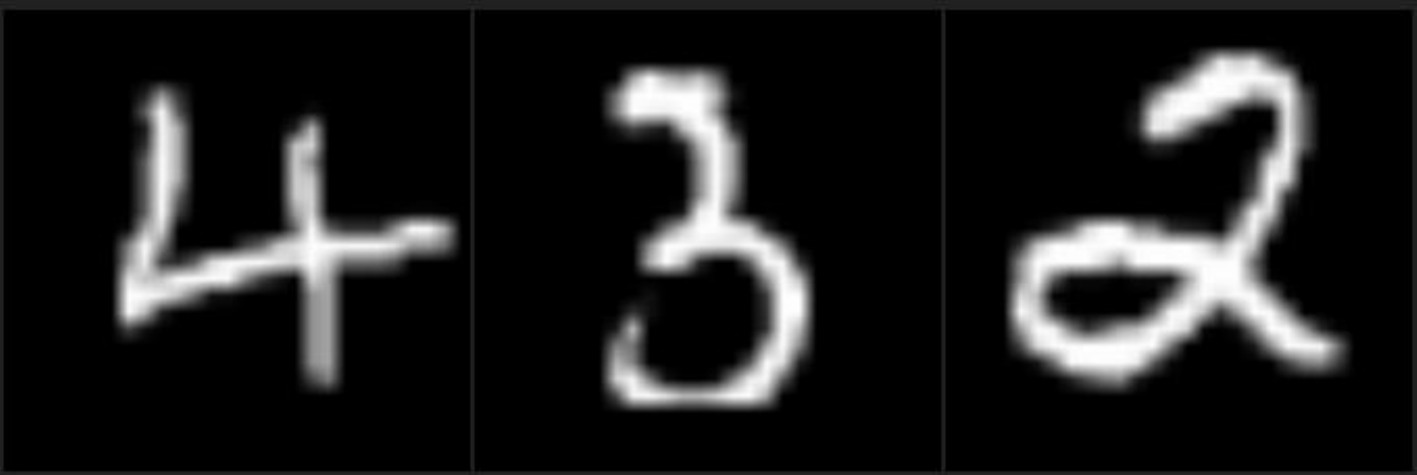}}
    	\end{minipage} & Digit 4, 3, 2 \\
        \multirow{4}{*}{Sogou} & Cifar-10 & 
            \begin{minipage}[b]{0.3\columnwidth}
    		\centering
    		\raisebox{-.5\height}{\includegraphics[width=\linewidth]{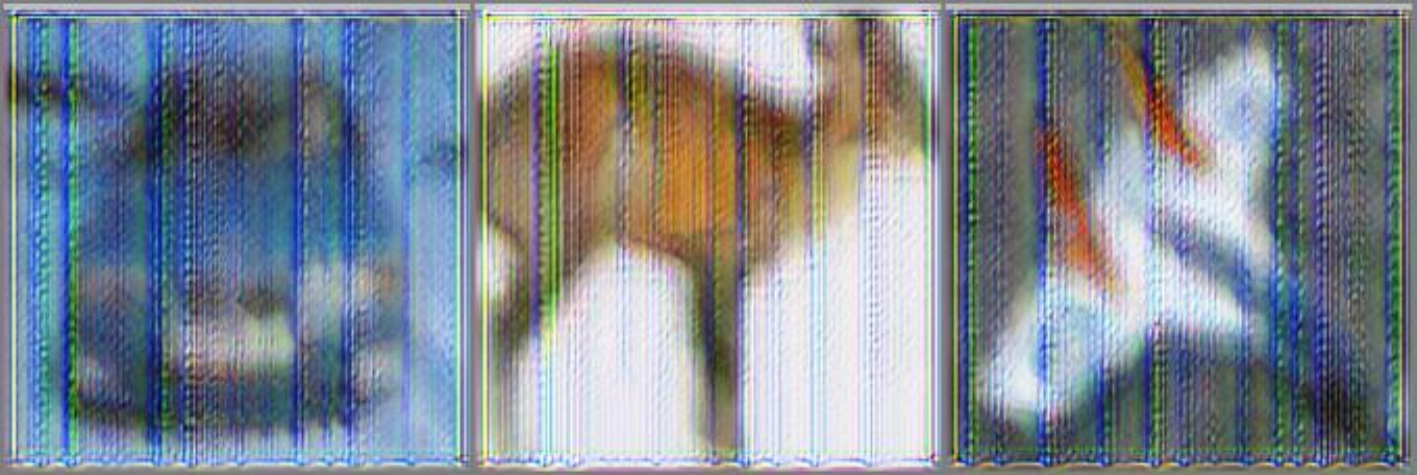}}
    	\end{minipage} & Airplane & \makecell{``2008nia2n 5 yue4 19\\ri4 , be3i ji1ng go1ng\\ye4 da4 xue2 ti3 yu4\\gua3n shi4 be3i ji1ng...''} & Sports & 
            \begin{minipage}[b]{0.3\columnwidth}
    		\centering
    		\raisebox{-.5\height}{\includegraphics[width=\linewidth]{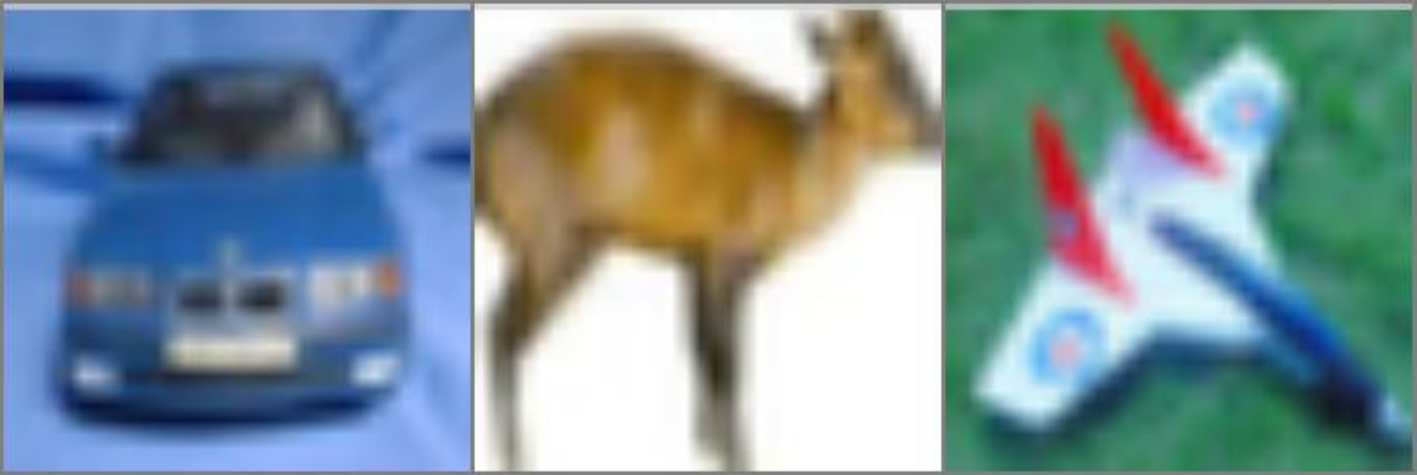}}
    	\end{minipage} & \makecell{Automobile,\\Deer, Airplane} \\
        & STL-10 & 
            \begin{minipage}[b]{0.3\columnwidth}
    		\centering
    		\raisebox{-.5\height}{\includegraphics[width=\linewidth]{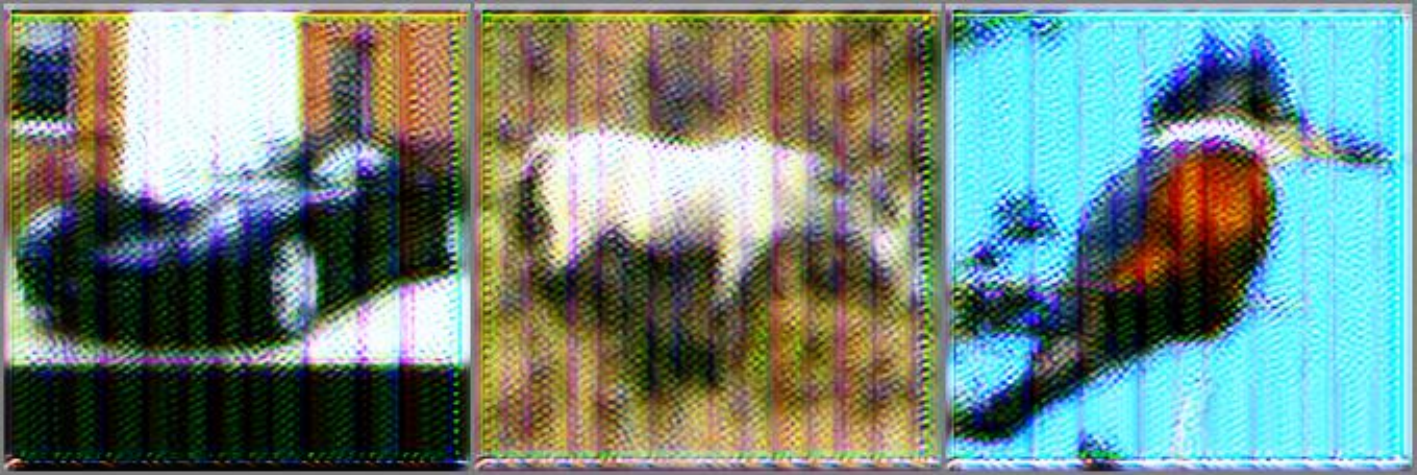}}
    	\end{minipage} & Bird & \makecell{``ya1n zha4o du1 shi4\\ba4o ju4 hu4 she1n\\zhe4ng qua4n jia1o yi4\\suo3 go1ng ka1i xi4n...''} & Finance & 
            \begin{minipage}[b]{0.3\columnwidth}
    		\centering
    		\raisebox{-.5\height}{\includegraphics[width=\linewidth]{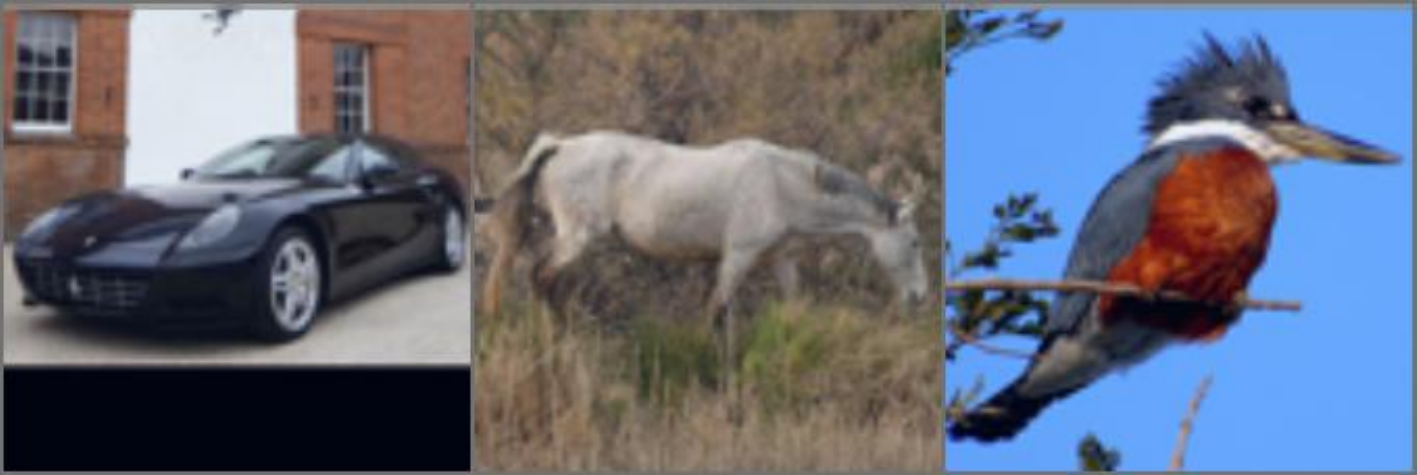}}
    	\end{minipage} & Car, Horse, Bird \\
        & \makecell{Tiny \\ ImageNet} & 
            \begin{minipage}[b]{0.3\columnwidth}
    		\centering
    		\raisebox{-.5\height}{\includegraphics[width=\linewidth]{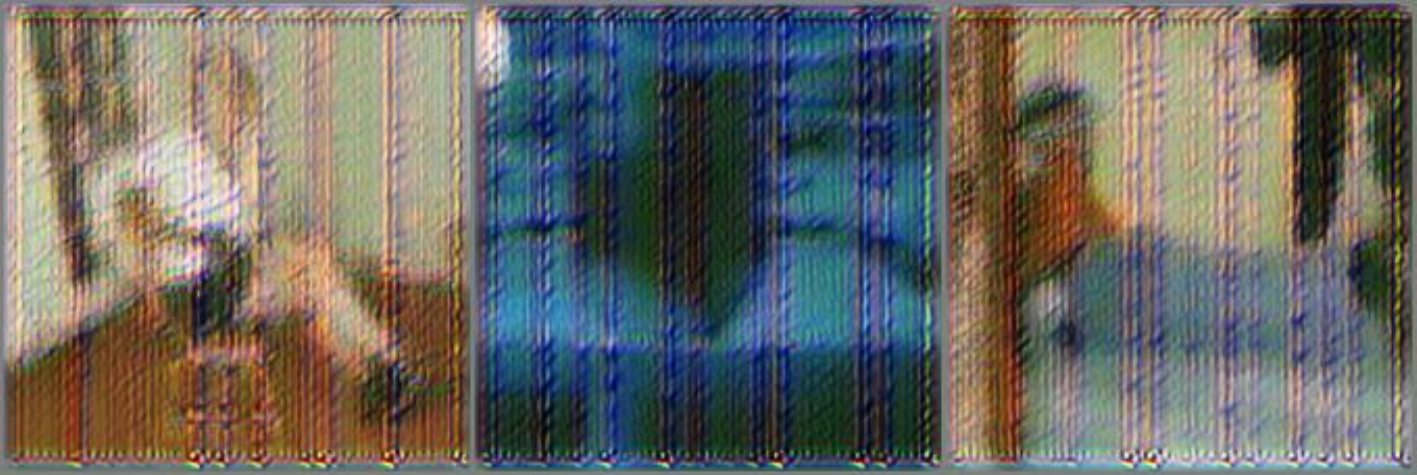}}
    	\end{minipage} & Lemon & \makecell{``te2ng xu4n ke1 ji4\\xu4n be3i ji1ng shi2\\jia1n 5 yue4 16 ri4 xia1o\\xi1 , ju4 guo2 wa4i...''} & Technology & 
            \begin{minipage}[b]{0.3\columnwidth}
    		\centering
    		\raisebox{-.5\height}{\includegraphics[width=\linewidth]{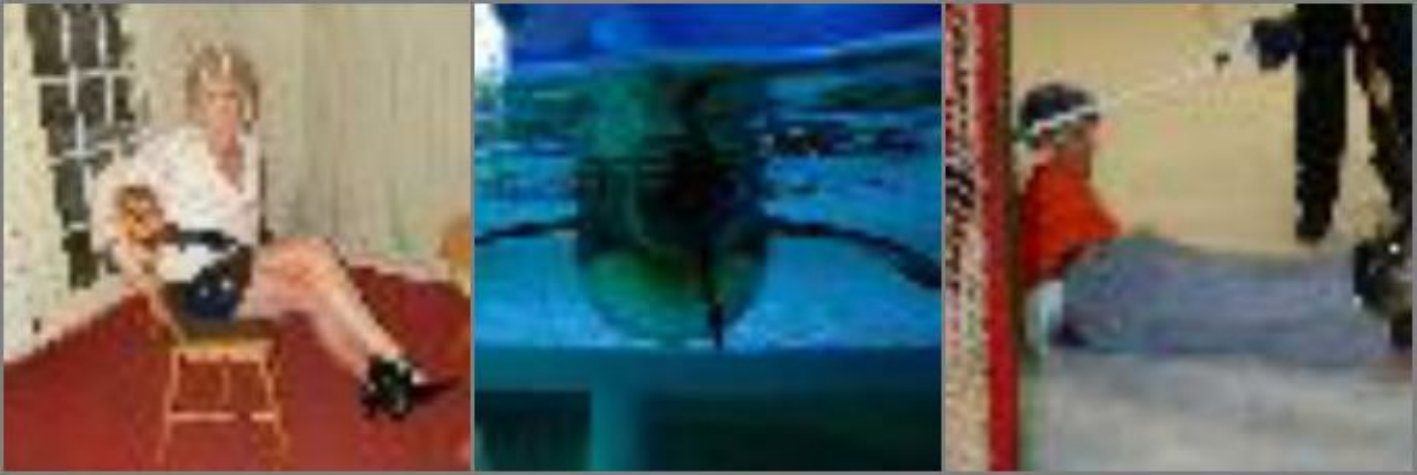}}
    	\end{minipage} & \makecell{Miniskirt,\\King penguin,\\Broom} \\
        \bottomrule
    \end{tabular}}
    \label{table:completevision appendix}
\end{table}

\begin{figure}[H]
\centering
\end{figure}
\begin{figure}[H]
\centering
\subfigure[Yelp\_MNIST\_fused.]{
\label{figure:yelp_mnist_fused}
\includegraphics[width=0.34\textwidth]{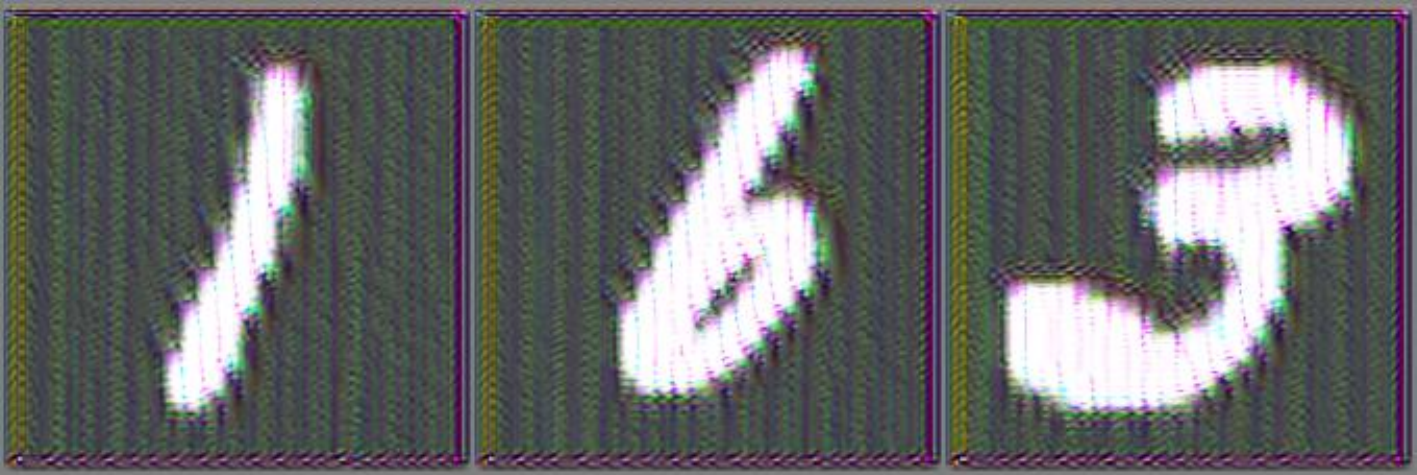}}
\subfigure[Yelp\_MNIST\_container.]{
\label{figure:yelp_mnist_container}
\includegraphics[width=0.34\textwidth]{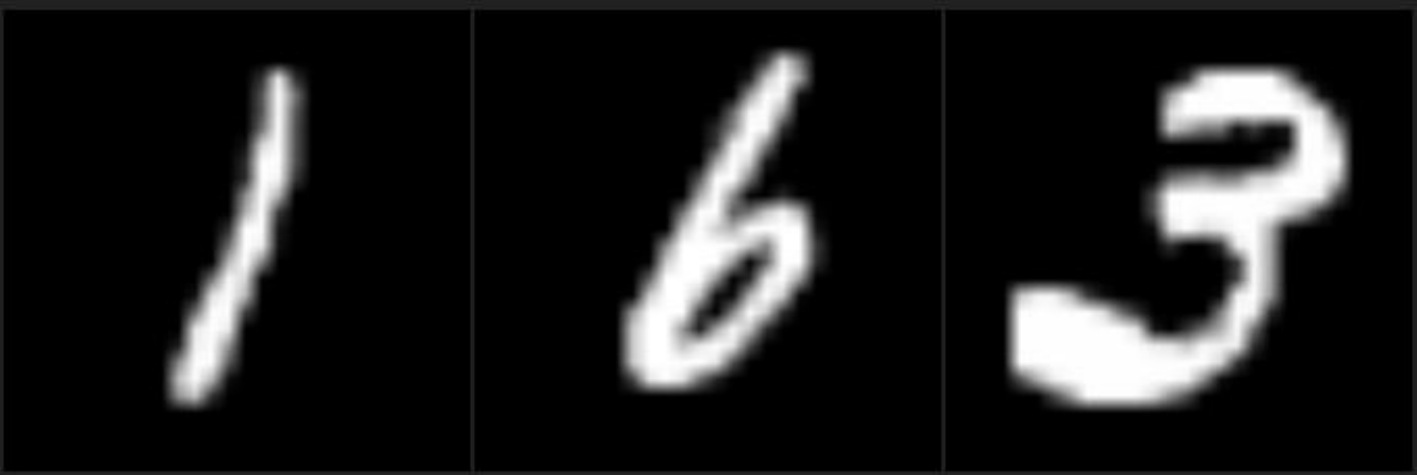}}
\end{figure}
\begin{figure}[H]
\centering
\subfigure[Sogou\_MNIST\_fused.]{
\label{figure:sogou_mnist_fused}
\includegraphics[width=0.34\textwidth]{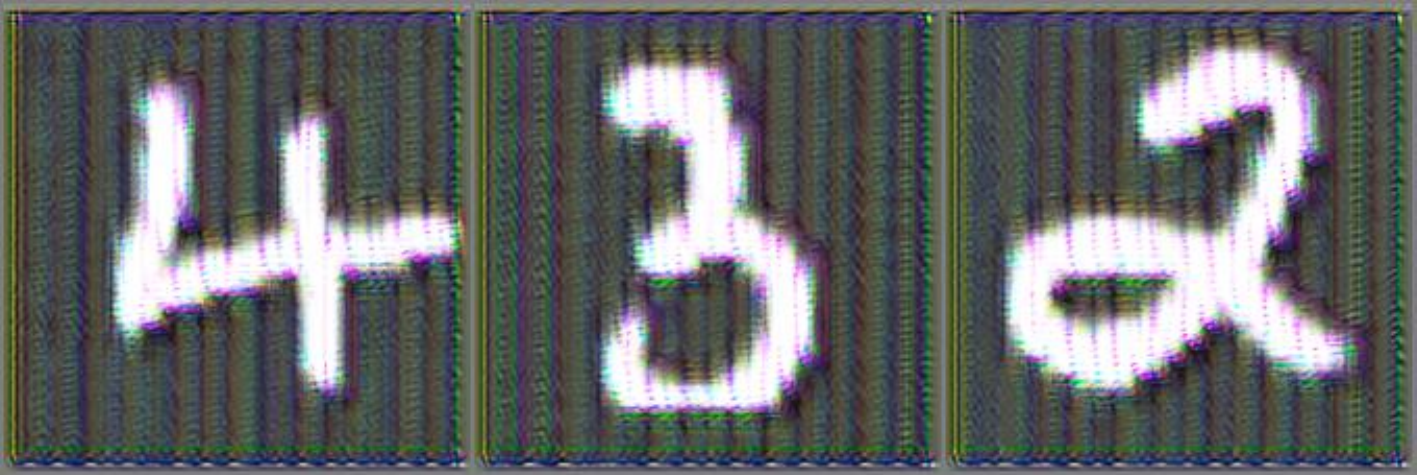}}
\subfigure[Sogou\_MNIST\_container.]{
\label{figure:sogou_mnist_container}
\includegraphics[width=0.34\textwidth]{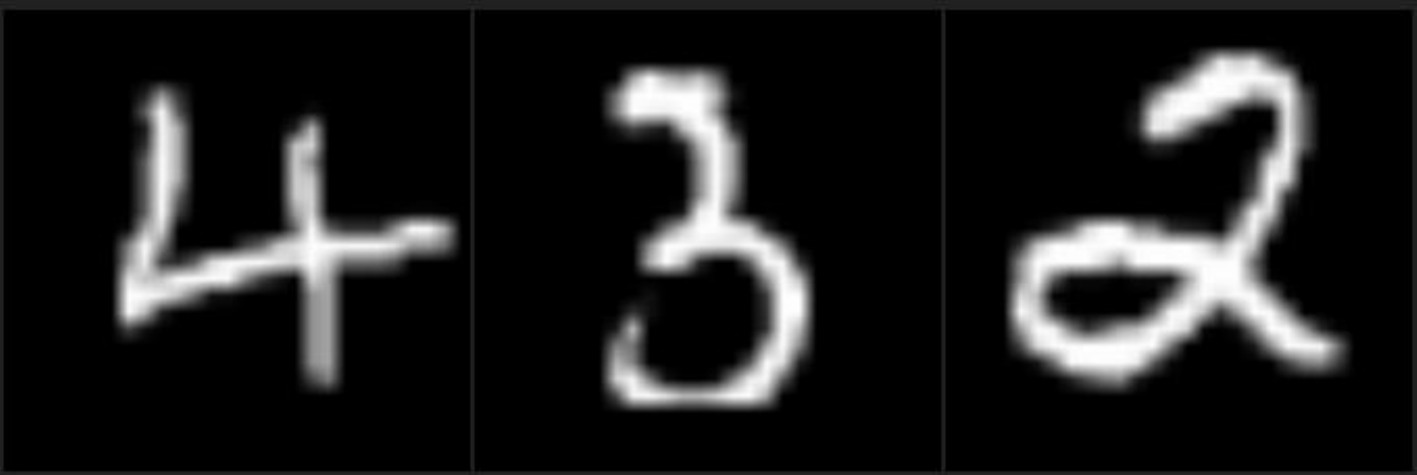}}
\end{figure}
\begin{figure}[H]
\centering
\subfigure[Yelp\_CIFAR-10\_fused.]{
\label{figure:yelp_cifar10_fused}
\includegraphics[width=0.34\textwidth]{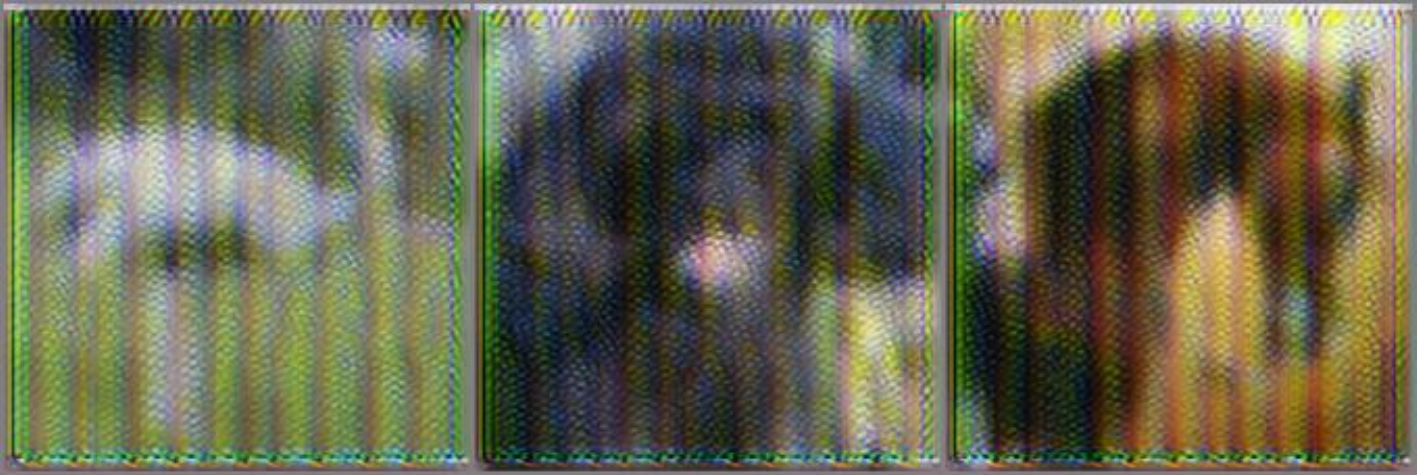}}
\subfigure[Yelp\_CIFAR-10\_container.]{
\label{figure:yelp_cifar10_container}
\includegraphics[width=0.34\textwidth]{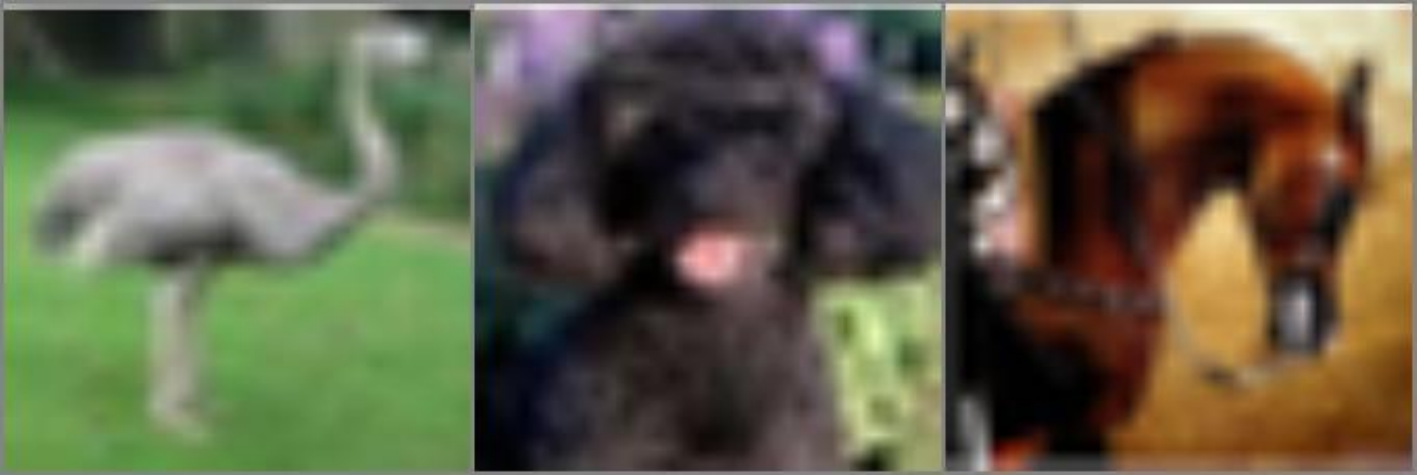}}
\end{figure}
\begin{figure}[H]
\centering
\subfigure[Sogou\_CIFAR-10\_fused.]{
\label{figure:sogou_cifar10_fused}
\includegraphics[width=0.34\textwidth]{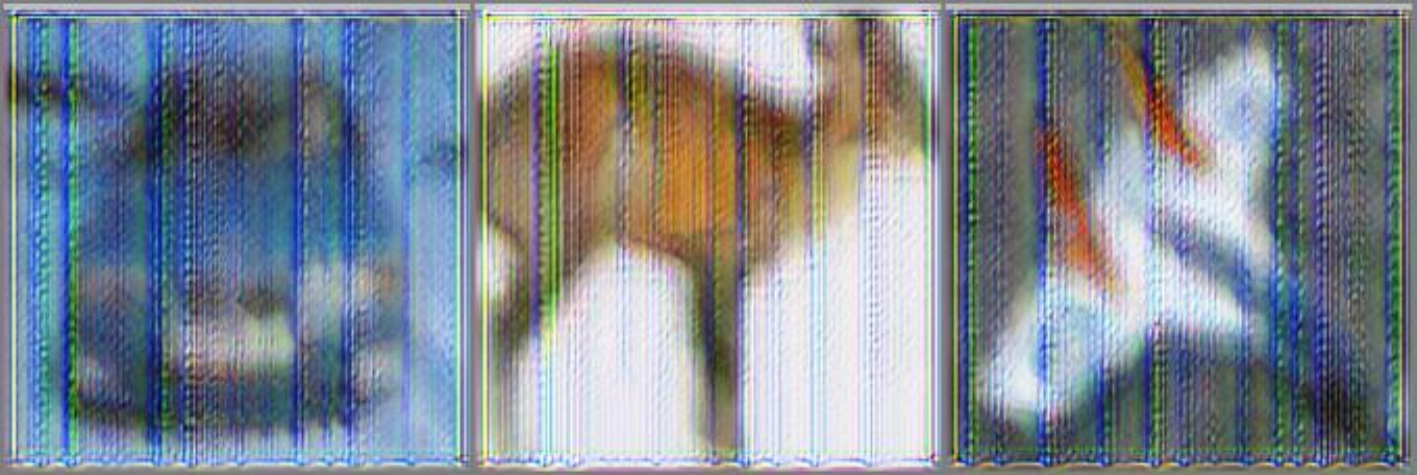}}
\subfigure[Sogou\_CIFAR-10\_container.]{
\label{figure:sogou_cifar10_container}
\includegraphics[width=0.34\textwidth]{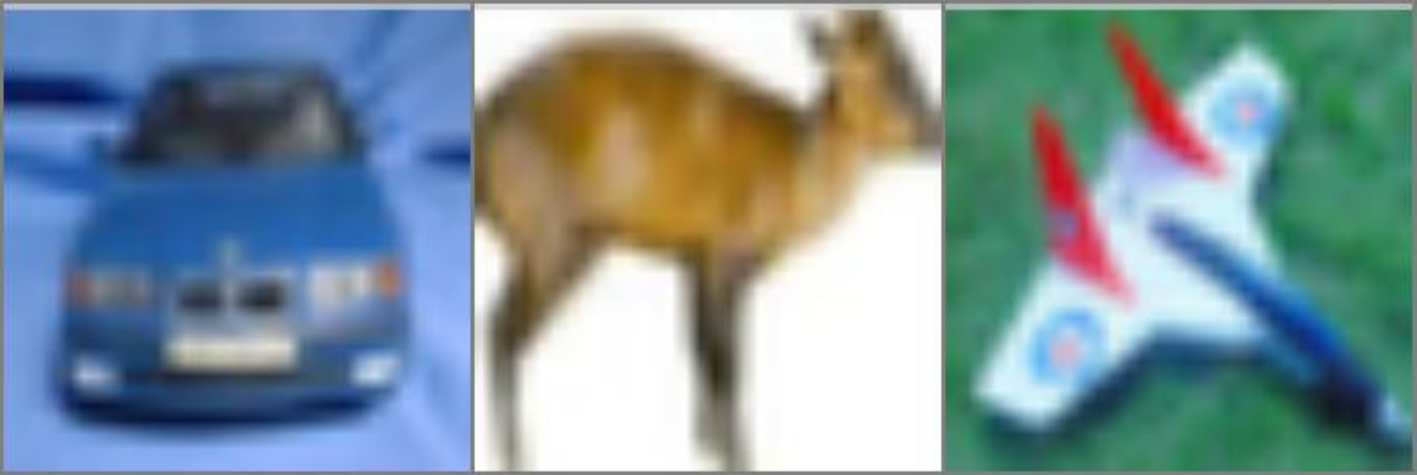}}
\end{figure}
\begin{figure}[H]
\centering
\subfigure[Yelp\_STL-10\_fused.]{
\label{figure:yelp_stl10_fused}
\includegraphics[width=0.34\textwidth]{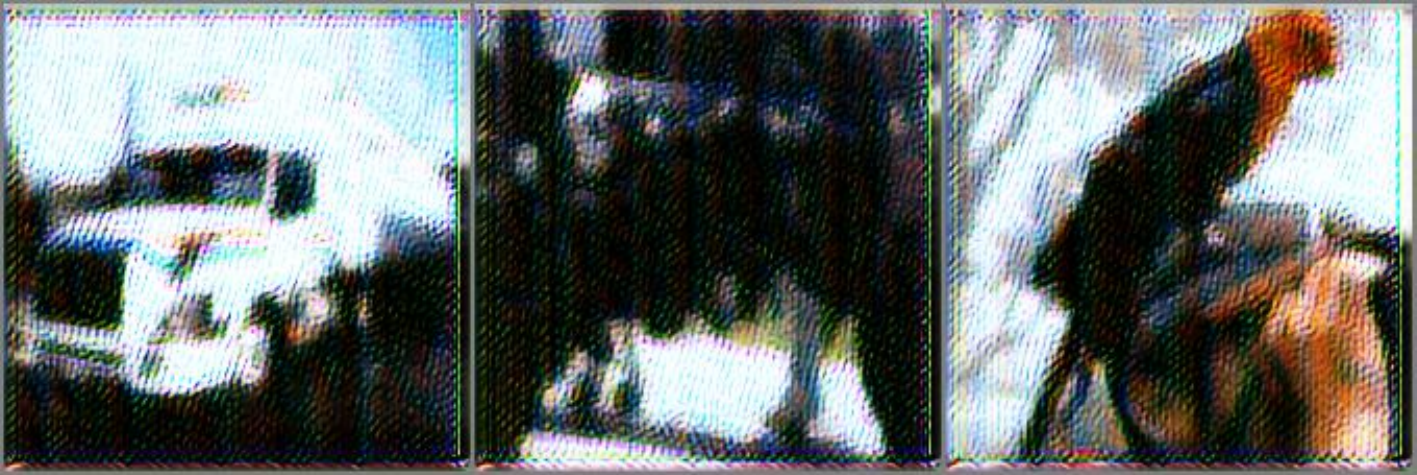}}
\subfigure[Yelp\_STL-10\_container.]{
\label{figure:yelp_stl10_container}
\includegraphics[width=0.34\textwidth]{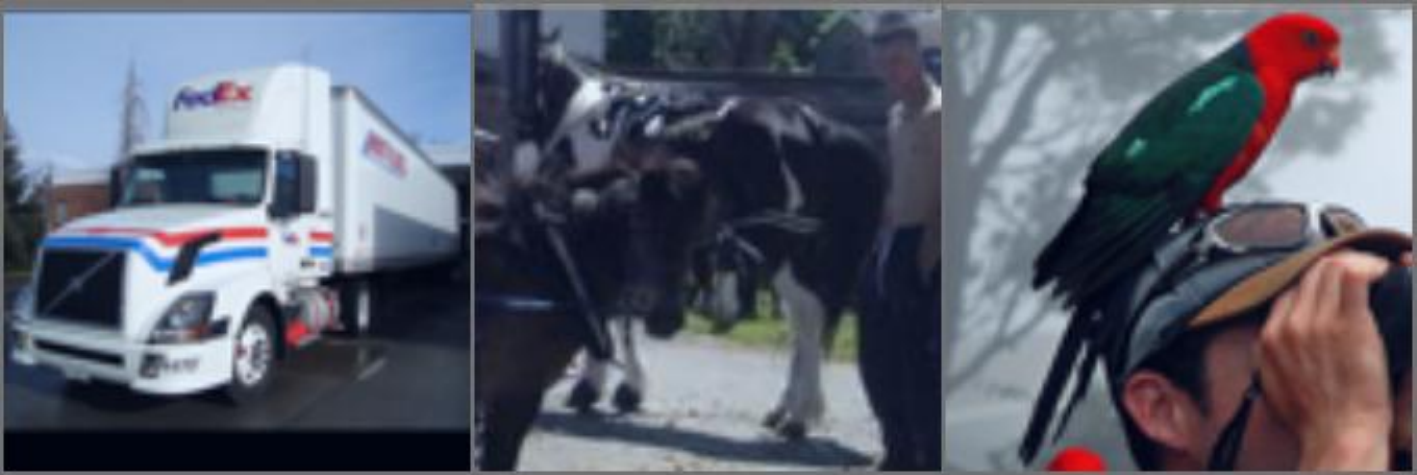}}
\end{figure}
\begin{figure}[H]
\centering
\subfigure[Sogou\_STL-10\_fused.]{
\label{figure:sogou_stl10_fused}
\includegraphics[width=0.34\textwidth]{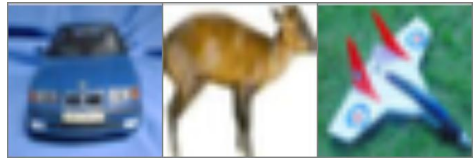}}
\subfigure[Sogou\_STL-10\_container.]{
\label{figure:sogou_stl10_container}
\includegraphics[width=0.34\textwidth]{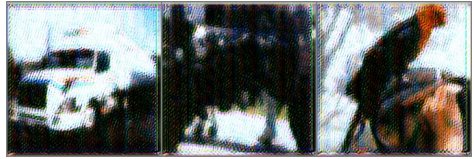}}
\caption{The scaled-up version of visual results in \autoref{table:completevision}.
We use $x\_y\_z$ notation to denote the hijacking dataset $x$ and the container dataset $y$, while $z$ indicates the image type, i.e., a fused or container image.}
\label{figure:completevision appendix}
\end{figure}

\end{document}